\documentclass[12pt,english]{article}
\usepackage{lmodern}
\usepackage[T1]{fontenc}
\usepackage[latin9]{inputenc}
\usepackage{amsmath}
\usepackage{amssymb}
\usepackage{esint}

\usepackage{geometry}
\makeatletter
\usepackage{float} 
\usepackage[colorlinks=true, linkcolor=ForestGreen, citecolor=blue]{hyperref}
\usepackage{cite}
\usepackage{babel}
\usepackage{mathdots}
\usepackage{amsfonts}
\usepackage{mathrsfs}
\usepackage{amsmath}
\usepackage{esint}
\usepackage{graphicx}
\usepackage{youngtab}
\usepackage{multicol}
\usepackage{multirow}
\usepackage{slashed}
\usepackage{simplewick}
\usepackage{braket}
\usepackage{bm}
\usepackage{relsize}
\usepackage[dvipsnames,table,xcdraw]{xcolor}
\usepackage{subfigure}
\usepackage{feynmp}
\usepackage{pifont}
\usepackage{cancel}
\allowdisplaybreaks

\makeatother

\usepackage[dvipsnames]{xcolor}

\begin{document}
\author{Wan-Zhe Feng\footnote{Email: vicf@tju.edu.cn},~~Ao Li\footnote{Email: leo\_6626@tju.edu.cn},~~Zong-Huan Ye\footnote{Email: y2953083702@tju.edu.cn},~~Zi-Hui Zhang\footnote{Email: zhangzh\_@tju.edu.cn}\\
\textit{\small{Center for Joint Quantum Studies and Department of Physics,}}\\
\textit{\small{School of Science, Tianjin University, Tianjin 300350, PR. China}}}
\title{Electroweak right-handed neutrino portal dark matter}

\date{}

\maketitle

\begin{abstract}

We study dark matter coupled to the Standard Model via electroweak scale right-handed neutrinos in a Type-I seesaw framework. We consider a minimal dark sector containing a fermion $\chi$ and a complex scalar $\phi$ whose only connection to the Standard Model is through renormalizable Yukawa interactions with right-handed Majorana neutrinos, thus realizing a neutrino portal after seesaw mixing. We discuss three representative realizations of electroweak right-handed neutrinos arising from the Type-I seesaw mechanism, spanning small, tiny, and ultraweak couplings to the Standard Model sector, so that the dark particles can either undergo secluded freeze-out or be produced via freeze-in. Instead of merely estimating the order of magnitude of the seesaw couplings, we use the Particle Swarm Optimization algorithm to obtain viable seesaw parameter sets consistent with neutrino data and other constraints, and then compute the coupled evolution of the dark particles and right-handed neutrinos, reproducing the observed dark matter relic abundance in representative benchmark scenarios. 
For freeze-out, dark matter depletion is controlled by coupled dark sector dynamics, requiring a full Boltzmann treatment for a reliable relic abundance. For freeze-in, internal dark interactions also alter the relic density: treating hidden particles as independent components with late decays added afterward can misestimate the abundance by $30\%$ or even $95\%$, depending on the interaction structure.
Electroweak right-handed neutrino portal dark matter thus provides a robust and predictive framework that tightly connects neutrino physics, heavy neutral lepton phenomenology, and the cosmological dark matter relic density, offering a well-motivated benchmark for complementary collider, neutrino, and cosmological probes at the high energy frontier.
%

\end{abstract}

\newpage{}

\hypersetup{colorlinks=black,linktocpage=true}
\tableofcontents
\vspace{0.5cm}

\section{Introduction}\label{sec:Intro}

The existence of non-baryonic dark matter (DM) is firmly established by a wide range
of astrophysical and cosmological observations, yet its particle nature remains unknown.
At the same time, the discovery of neutrino oscillations has demonstrated
that neutrinos are massive and mix, providing clear evidence for physics beyond the Standard Model (SM).
A natural question is whether the origin of neutrino masses
and the nature of dark matter can be linked in a unified framework.

From a theoretical perspective, the Type-I seesaw mechanism~\cite{Minkowski:1977sc,Ramond:1979py,Gell-Mann:1979vob,Yanagida:1979as,Mohapatra:1979ia}
is the simplest realization of the effective Weinberg dimension-5 operator for the neutrino mass generation:
light neutrino masses are generated by integrating out heavy right-handed Majorana neutrinos,
which are singlets of the SM gauge group,
without introducing additional $SU(2)_L$ multiplets.
The chargeless nature of heavy Majorana right-handed neutrinos allows for lepton number violation and provides a natural
setting for leptogenesis, while at the same time these singlet fermions can
serve as portals to dark sectors.
In contrast, Type-II and Type-III seesaw mechanisms require an extra scalar $SU(2)_L$ triplet Higgs
and a fermionic $SU(2)_L$ triplet, respectively.

If the Majorana masses of the right-handed neutrinos lie close to the electroweak scale,
rather than at some extremely high-energy scale, e.g., the GUT scale,
it is plausible that the origins of both the electroweak scale and the seesaw scale are interconnected.
From an experimental standpoint, this mass range is particularly attractive:
electroweak to TeV scale right-handed neutrinos are potentially accessible at high-energy colliders,
and the corresponding signatures, such as displaced vertices and lepton number violation~\cite{CMS:2015qur,CMS:2018iaf,CMS:2018jxx,CMS:2022hvh,ATLAS:2023tkz,CMS:2024xdq,ATLAS:2024rzi},
as well as deviations from unitarity in the lepton mixing matrix~\cite{Kozynets:2024xgt}, can be probed in current and future facilities.
Moreover, constraints on TeV scale right-handed neutrinos are often less stringent than those on very light
sterile neutrinos, which are tightly constrained by cosmology~\cite{Dolgov:2000jw,Zelko:2022tgf,Roach:2022lgo,Du:2025iow}, meson decays~\cite{NA62:2020mcv,NA62:2021bji,NA62:2025csa}, and precision low-energy experiments~\cite{Kollenberger:2024kfs,KATRIN:2025lph}.
Electroweak scale implementations of the seesaw mechanism are therefore both theoretically well motivated and phenomenologically interesting.

In parallel, the persistent absence of signals in direct detection experiments
challenges the standard WIMP picture, in which dark matter couples appreciably to quarks.
Neutrino portal scenarios offer a compelling alternative:
dark matter resides in a hidden sector and communicates with the SM almost
exclusively through right-handed neutrinos.
In such models, dark matter does not automatically couple to baryons at tree level, and the
elastic scattering cross section off nuclei can naturally lie far below the
sensitivity of current and near-future direct detection experiments.
At the same time, the same portal fields control both the neutrino
mass generation and the dark sector phenomenology, leading to a tightly interconnected structure.
Depending on the strength of the right-handed neutrino coupling to the SM and to the dark particles,
the hidden sector may be weakly or ultraweakly coupled to the SM~\cite{Li:2022bpp},
leading either to freeze-out evolution~\cite{Escudero:2016tzx,Escudero:2016ksa,Tang:2016sib,Bandyopadhyay:2018qcv,Blennow:2019fhy,Borah:2021pet,Coito:2022kif,Borah:2025fkd}
or to freeze-in production~\cite{Becker:2018rve,Bian:2018mkl,Cosme:2020mck,Du:2020avz,Chianese:2020khl,Liu:2022rst,Barman:2022scg,Liu:2023zah}.

In this work we revisit the Type-I seesaw mechanism with an emphasis on
electroweak scale right-handed neutrinos and their role as portals to a dark sector.
We consider a minimal dark sector containing a fermion $\chi$ and a complex scalar $\phi$,
stabilized by an appropriate symmetry,
and assume that its only renormalizable connection to the SM is provided by Yukawa couplings to right-handed neutrinos.
After seesaw mixing, dark sector particles interact with SM particles
through light and heavy neutrinos in the mass eigenstates.
Depending on the size of the Yukawa couplings and the seesaw parameters,
the dark sector may thermalize with the SM bath and undergo a conventional freeze-out,
or it may remain out of equilibrium and be populated via freeze-in.
In the latter case, both $\chi$ and $\phi$ can in principle contribute to the present-day DM abundance.

To organize the discussion, we identify three representative realizations of
electroweak scale right-handed neutrinos: the regular seesaw (Case-RS), the
structure cancellation scenario (Case-SC), and the split seesaw (Case-SS).
These cases span small, tiny, and ultraweak couplings of the
electroweak right-handed neutrinos to the SM, while all reproduce the observed
light neutrino masses and mixings. Using the Particle Swarm Optimization (PSO) algorithm,
we determine viable seesaw parameter sets for each case and construct benchmark models
compatible with neutrino oscillation data and other relevant constraints.
In this way, the right-handed neutrino couplings to the SM are determined precisely in each benchmark,
rather than being roughly estimated only at the order-of-magnitude level as in much of the existing literature.
We find that the evolution of all dark particles can be completely
determined by the seesaw mechanism together with the right-handed neutrino couplings to the dark sector.

We analyze freeze-out and freeze-in production in Cases RS, SC, and SS.
For the freeze-out case, we show that the dark matter abundance is typically
controlled by $\chi\overline{\chi}\to \mathbb{N}\mathbb{N}$,
or by resonant coannihilation $\chi\phi\to LH$,
and that a full coupled Boltzmann treatment of the dark sector is essential for obtaining the correct relic density.
For the freeze-in case,
we compare our full calculation with a simplified treatment
in which dark sector internal interactions are neglected and the various species are treated as independent freeze-in components.
We explicitly show that this approximation can misestimate the final dark matter abundance by about $30\%$
when the hidden sector possesses sufficiently strong three-point internal interactions,
and by more than $95\%$ when such interactions are absent,
demonstrating that a careful treatment of hidden sector dynamics is essential in right-handed neutrino portal freeze-in models.
A key aspect of our analysis is the treatment of the visible sector and the dark sector as a multi-temperature system~\cite{Aboubrahim:2020lnr,Aboubrahim:2021ycj,Feng:2023ubl,Feng:2024pab,Feng:2024nkh}.
When hidden sector interactions are strong enough,
$\chi$ and $\phi$ can reach internal thermal equilibrium at a hidden temperature
$T_h$ distinct from the visible temperature $T_v$, and the evolution must be
described by coupled Boltzmann equations for the comoving number densities and
the temperature ratio $\eta \equiv T_v/T_h$. When hidden sector interactions are
too weak, the dark sector never equilibrates and the standard freeze-in treatment can apply.

The remainder of this paper is organized as follows.
In Section~\ref{Sec:RHN}, we review the Type-I seesaw mechanism and its electroweak scale realizations,
and introduce the three representative cases considered in this work: Cases RS, SC, and SS.
In Section~\ref{Sec:PSO}, we summarize the relevant neutrino data and experimental constraints,
and describe how the PSO algorithm is used to determine viable seesaw parameter sets consistent with the observed light neutrino masses and mixings.
In Section~\ref{Sec:RHNP}, we introduce the right-handed neutrino portal dark sector and classify the relevant interactions.
In Section~\ref{Sec:Pheno}, we study the cosmological evolution of the dark sector particles for both freeze-out and freeze-in production, emphasizing the role of coupled dark sector dynamics in obtaining the correct relic abundance.
Finally, we present our conclusions and discuss future directions in Section~\ref{Sec:Con}.
Additional technical details are collected in the appendices.

\section{Type-I seesaw mechanism and generalizations}\label{Sec:RHN}

In this section, we provide a brief review of the Type-I seesaw mechanism and generalizations.
We concentrate on the simplest Type-I seesaw mechanism and its variations,
featuring three right-handed Majorana neutrinos,
with at least one residing in the electroweak region.
Electroweak scale right-handed neutrinos are intriguing due to their relatively
low mass, which increases the likelihood of detection at colliders,
and current constraints remain comparatively weak.
Electroweak right-handed neutrinos can also act as portals to dark sectors with $\mathcal{O}(100~{\rm GeV})$
dark matter candidates, placing them among the most promising targets for dark matter searches.

We begin with a general discussion of perspectives on new physics model building involving the Type-I seesaw mechanism.
Our notation and conventions are summarized in Appendix~\ref{App:Conven},
followed by a detailed one-generation derivation of the Type-I seesaw mixing using two-component spinor formalism in Appendix~\ref{App:SSdetail}.
We then discuss three distinct regions of Type-I seesaw parameter space in which
electroweak scale right-handed neutrinos have different coupling strengths to the SM.

In this paper, we use the PSO algorithm to determine viable seesaw parameter sets
for each case and construct benchmark models compatible with neutrino oscillation data and other relevant constraints,
and thus the right-handed neutrino couplings to the SM are determined precisely in each benchmark.

\subsection{A consistent Type-I seesaw mechanism revisit}

The Type-I seesaw Lagrangian is given by~\cite{Minkowski:1977sc,Ramond:1979py,Gell-Mann:1979vob,Yanagida:1979as,Mohapatra:1979ia}
\begin{equation}
\mathcal{L} \supset -y_{\alpha i}^{\nu}\,\overline{L_{\alpha}}\widetilde{H}P_{R}\mathbb{N}_{i}
- y_{\alpha i}^{\nu\,*}\, \overline{\mathbb{N}_{i}} P_L \widetilde{H}^\dagger L_\alpha
-\frac{M_{i}}{2}\,\overline{\mathbb{N}_{i}^{c}}\,\mathbb{N}_{i}\,,
\end{equation}
where $\widetilde{H}={\rm i}\sigma_{2}H$, and $\mathbb{N}_{i}$ are the Majorana right-handed neutrinos.
The corresponding quantum numbers are 
\begin{center}
\begin{tabular}{|c|c|c|c|}
\hline
Fields & $L$ & $H$ & $\mathbb{N}$\tabularnewline
\hline
$B-L$ & $-1$ & $0$ & $0$\tabularnewline
\hline
\end{tabular}
\par\end{center}

Under this assignment, Yukawa interactions $\overline{L_{\alpha}}\widetilde{H}P_{R}\mathbb{N}_{i} +h.c.$ violate $B-L$.
Hence, $B-L$ cannot be imposed as an exact symmetry, either global or local.
From this perspective, since the heavy Majorana fermions $\mathbb{N}_i$
carry no quantum numbers at all, it is natural to consider the possibility
that multiple (potentially far more than three) such heavy states
may exist at high energies, each with distinct couplings to the $LH$ operator.

An alternative interpretation of the Type-I seesaw Lagrangian involves
imposing $B-L$ as a gauge symmetry. In this case, the original Lagrangian
takes the form
\begin{equation}
\mathcal{L}\supset-y_{\alpha \beta}^{\nu}\,\overline{L_{\alpha}}\widetilde{H}P_{R}N_{\beta}
- y_{\alpha\beta}^{\nu\,*}\, \overline{{N}_{\beta}} P_L \widetilde{H}^\dagger L_\alpha
-y_{\alpha}^{N}\Phi\overline{N_{\alpha}^{c}}N_{\alpha}\,,
\end{equation}
where $N_{\alpha=1,2,3}$ are Dirac fermions carrying $B-L$ quantum number,
and $\Phi$ is the $U(1)_{B-L}$ Higgs field responsible for spontaneously
breaking the $B-L$ gauge symmetry at a high scale, commonly assumed
to be near the GUT scale. In this scenario, the charge assignments
under $U(1)_{B-L}$ are summarized as follows
\begin{center}
\begin{tabular}{|c|c|c|c|c|}
\hline
Fields & $L$ & $H$ & $N$ & $\Phi$\tabularnewline
\hline
$B-L$ & $-1$ & $0$ & $-1$ & $+2$\tabularnewline
\hline
\end{tabular}
\par\end{center}

From this perspective, Yukawa interactions $\overline{L_{\alpha}}\widetilde{H}P_{R}N_{\beta}+h.c.$
are $U(1)_{B-L}$ gauge invariant, while the Majorana mass term for $N$ arises
from the last term when $\Phi$ acquires its vacuum expectation value.
After the spontaneous breaking of the $U(1)_{B-L}$ symmetry, $N_{\alpha=1,2,3}$ become
the massive Majorana particles and, as a result, carry no quantum
numbers. In this scenario, the theory requires exactly 3 such right-handed
neutrinos, each originally carrying a $B-L$ charge $-1$,
to cancel the $U(1)_{B-L}$ gauge anomaly. It is natural to expect
that all three $N_{i}$ lie around the same mass scale, determined
by the vacuum expectation value of the $U(1)_{B-L}$ Higgs field.
Given this structure, any additional mixing between $LH$ and
other Majorana fermions appears unlikely.

In summary, the right-handed neutrino with a Majorana mass term can
be interpreted in two distinct frameworks:
\begin{enumerate}
\item \textbf{As a truly neutral Majorana fermion} with no conserved charges.
In this interpretation, no lepton-number-related symmetry, whether
global or gauged, can be imposed, since $\overline{L_{\alpha}}\widetilde{H}P_{R}\mathbb{N}_{i}$
explicitly violates both lepton number and $B-L$.
In general, \textbf{more than three} $\mathbb{N}_i$ fields may couple to $LH$.
\item \textbf{As a $(B-L)$-charged Dirac fermion} prior to the $B-L$ symmetry
breaking. In this case, the $B-L$ gauge symmetry is preserved at
very high energies, and the Majorana mass terms for $N_\alpha$ arise only
after spontaneous breaking of $U(1)_{B-L}$ symmetry, typically via
the vacuum expectation value of a scalar field.
\textbf{Precisely three} right-handed neutrinos are needed to cancel the $U(1)_{B-L}$ anomaly.
\end{enumerate}

It would be theoretically inconsistent to impose a low-energy $U(1)_{B-L}$ gauge symmetry while simultaneously allowing explicit Majorana mass terms for the right-handed neutrinos together with the Yukawa interaction $\overline{L}\widetilde{H}N$.
The Majorana mass terms require the right-handed neutrinos to be neutral under the $U(1)_{B-L}$ symmetry,
in which case the above Yukawa interactions explicitly violate $B-L$.
This construction is self-contradictory, since the Yukawa interactions break the very symmetry that the low-energy theory is intended to preserve.

A string theory interpretation of the seesaw mechanism is particularly
fruitful in the context of D-brane model building.
In this framework,
right-handed neutrinos can be realized either as chargeless Majorana
fermions arising from closed string modes~\cite{FWZ},
or as bi-fundamental Dirac fermions originating from open strings stretched between different
stacks of D-branes~\cite{Ibanez:2006da,Cvetic:2007ku,Antusch:2007jd,Cvetic:2008hi,Ibanez:2008my,Cvetic:2010mm}.
In D-brane constructions, the $U(1)$ charges
associated with individual D-branes can be consistently canceled through
the inclusion of appropriate instantons localized at brane intersections.
As a result, it is often possible to construct a variety of gauge-invariant
interactions without explicitly breaking the underlying symmetries.
Within string theory constructions,
the number of right-handed neutrinos is thus not restricted.

\subsection{SM neutrino masses from seesaw mixings}

With an illustrative detailed derivation for one-generation neutrino seesaw mixing given in Appendix~\ref{App:SSdetail},
the realistic seesaw scenario includes three generations of left-handed neutrinos,
with three right-handed Majorana neutrinos added to complete the model.\footnote{
The case where two right-handed Majorana neutrinos are added is also
discussed in the literature~\cite{Guo:2003cc,Mei:2003gn,Guo:2006qa,Nath:2018hjx,Chianese:2018dsz,Chianese:2019epo,Xing:2020ald},
leading to the generation of two light neutrinos and one exactly massless neutrino.
We will not discuss this case in this paper.}
The most general Lagrangian is written in the flavor eigenbasis as
\begin{align}
\mathcal{L}_{{\rm mass}} & =-y_{\alpha\beta}^{\nu}\overline{L_{\alpha}}\widetilde{H}P_{R}\mathbb{N}_{\beta}^{0} +h.c. -\frac{M_{\alpha}}{2}\overline{\mathbb{N}_{\alpha}^{0}}\mathbb{N}_{\alpha}^{0}
=\frac{m_{i}}{2}\overline{\mathbb{V}_{i}}\mathbb{V}_{i}\,,
\end{align}
where $\mathbb{V}_{i=1-6}$ denote mass eigenstates.
The masses $m_{1,2,3}$ are the light neutrino masses $\lesssim10^{-2}$~eV
which should match the observed neutrino data as will be detailed
in Section~\ref{sec:ex}, and $m_{4,5,6}$ are the heavy neutrino
masses with $m_{4}\approx M_{1},m_{5}\approx M_{2},m_{6}\approx M_{3}$.

The seesaw mixing transfers neutrinos from the flavor eigenbasis to the mass eigenbasis.
The six flavor eigenstates $\mathbb{V}_{i=1-6}^{0}$ are a combination
of 4-component Majorana fermions made by SM neutrinos
\begin{equation}
\mathbb{V}_{1}^{0}=\left(\begin{array}{c}
\nu_{e}\\
{\rm i}\sigma_{2}\nu_{e}^{*}
\end{array}\right)\,,\qquad\mathbb{V}_{2}^{0}=\left(\begin{array}{c}
\nu_{\mu}\\
{\rm i}\sigma_{2}\nu_{\mu}^{*}
\end{array}\right)\,,\qquad\mathbb{V}_{3}^{0}=\left(\begin{array}{c}
\nu_{\tau}\\
{\rm i}\sigma_{2}\nu_{\tau}^{*}
\end{array}\right)\,,
\end{equation}
and right-handed Majorana fermions
\begin{equation}
\mathbb{V}_{4}^{0}=\mathbb{N}_{1}^{0}=\left(\begin{array}{c}
-{\rm i}\sigma_{2}{N_{1}^{0}}^{*}\\
N_{1}^{0}
\end{array}\right)\,,\qquad\mathbb{V}_{5}^{0}=\mathbb{N}_{2}^{0}=\left(\begin{array}{c}
-{\rm i}\sigma_{2}{N_{2}^{0}}^{*}\\
N_{2}^{0}
\end{array}\right)\,,\qquad\mathbb{V}_{6}^{0}=\mathbb{N}_{3}^{0}=\left(\begin{array}{c}
-{\rm i}\sigma_{2}{N_{3}^{0}}^{*}\\
N_{3}^{0}
\end{array}\right)\,,
\end{equation}
transform to the mass eigenstates $\mathbb{V}_{i=1-6}$
using a $6\times 6$ unitary matrix $\mathbf{U}$ and vice versa:
\begin{equation}
\mathbb{V}_{i}=\mathbf{U}^{-1}\mathbb{V}_{i}^{0}\,,\qquad{\rm or}\qquad\mathbb{V}_{i}^{0}=\mathbf{U}\mathbb{V}_{i}\,,\label{eq:VTran}
\end{equation}
and the upper-left $3\times3$ block matrix of the matrix $\mathbf{U}$
should reproduce the neutrino oscillation data, conventionally encoded in the Pontecorvo-Maki-Nakagawa-Sakata (PMNS) matrix,\footnote{
The unitary PDG parameterization of $U_{\rm PMNS}$ is given in Appendix~\ref{App:PMNS}.} up to allowed non-unitarity effects,
\begin{equation}
(\begin{array}{ccc}
\nu_{e} & \nu_{\mu} & \nu_{\tau}\end{array})^{T}=U_{{\rm PMNS}}(\begin{array}{ccc}
\nu_{1} & \nu_{2} & \nu_{3}\end{array})^{T}\,.
\end{equation}
Thus the neutrinos in the mass eigenbasis are written explicitly in
terms of 4-component Majorana spinors
\begin{equation}
\mathbb{V}_{1}=\left(\begin{array}{c}
\nu_{1}\\
{\rm i}\sigma_{2}\nu_{1}^{*}
\end{array}\right)\,,\qquad\mathbb{V}_{2}=\left(\begin{array}{c}
\nu_{2}\\
{\rm i}\sigma_{2}\nu_{2}^{*}
\end{array}\right)\,,\qquad\mathbb{V}_{3}=\left(\begin{array}{c}
\nu_{3}\\
{\rm i}\sigma_{2}\nu_{3}^{*}
\end{array}\right)\,,\label{eq:LNu}
\end{equation}
which correspond to the three light active neutrino mass eigenstates,
although their Majorana nature has not yet been experimentally confirmed; and
\begin{equation}
\mathbb{V}_{4}\equiv \mathbb{N}_{1}=\left(\begin{array}{c}
N_{1}\\
{\rm i}\sigma_{2}N_{1}^{*}
\end{array}\right)\,,\qquad\mathbb{V}_{5}\equiv \mathbb{N}_{2}=\left(\begin{array}{c}
N_{2}\\
{\rm i}\sigma_{2}N_{2}^{*}
\end{array}\right)\,,\qquad\mathbb{V}_{6}\equiv \mathbb{N}_{3}=\left(\begin{array}{c}
N_{3}\\
{\rm i}\sigma_{2}N_{3}^{*}
\end{array}\right)\,,\label{eq:HNu}
\end{equation}
are the (not-yet-observed) heavy Majorana neutrinos.

After the seesaw mixing, charged current interactions are modified:
\begin{align}
\mathcal{L}_{W^{+}} & =\frac{g_{2}}{\sqrt{2}}W_{\mu}^{+}\overline{\nu_{\alpha L}}\gamma^{\mu}\ell_{\alpha L}
=\frac{g_{2}}{\sqrt{2}}W_{\mu}^{+}\overline{V_{{\rm SM}\,\alpha}}P_{R}\gamma^{\mu}\ell_{\alpha L}
=\frac{g_{2}}{\sqrt{2}}W_{\mu}^{+}\overline{\mathbb{V}_{\alpha}^{0}}P_{R}\gamma^{\mu}\ell_{\alpha L}\nonumber\\
 & =\frac{g_{2}}{\sqrt{2}}\sum_{i=1}^{6}U_{\alpha i}\ W_{\mu}^{+}\overline{\mathbb{V}_{i}}\gamma^{\mu}P_{L}\ell_{\alpha L} \label{eq:SMWp}\,,\\
\mathcal{L}_{W^{-}} & =\frac{g_{2}}{\sqrt{2}}W_{\mu}^{-}\overline{\ell_{\alpha L}}\gamma^{\mu}\nu_{ \alpha L}
=\frac{g_{2}}{\sqrt{2}}W_{\mu}^{-}\overline{\ell_{\alpha L}}\gamma^{\mu}P_{L}V_{{\rm SM}\,\alpha}
=\frac{g_{2}}{\sqrt{2}}W_{\mu}^{-}\overline{\ell_{\alpha L}}\gamma^{\mu}P_{L}\mathbb{V}_{\alpha}^{0}\nonumber\\
 & =\frac{g_{2}}{\sqrt{2}}\sum_{i=1}^{6}U_{\alpha i}\ W_{\mu}^{-}\overline{\ell_{\alpha L}}\gamma^{\mu}P_{L}\mathbb{V}_{i}\,,\label{eq:SMWm}
\end{align}
and the neutral current coupling is also modified as
\begin{align}
\mathcal{L}_{Z} & =\frac{g_{2}}{2\cos\theta_{W}}Z_{\mu}\overline{\nu_{\alpha L}}\gamma^{\mu}\nu_{\alpha L}
=\frac{g_{2}}{2\cos\theta_{W}}Z_{\mu}\overline{V_{{\rm SM}\,\alpha}}\gamma^{\mu}P_{L}V_{{\rm SM}\,\alpha}
=\frac{g_{2}}{2\cos\theta_{W}}Z_{\mu}\overline{\mathbb{V}_{\alpha}^{0}}\gamma^{\mu}P_{L}\mathbb{V}_{\alpha}^{0}\nonumber \\
 & =\frac{g_{2}}{2\cos\theta_{W}}Z_{\mu}\left(\sum_{i=1}^{6}U_{\alpha i}\overline{\mathbb{V}_{i}}\right)\gamma^{\mu}P_{L}\left(\sum_{j=1}^{6}U_{\alpha j}\mathbb{V}_{j}\right)\nonumber \\
 & =\frac{g_{2}}{2\cos\theta_{W}}\sum_{i=1}^{6}\sum_{j=1}^{6}U_{\alpha i}U_{\alpha j}\ Z_{\mu}\overline{\mathbb{V}_{i}}\gamma^{\mu}P_{L}\mathbb{V}_{j}\,.
\label{eq:SMZ}
\end{align}
The Higgs couplings involving neutrinos are written as:
\begin{align}
\mathcal{L}_{H} & =-\frac{1}{\sqrt{2}}y_{\alpha \beta}^{\nu}h\,\big(\nu_{\alpha}^{\dagger}N_{\beta}+N_{\beta}^{\dagger}\nu_{\alpha}\big)
=-\frac{1}{\sqrt{2}}y_{\alpha \beta}^{\nu}h\,
\Big(\overline{\mathbb{V}_{\alpha}^{0}}P_{R}\mathbb{N}_{\beta}^{0}
+\overline{\mathbb{N}_{\beta}^{0}}P_{L}\mathbb{V}_{\alpha}^{0}\Big)\nonumber \\
 & =-\frac{1}{\sqrt{2}}y_{\alpha \beta}^{\nu}h\,
 \Big(\sum_{i=1}^6 U_{\alpha i}\overline{\mathbb{V}_{i}}\,P_{R} \sum_{j=1}^6 U_{\beta+3, j}\mathbb{V}_{j}
 +\sum_{j=1}^6 U_{\beta+3, j}\overline{\mathbb{V}_{j}} \,P_{L} \sum_{i=1}^6 U_{\alpha i}\mathbb{V}_{i} \Big)\nonumber \\
 & =-\frac{1}{\sqrt{2}}y_{\alpha \beta}^{\nu}h\,
 \Big(\sum_{i=1}^6 \sum_{j=1}^6 U_{\alpha i} U_{\beta+3, j} \overline{\mathbb{V}_{i}}\,P_{R} \mathbb{V}_{j}
 + \sum_{i=1}^6 \sum_{j=1}^6 U_{\alpha i} U_{\beta+3, j} \overline{\mathbb{V}_{j}} \,P_{L} \mathbb{V}_{i} \Big)\,.
\label {eq:SMH}
\end{align}
Here and in what follows, $U$ with Latin indices running from $1$ to $6$ denotes the matrix elements of
the $6\times 6$ unitary transformation matrix $\mathbf{U}$.
Greek indices $\alpha,\beta$ denote flavors running from $1$ to $3$.

\subsection{Electroweak right-handed neutrino from Type-I seesaw}\label{Sec:Seesaw}

In this subsection, we summarize three simple realizations of the Type-I seesaw mechanism involving only three additional electroweak scale right-handed Majorana neutrinos. The corresponding Yukawa couplings $y^\nu$ can span ultraweak, tiny, and small values, leading to quantitatively different thermal behaviors of the electroweak right-handed neutrinos.

\subsubsection{Case-RS: right-handed neutrino from Regular Type-I seesaw}\label{sec:CaseRS}

The relevant Lagrangian for the regular seesaw mass mixing reads
\begin{equation}
\mathcal{L}_{{\rm mass}}=-y_{\alpha \beta}^{\nu}\overline{L_{\alpha}}\widetilde{H}P_{R}\mathbb{N}_{\beta}^0+h.c.-\frac{M_{\alpha}}{2}\overline{\mathbb{N}_{\alpha}^0}\mathbb{N}_{\alpha}^0\,,\label{eq:RSL}
\end{equation}
Hereafter, boldface symbols denote matrices.
In the basis of two-component spinors
\begin{equation}
\left(\begin{array}{cc}
\mathbf{V_{0}}^{T} & {\bf \mathbf{N}}^{T}\end{array}\right)=\left(\begin{array}{cccccc}
\nu_{e} & \nu_{\mu} & \nu_{\tau} & {N}_{1}^0 & {N}_{2}^0 & {N}_{3}^0\end{array}\right)
\end{equation}
where $\mathbf{V_{0}}\equiv(\begin{array}{ccc}
\nu_{e} & \nu_{\mu} & \nu_{\tau}\end{array})^{T}$ and ${\bf \mathbf{N}}\equiv(\begin{array}{ccc}
{N}_{1}^0 & {N}_{2}^0 & {N}_{3}^0\end{array})^{T}$ are $3\times1$ column matrices consisting of two-component spinors
in the original flavor basis (left-handed SM neutrinos and right-handed
component of the Majorana neutrinos), the mass terms are expressed
as
\begin{equation}
-\mathcal{L}_{{\rm mass}}=\frac{1}{2}(\mathbf{V_{0}}^{\dagger}\mathbf{M}_{D}{\bf \mathbf{N}}+\mathcal{{\bf \mathbf{N}}}^{\dagger}\mathbf{M}_{D}^{T}\mathbf{V_{0}}+\mathbf{V_{0}}^{c\dagger}\mathbf{M}_{D}\mathcal{{\bf \mathbf{N}}}^{c}+\mathcal{{\bf \mathbf{N}}}^{c\dagger}\mathbf{M}_{D}^{T}\mathbf{V_{0}}^{c})+\frac{1}{2}\mathbf{M}_{N}(\mathcal{{\bf \mathbf{N}}}^{c\dagger}\mathcal{{\bf \mathbf{N}}}+\mathcal{{\bf \mathbf{N}}}^{\dagger}\mathcal{{\bf \mathbf{N}}}^{c})\,,\label{eq:RSMF}
\end{equation}
where the $3\times3$ mass matrix $\mathbf{M}_{D}$ is defined as
$\mathbf{M}_{D\,\alpha \beta}=\frac{1}{\sqrt{2}}y^\nu_{\alpha \beta}v$, and $\mathbf{M}_{N}={\rm diag}\{M_{\alpha}\}$.
The seesaw mixing matrix is thus written as
\begin{equation}
\left(\begin{array}{cc}
0 & \mathbf{M}_{D}\\
\mathbf{M}_{D}^{T} & \mathbf{M}_{N}
\end{array}\right)_{6\times6}\,,\label{eq:RSMM}
\end{equation}
which is a direct generalization of Eq.~(\ref{eq:1GMM}) for the one-generation
neutrino model. For Case-RS, the sub-eV light neutrino masses can be roughly estimated as
\begin{equation}
m_{\alpha}\sim\frac{(y_\alpha^{\nu})^2\,v^{2}}{2M_{\alpha}} \label{eq:RSYuest}
\end{equation}
and two cases are most commonly discussed
\begin{itemize}
\item The Majorana masses $M_{\alpha}$ lie near the GUT scale $\sim10^{14-15}$ GeV,
with Yukawa couplings of $\mathcal{O}(1)$.
This is commonly regarded as the prototype of the Type-I seesaw mechanism, featuring natural $\mathcal{O}(1)$ Yukawa couplings at the GUT scale.
\item The Majorana masses $M_\alpha$ lie at the electroweak--TeV scale,
potentially yielding collider-accessible right-handed neutrinos,
with Yukawa couplings to the SM of order $10^{-7} - 10^{-6}$.
Such small couplings bring the right-handed neutrinos into equilibrium with the SM
only when the cosmic temperature drops to below TeV.
\end{itemize}
We will focus on the electroweak scale right-handed neutrinos in this paper.

\subsubsection{Case-SC: right-handed neutrino from Structure Cancellation}

This case has exactly the same form of the mass matrix as Case-RS,
i.e., Eq.~(\ref{eq:RSMM}), while the $3\times3$ mixing matrix, relabeled
as $\widetilde{\mathbf{M}}_{D}$, has a specific structure~\cite{Ingelman:1993ve,Kersten:2007vk}
\begin{equation}
\widetilde{\mathbf{M}}_{D}\equiv\frac{v}{\sqrt{2}}\mathbf{Y}+\epsilon\mathbf{X}=\frac{v}{\sqrt{2}}\left(\begin{array}{ccc}
Y_{1} & Y_{2} & Y_{3}\\
A Y_{1} & A Y_{2} & A Y_{3}\\
B Y_{1} & B Y_{2} & B Y_{3}
\end{array}\right)+\epsilon\mathbf{X}\,,\label{eq:MMCRS}
\end{equation}
where $\epsilon$ is a tiny real parameter, $A,B$ are real
numbers, and $\mathbf{X}$ is a $3\times3$ matrix with entry values
within $[-1,1]$. $Y_{\alpha}$ are the neutrino Yukawa couplings
which need to satisfy a further constraint:
\begin{equation}
{\sum}\frac{Y_{\alpha}^{2}}{M_{\alpha}}=0\,. \label{eq:SCCon}
\end{equation}
With the above condition, the rank-1 Yukawa matrix $\mathbf{Y}$ renders the $6\times6$ seesaw mixing matrix rank-3,
implying three exactly massless neutrinos, if the $\epsilon\mathbf{X}$ term in Eq.~(\ref{eq:MMCRS}) is absent.
Turning on $\epsilon\mathbf{X}$ lifts these zero modes and generates tiny masses for the three light neutrinos.

Although fine-tuning is required to obtain sub-eV light neutrino masses,
this is the only case in which electroweak scale right-handed neutrinos can have sizable couplings of order $\mathcal{O}(10^{-3})$ to the SM.

\subsubsection{Case-SS: right-handed neutrino from Split Seesaw}

The minimal seesaw~\cite{Guo:2003cc,Mei:2003gn,Guo:2006qa,Nath:2018hjx,Chianese:2018dsz,Chianese:2019epo,Xing:2020ald} contains only two heavy right-handed Majorana neutrinos
and can already produce the observed neutrino masses and mixings,
leaving the lightest neutrino exactly massless.
It is possible, however, to add a third right-handed Majorana (assumed to be the first-generation right-handed neutrino $\mathbb{N}_{1}$)
with an arbitrary mass,
from eV up to above the TeV scale, while coupling ultraweakly to the SM through a Yukawa interaction $y_{1}^{\nu}\,\overline{\mathbb{N}_{1}}\,L H$, with $y_{1}^{\nu}$ much smaller than in the conventional seesaw estimated by Eq.~\eqref{eq:SeesawY}.
Although $\mathbb{N}_{1}$ couples ultraweakly  to the SM, it endows the lightest neutrino with a nonzero mass.
We refer to this as the \textbf{split seesaw} (Case-SS).
The Lagrangian remains that of Eq.~(\ref{eq:RSL}), but with all $y_{i1}$ (the couplings of $\mathbb{N}_{1}$ to $L H$) taken to be small.
For simplicity, we assume that only the lightest $\mathbb{N}_{1}$, the one ultraweakly coupled to the SM,
acts as the portal to the dark sector,\footnote{
In this scenario, $\mathbb{N}_{1}$ may itself constitute the dark matter if its couplings to the SM are sufficiently tiny,
as explored in~\cite{Lucente:2021har,Datta:2021elq,Abada:2025gvc}.}
while $\mathbb{N}_{2}, \mathbb{N}_{3}$ do not couple to the dark sector.

This case is equivalent to what is often termed the minimal extended seesaw~\cite{Barry:2011wb,Zhang:2011vh,Jung:2022bgu}:
one adds a sterile gauge-singlet Majorana field $\mathbb{S}^{T}=\bigl(-{\rm i}\sigma_{2}S^{*},\,S\bigr)$ with mass $m_{S}$.
The motivation is to obtain a sterile neutrino $\mathbb{S}$ with essentially arbitrary mass (from eV up to above the TeV scale)
in addition to the two or three heavy right-handed Majorana neutrinos of the ordinary seesaw,
while keeping its couplings to the SM minimal.
In the minimal extended seesaw, $\mathbb{S}$ does not couple directly to $LH$ (the Yukawa $\overline{\mathbb{S}}\,LH$ is set to zero),
but it mixes with the seesaw states $\mathbb{N}_i$ through mass terms.
Thus, in addition to Eq.~(\ref{eq:RSL}), the Lagrangian contains
\begin{equation}
\mathcal{L}_{{\rm mass}}^{{\rm ME}}
=-\frac{M_{Si}}{2}\,\overline{\mathbb{N}_{i}}\mathbb{S}-\frac{M_{Si}}{2}\,\overline{\mathbb{S}}\mathbb{N}_{i}
-\frac{m_{S}}{2}\,\overline{\mathbb{S}}\mathbb{S}\,.
\end{equation}
where the original minimal extended seesaw often takes $m_{S}=0$, whereas here we allow $m_{S}\neq 0$.
In the basis of two-component spinors
\begin{equation}
\left(\begin{array}{ccc}
\mathbf{V_{0}}^{T} & {\bf \mathbf{N}}^{T} & S\end{array}\right)=\left(\begin{array}{cccccc}
\nu_{e} & \nu_{\mu} & \nu_{\tau} & {N}_{1}^0 & {N}_{2}^0 & S\end{array}\right)\,,
\end{equation}
the mass terms are expressed as
\begin{align*}
-\mathcal{L}_{{\rm mass}}^{{\rm ME}}= & \frac{1}{2}(S^{\dagger}\mathbf{M}_{S}^{T}{\bf \mathbf{N}}^{c}+\mathcal{{\bf \mathbf{N}}}^{\dagger}\mathbf{M}_{S}S^{c}+S^{c\dagger}\mathbf{M}_{S}^{T}{\bf \mathbf{N}}+\mathcal{{\bf \mathbf{N}}}^{c\dagger}\mathbf{M}_{S}S)+\frac{1}{2}m_{S}(S^{c\dagger}S+S^{\dagger}S^{c})\,,
\end{align*}
giving rise to the seesaw mixing matrix
\begin{equation}
\left(\begin{array}{ccc}
0 & \mathbf{M}_{D} & 0\\
\mathbf{M}_{D}^{T} & \mathbf{M}_{N} & \mathbf{M}_{S}\\
0 & \mathbf{M}_{S}^{T} & m_{S}
\end{array}\right)_{6\times6}\,,
\end{equation}
where $\mathbf{M}_{S}=(\begin{array}{cc}
M_{S1} & M_{S2}\end{array})^{T}$ is a $2\times1$ column matrix.

In such scenarios, after the diagonalization of the mass matrix of
all Majorana fermions, including all $\mathbb{N}_{i}$ and $\mathbb{S}$,
the case goes back to the Split Seesaw.
In the mass eigenbasis, the mostly-$\mathbb{S}$ Majorana fermion couples to the SM only with tiny coupling,
proportional to $M_{Si}/M_{i}$ where $M_{i}$ are the masses of $\mathbb{N}_{i}$.

\subsubsection{Seesaw parameter sets}\label{sec:SSP}

As shown in the detailed derivation in Appendix~\ref{App:SSdetail}, and as discussed in Section~\ref{sec:CaseRS}, obtaining a light neutrino mass $m_\nu\sim\mathcal{O}(10^{-2}~\mathrm{eV})$ within the \emph{regular Type-I seesaw mechanism} leads to the rough estimate, c.f., Eq.~\eqref{eq:RSYuest},
\begin{equation}
y^\nu_\alpha \approx \sqrt{2 m_\nu M_\alpha} \big/ v
\approx \sqrt{3\times M_\alpha [{\rm GeV}]}\times 10^{-8}\,,
\label{eq:SeesawY}
\end{equation}
where $M_\alpha$ denotes the mass of the heavy right-handed neutrino and $y^\nu_\alpha$ its Yukawa coupling to the SM.
This implies that GUT scale right-handed neutrinos typically correspond to Yukawa couplings of order $\mathcal{O}(1)$,
whereas electroweak scale right-handed neutrinos typically correspond to Yukawa couplings of order $\mathcal{O}(10^{-7})$.

The seesaw mixing matrix can be block-diagonalized using the Schur
complement
\begin{align}
\left(\begin{array}{cc}
0 & \mathbf{M}_{D}\\
\mathbf{M}_{D}^{T} & \mathbf{M}_{N}
\end{array}\right) & \rightarrow\left(\begin{array}{cc}
\mathbf{1}_{3\times3} & -\mathbf{M}_{D}\mathbf{M}_{N}^{-1}\\
0 & \mathbf{1}_{3\times3}
\end{array}\right)\left(\begin{array}{cc}
0 & \mathbf{M}_{D}\\
\mathbf{M}_{D}^{T} & \mathbf{M}_{N}
\end{array}\right)\left(\begin{array}{cc}
\mathbf{1}_{3\times3} & 0\\
-\mathbf{M}_{N}^{-1}\mathbf{M}_{D}^{T} & \mathbf{1}_{3\times3}
\end{array}\right)\nonumber \\
 & =\left(\begin{array}{cc}
-\mathbf{M}_{D}\mathbf{M}_{N}^{-1}\mathbf{M}_{D}^{T} & 0\\
0 & \mathbf{M}_{N}
\end{array}\right)\,, \label{eq:blockdiag}
\end{align}
for the seesaw limit, where the characteristic scale of $\mathbf{M}_N$ is much larger than that of $\mathbf{M}_D$
(schematically, $\mathbf{M}_D \mathbf{M}_N^{-1} \ll 1$).
The Schur complement of the seesaw mixing matrix is approximately unitary
\begin{align}
&\left(\begin{array}{cc}
\mathbf{1}_{3\times3} & -\mathbf{M}_{D}\mathbf{M}_{N}^{-1}\\
0 & \mathbf{1}_{3\times3}
\end{array}\right)\left(\begin{array}{cc}
\mathbf{1}_{3\times3} & 0\\
-\mathbf{M}_{N}^{-1}\mathbf{M}_{D}^{T} & \mathbf{1}_{3\times3}
\end{array}\right) \nonumber \\
& = \left(\begin{array}{cc}
\mathbf{1}_{3\times3}+\mathbf{M}_{D}\mathbf{M}_{N}^{-1}\mathbf{M}_{N}^{-1}\mathbf{M}_{D}^{T} & -\mathbf{M}_{D}\mathbf{M}_{N}^{-1}\\
-\mathbf{M}_{N}^{-1}\mathbf{M}_{D}^{T} & \mathbf{1}_{3\times3}
\end{array}\right) \approx \mathbf{1}_{6\times6}
\end{align}
The top-left $3\times3$ $\mathbf{M}_{D}\mathbf{M}_{N}^{-1}\mathbf{M}_{D}^{T}$ will be further diagonalized by
\begin{align}
\left(\begin{array}{cc}
-\mathbf{M}_{D}\mathbf{M}_{N}^{-1}\mathbf{M}_{D}^{T} & 0\\
0 & \mathbf{M}_{N}
\end{array}\right) & \rightarrow\left(\begin{array}{cc}
U^{T} & 0\\
0 & \mathbf{1}_{3\times3}
\end{array}\right)\left(\begin{array}{cc}
\mathbf{M}_{D}\mathbf{M}_{N}^{-1}\mathbf{M}_{D}^{T} & 0\\
0 & \mathbf{M}_{N}
\end{array}\right)\left(\begin{array}{cc}
U & 0\\
0 & \mathbf{1}_{3\times3}
\end{array}\right)\nonumber \\
 &  
 =\left(\begin{array}{cc}
\mathbf{M}_{\nu} & 0\\
0 & \mathbf{M}_{N}
\end{array}\right)_{6\times6}\,,
\end{align}
where the minus sign is absorbed into the fields, and $\mathbf{M}_{\nu}$ is the diagonal light neutrino mass matrix.
In this way, one obtains semi-analytic expressions for the light neutrino mass eigenvalues
and can thereby determine the allowed ranges of all input parameters.

Using the seesaw relation in Eq.~\eqref{eq:SeesawY},
and imposing that the three right-handed neutrinos lie near the electroweak scale,
we estimate the parameter regions to be used in the PSO scan for the three distinct cases as follows:
\begin{itemize}
\item For Case-RS, there are in total \textbf{12 input parameters:}
\begin{equation}
\left(\begin{array}{cc}
y^\nu_{\alpha \beta} & M_{\alpha}\end{array}\right)\label{PCaseRS}
\end{equation}
with the diagonal Yukawa $y^\nu_{\alpha \alpha}\in[10^{-7},10^{-6}]$, off-diagonal
Yukawa $y^\nu_{\alpha \beta}\in[0,10^{-6}]$, and the 3 Majorana masses $M_{\alpha}\in[100,600]$~GeV.

\item For Case-SC, there are in total \textbf{17 input parameters:}
\begin{equation}
\left(\begin{array}{cccccccc}
Y_{i=1,2} & A & B & \mathbf{X}_{\alpha \beta} & \epsilon & M_{1} & M_{2} & M_{3}\end{array}\right)\label{PCaseSC}
\end{equation}
where the third Yukawa coupling is determined by the condition Eq.~(\ref{eq:SCCon}).
The Yukawa coupling $Y_{i=1,2}\in[0.001,1]$, $A,B\in[0.1,25]$,
$\mathbf{X}_{\alpha \beta}\in[-1,1]$, $\epsilon\in[10^{-10},10^{-6}]$, $M_{\alpha}\in[100,600]$~GeV.

\item For Case-SS, there are also in total \textbf{12 input parameters} similar to
Case-RS:
\begin{equation}
\left(\begin{array}{cc}
y^\nu_{\alpha \beta} & M_{\alpha}\end{array}\right)\label{PCaseSS}
\end{equation}
with the mass of $\mathbb{N}_{1}$ around the electroweak scale,
and $\mathbb{N}_{1}$ is ultraweakly coupled to the SM with $y^\nu_{\alpha 1}\ll 10^{-7}$.
In Case-SS, $\mathbb{N}_{2}$ and $\mathbb{N}_{3}$ are responsible for generating the light neutrino masses and mixings,
and are assumed not to couple to the dark sector. Their parameters may therefore span a wide range.
Two representative benchmark choices are as follows:
(1) $\mathbb{N}_{2},\mathbb{N}_{3}$
are also around the electroweak scale, with Majorana masses set in
the region $M_{2},M_{3}\in[100,10^{4}]$~GeV, and the diagonal Yukawa
$y^\nu_{22},y^\nu_{33}\in[10^{-9},10^{-6}]$, off-diagonal Yukawa $y^\nu_{\alpha 2},y^\nu_{\alpha 3}\in[0,10^{-6}]$;
or (2) $\mathbb{N}_{2},\mathbb{N}_{3}$ are at the GUT scale, with
Majorana masses within $M_{2},M_{3}\in[10^{14},10^{15}]$~GeV, and the
diagonal Yukawa $y^\nu_{\alpha 2},y^\nu_{\alpha 3}\in[0.01,1]$.
In both cases, $\mathbb{N}_{2},\mathbb{N}_{3}$ do not play a role in the dark sector evolution.
\end{itemize}

\section{Determination of seesaw parameters with PSO}\label{Sec:PSO}

The PSO algorithm provides an exact numerical framework for identifying viable parameter sets and can efficiently explore the multidimensional parameter space of the seesaw model in search of solutions consistent with neutrino experimental data. In this section, we first summarize the relevant experimental data on neutrinos, together with the main constraints on heavy right-handed neutrinos. We then present a detailed introduction to the PSO algorithm, including its basic setup and the fitness functions adopted in our analysis.
We compare the PSO approach with the Casas--Ibarra (CI) formalism~\cite{Casas:2001sr} and highlight the advantages of using PSO in the present study in Section~\ref{sec:PSOCI}.

\subsection{Neutrino experimental data}\label{sec:ex}

The latest neutrino experimental results are summarized in Table~\ref{Tab:NuEX}.
An additional constraint on the neutrino mass spectrum is given by the upper bound on the sum of neutrino masses~\cite{DES:2021wwk},
\begin{equation}
\sum_{\alpha} m_{\nu \alpha} \lesssim 0.12~{\rm eV}\,.
\end{equation}

\begin{table}[htbp]
\begin{tabular}{lcc}
\hline\hline
Parameter & Best fit $\pm 1\sigma$ & $3\sigma$ range \\
\hline
$\sin^2\theta_{12}$ & $0.308^{+0.012}_{-0.011}$ & $0.275 - 0.345$ \\

$\theta_{12}/^\circ$ & $33.68^{+0.73}_{-0.70}$ & $31.63 - 35.95$ \\

$\sin^2\theta_{23}$ & $0.470^{+0.017}_{-0.013}$ & $0.435 - 0.585$ \\

$\theta_{23}/^\circ$ & $43.3^{+1.0}_{-0.8}$ & $41.3 - 49.9$ \\

$\sin^2\theta_{13}$ & $0.02215^{+0.00056}_{-0.00058}$ & $0.02030 - 0.02388$ \\

$\theta_{13}/^\circ$ & $8.56^{+0.11}_{-0.11}$ & $8.19 - 8.89$ \\

$\delta_{\rm CP}/^\circ$ & $212^{+26}_{-41}$ & $124 - 364$ \\

$\Delta m^2_{21}/(10^{-5}\,\mathrm{eV}^2)$ & $7.49^{+0.19}_{-0.19}$ & $6.92 - 8.05$ \\

$\Delta m^2_{31}/(10^{-3}\,\mathrm{eV}^2)$ & $2.513^{+0.021}_{-0.019}$ & $2.451 - 2.578$ \\
\hline\hline
\end{tabular}
\centering
\caption{Latest global-fit neutrino oscillation parameters for normal ordering, taken from NuFIT 6.0 (2024), using the ``IC24 with SK-atm'' dataset combination~\cite{Esteban:2024eli}.}
\label{Tab:NuEX}
\end{table}

\subsubsection{Non-unitarity constraint}\label{App:NU}

The light neutrino mixing matrix is exactly unitary if the SM neutrinos do not mix with the right-handed neutrinos,
where the full rotation matrix $\mathbf{U}$ takes the simplified form
\begin{equation}
\mathbf{U}_{0}=\left(\begin{array}{cc}
U_{0} & 0\\
0 & \mathbf{1}_{3\times3}
\end{array}\right)\,,
\end{equation}
and thus  the SM neutrino mixing matrix $U_{0}$ is unitary.

Under the seesaw mechanism, however, the SM neutrinos mix with heavy Majorana right-handed neutrinos.
The full rotation matrix then becomes a $6\times6$ unitary matrix, which can be written as
\begin{equation}
\mathbf{U}=\left(\begin{array}{cc}
U & U_{1}\\
U_{2} & U_{N}
\end{array}\right)\,,\label{eq:U66}
\end{equation}
Although the full $6\times6$ matrix $\mathbf{U}$ remains unitary,
the upper-left $3\times3$ block $U$ is no longer exactly unitary.
This block should reproduce the neutrino oscillation data, conventionally encoded in the PMNS matrix,
while its deviation from unitarity is constrained by electroweak precision data and therefore must remain within the allowed limits.

The upper-right $3\times3$ block $U_{1}$ is related to $U$ by
\begin{equation}
UU^{\dagger}+U_{1}U_{1}^{\dagger}=\mathbf{1}_{3\times3}\,,
\end{equation}
since the $6\times6$ rotation matrix $\mathbf{U}$ is unitary. If
the mixing is small such that all entries of the matrix $U_{1}$ are
much less than 1, one has
\begin{align}
UU^{\dagger} & \approx \big(\mathbf{1}_{3\times3}-U_{1}U_{1}^{\dagger}\big)+\frac{1}{4}U_{1}U_{1}^{\dagger}U_{1}U_{1}^{\dagger}\nonumber \\
 & =\big(1-\frac{1}{2}U_{1}U_{1}^{\dagger}\big)\cdot\big(1-\frac{1}{2}U_{1}U_{1}^{\dagger}\big)\nonumber \\
 & =\Big[\big(1-\frac{1}{2}U_{1}U_{1}^{\dagger}\big)U_{0}\Big]\cdot\Big[\big(1-\frac{1}{2}U_{1}U_{1}^{\dagger}\big)U_{0}\Big]^{\dagger}\,,
\end{align}
and thus the departure from unitarity of the matrix $U$ can be approximately characterized
by a $3\times3$ matrix $\eta\equiv\frac{1}{2}U_{1}U_{1}^{\dagger}$ as
\begin{equation}
U=\big(1-\frac{1}{2}U_{1}U_{1}^{\dagger}\big)U_{0}=(1-\eta)U_{0}\,.
\end{equation}
Conventionally, the matrix $\eta$ is written as~\cite{Blennow:2016jkn}
\begin{equation}
\eta=\left(\begin{array}{ccc}
\eta_{ee} & \eta_{e\mu} & \eta_{e\tau}\\
\eta_{e\mu}^{*} & \eta_{\mu\mu} & \eta_{\mu \tau}\\
\eta_{e\tau}^{*} & \eta_{\mu\tau}^{*} & \eta_{\tau\tau}
\end{array}\right)
\end{equation}
with diagonal elements real. Since $U$ should match the experimental
fit of the PMNS matrix, the entries of the $\eta$ matrix can be constrained
by experimental data.

The neutrino oscillation probability for $\nu_{\alpha}\to\nu_{\beta}$, c.f.,~Eq.~(\ref{Eq:NTProb}), can be expressed in terms of the entries of the $\eta$ matrix in both short-baseline and long-baseline experiments. In this way, experimental data constrain the individual elements of $\eta$, thereby imposing additional bounds on the mixing between the SM neutrinos and the right-handed neutrinos in the seesaw framework. These bounds are commonly referred to as \emph{non-unitarity constraints}. Short-baseline experiments such as NOMAD~\cite{NOMAD:2003mqg} and NuTeV~\cite{NuTeV:2002daf} have comparable sensitivities to the zero-distance appearance probability, yielding $P_{\mu e}^{{\rm eff,SBL}}<6\times10^{-4}$ at 90\% C.L., while the combined bound becomes $P_{\mu e}^{{\rm eff,SBL}}<4\times10^{-4}\,(6\times10^{-4})$ at 90\% (99\%) C.L.~\cite{Forero:2021azc}. In the same analysis, long-baseline data from T2K~\cite{T2K:2021xwb} and NOvA~\cite{NOvA:2018gge} are treated by relating the spectra measured at the far detectors to those at the near detectors, such that the effective appearance and disappearance probabilities must be corrected for zero-distance effects. In addition, the MINOS/MINOS+ analysis~\cite{MINOS:2017cae} includes neutral-current (NC) events together with charged-current (CC) events. The NC sample is sensitive to a combination of the muon-neutrino survival probability and the electron- and tau-neutrino appearance probabilities. Constraints obtained from the combination of short-baseline and long-baseline oscillation data were derived in Ref.~\cite{Blennow:2016jkn}.

However, for seesaw scenarios with heavy right-handed neutrinos, the strongest current bounds typically arise not from oscillation data alone, but from global fits including electroweak precision observables ($M_W$, $s_{\rm eff}^2$, and $Z$-pole data), tests of lepton flavor universality, CKM-related weak decay observables, and charged lepton flavor violation processes.
These bounds are generally more restrictive than those obtained from oscillation data alone~\cite{deBlas:2013gla,Cheung:2020buy},
and in this work we adopt the current strongest limits from~\cite{Blennow:2023mqx}  summarized in Table~\ref{Tab:NUC}.
\begin{table}[h!]
\begin{center}
\begin{tabular}{ccccccc}
\hline
Parameter & $\eta_{ee}$ & $\eta_{\mu\mu}$ & $\eta_{\tau\tau}$ & $\eta_{e\mu}$ & $\eta_{e\tau}$ & $\eta_{\mu\tau}$\tabularnewline
\hline
\hline
95\% C.L. & $1.3\times 10^{-3}$ & $1.1\times 10^{-5}$ & $1\times 10^{-3}$ & $1.2\times 10^{-5}$ & $9\times 10^{-4}$ & $5.7\times 10^{-5}$\tabularnewline
\hline
\end{tabular}
\caption{Upper bounds of the non-unitarity parameters at 95\% C.L..} \label{Tab:NUC}
\end{center}
\end{table}

\subsubsection{Experimental constraints on right-handed neutrinos}

\begin{figure}[h!]
		\centering
		\includegraphics[width=8cm]{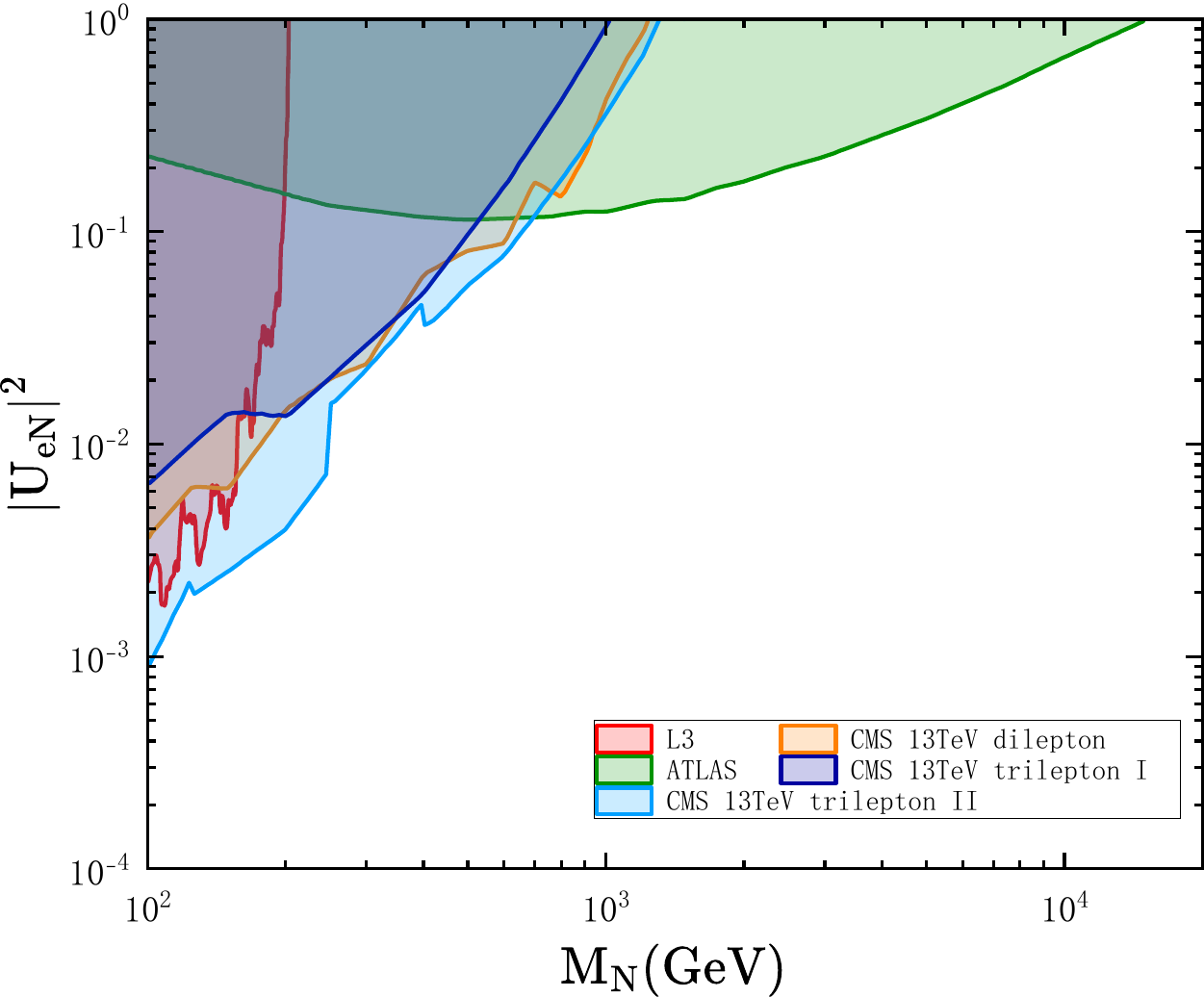}
		\caption{Exclusion limits at the 95\% confidence level on the active--sterile mixing parameter $|U_{e{N}}|^2$ are presented. The upper bounds are derived from various sources, including L3's search via $e^{+}e^{-} \to {N}\nu$~\cite{L3:2001zfe}, ATLAS same-sign $WW$ scattering~\cite{ATLAS:2023tkz}, and several CMS analyses at $\sqrt{s} = 13~\mathrm{TeV}$: same-sign dilepton~\cite{CMS:2018jxx}, trilepton~\cite{CMS:2018iaf,CMS:2024xdq}. }
		\label{Fig:eN}
	\end{figure}

\begin{figure}[h!]
		\centering
		\includegraphics[width=8cm]{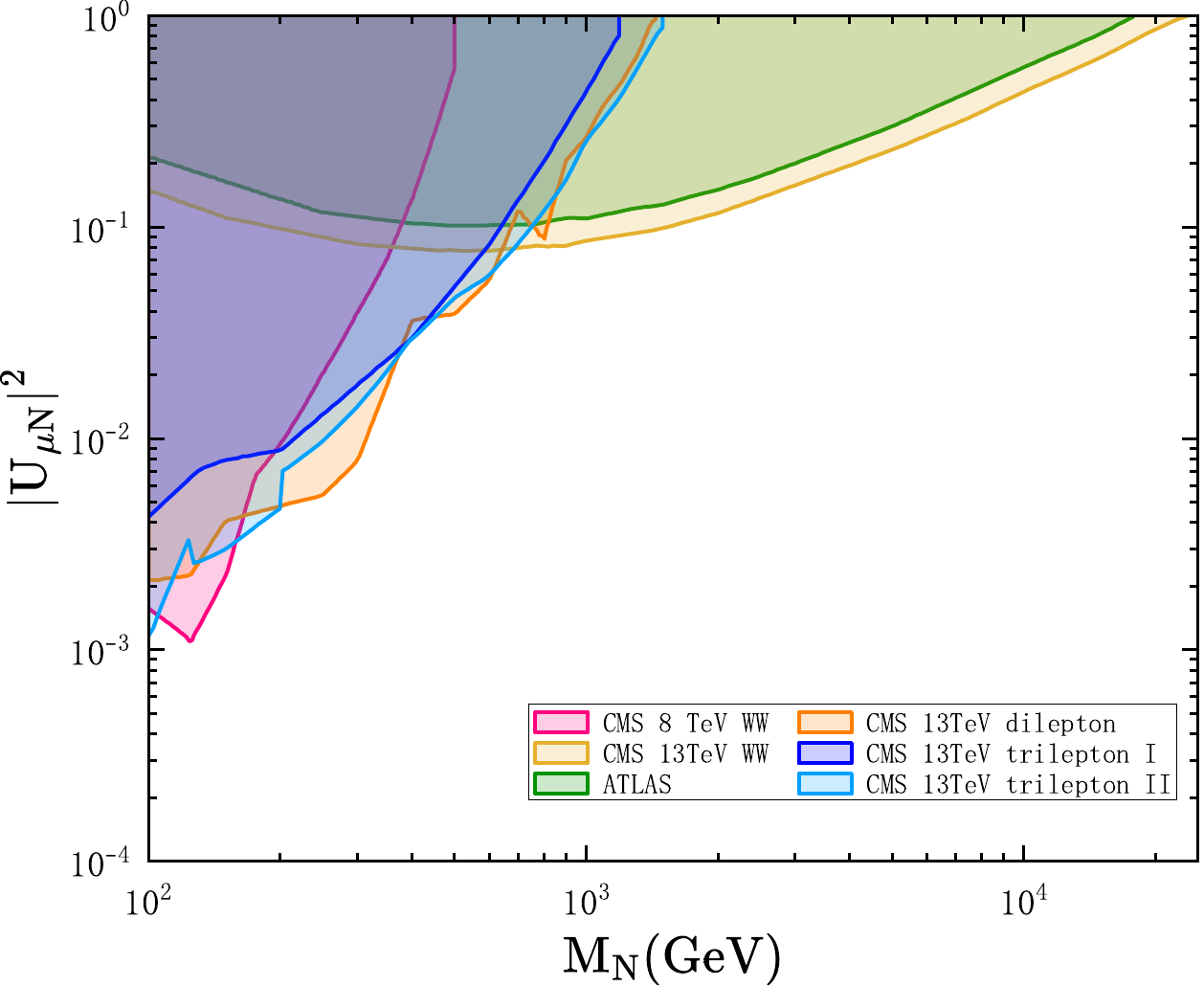}
		\caption{Exclusion limits at the 95\% confidence level on the active--sterile mixing parameter
  $|U_{\mu {N}}|^2$ are presented. The upper bounds are obtained from several sources, including ATLAS same-sign $WW$ scattering~\cite{ATLAS:2024rzi}, CMS same-sign dilepton at $\sqrt{s} = 13~\mathrm{TeV}$~\cite{CMS:2018jxx}, CMS trilepton analysis~\cite{CMS:2018iaf,CMS:2024xdq}, CMS same-sign $WW$ scattering~\cite{CMS:2022hvh}, CMS same-sign dilepton at $\sqrt{s} = 8~\mathrm{TeV}$~\cite{CMS:2015qur}.}
		\label{Fig:muN}
	\end{figure}

\begin{figure}[h!]
		\centering
		\includegraphics[width=8cm]{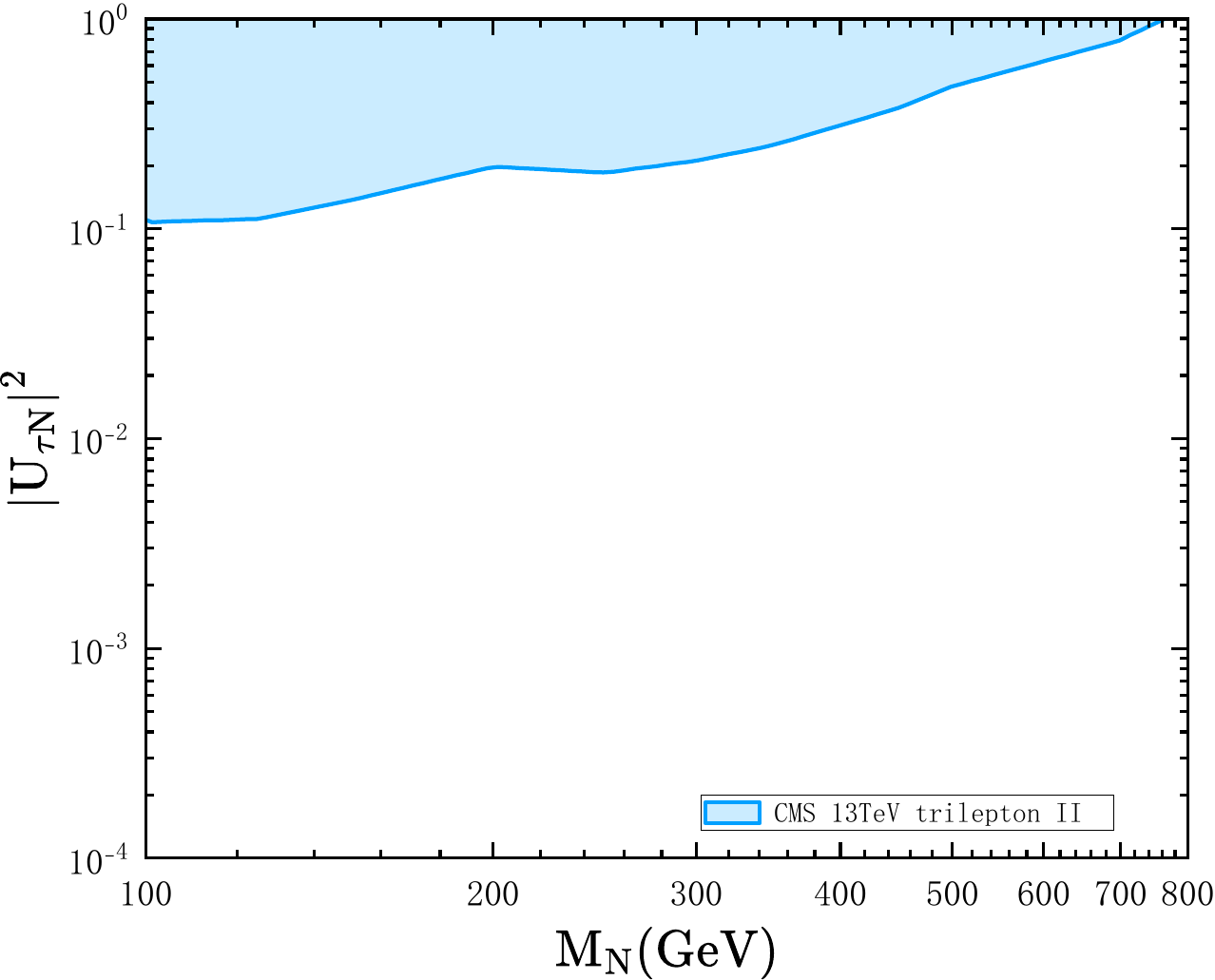}
		\caption{Exclusion limits at the 95\% confidence level on the active--sterile mixing parameter
  $|U_{\tau {N}}|^2$ are presented. The upper bounds are derived from the CMS trilepton analysis at $\sqrt{s} = 13~\mathrm{TeV}$~\cite{CMS:2024xdq}.}
		\label{Fig:tauN}
	\end{figure}

The CMS and ATLAS Collaborations at the LHC have analyzed proton-proton collision data from Drell-Yan and vector boson fusion (VBF) processes~\cite{Rauch:2016pai}, placing a broad range of constraints on heavy neutral leptons with masses from the electroweak scale up to the TeV scale~\cite{CMS:2018jxx,CMS:2018iaf,ATLAS:2024rzi,ATLAS:2023tkz,CMS:2022hvh,CMS:2015qur,CMS:2024xdq}.
These searches probe the final-state signatures induced by the mixing of heavy neutral leptons with the active neutrinos,
and the corresponding bounds are typically presented in terms of the effective active--sterile mixing parameters,
under the simplified assumption that a single heavy neutral lepton, denoted by $N$, dominates and mixes with only one SM lepton flavor at a time.
In the seesaw framework considered here, these bounds can therefore be applied to the heavy Majorana states $\mathbb{N}_i$
and constrain their mixing with light neutrinos.
Through this mixing, $\mathbb{N}_i$ participate in weak interactions
and may appear either as on-shell final-state particles or as internal propagators in Drell-Yan and VBF processes.
In addition, the L3 Collaboration analyzed electron-positron collision data at LEP and placed bounds on heavy neutral leptons
with masses around 100~GeV~\cite{L3:2001zfe}.
The combined experimental bounds are summarized in Figs.~\ref{Fig:eN}, \ref{Fig:muN}, and~\ref{Fig:tauN}.

\paragraph{ATLAS}
The ATLAS Collaboration has searched for events featuring two muons and hadronic jets in proton-proton collisions at $\sqrt{s} = 13~\mathrm{TeV}$, using data collected during 2015--2018 with an integrated luminosity of $140~\mathrm{fb^{-1}}$. Constraints were placed on the mixing parameter $|U_{\mu {N}}|^2$ for a heavy neutral lepton $N$ with masses ranging from $50~\mathrm{GeV}$ to $20~\mathrm{TeV}$.
The analysis targets same-sign $\mu^+\mu^+$  production via vector boson fusion in $W^+W^+$ scattering, with
$N$ appearing as an intermediate state.
The signal region requires events containing same-sign muon pairs along with at least two hadronic jets that are well separated in rapidity.
Using the same dataset, ATLAS has also analyzed final states with $\mu e$ and $ee$ leptons~\cite{ATLAS:2024rzi}, placing constraints on $|U_{e {N}}|^2$ and the product $|U_{e {N}}U_{\mu {N}}^*|$.

\paragraph{CMS}


The CMS Collaboration has searched for signatures of same-sign dileptons accompanied by jets in proton-proton collisions at $\sqrt{s} = 13~\mathrm{TeV}$, using the 2016 dataset with an integrated luminosity of $35.9~\mathrm{fb^{-1}}$. Constraints were placed on the mixing parameters $|U_{e{N}}|^2$, $|U_{\mu {N}}|^2$, and $|U_{e{N}}U_{\mu {N}}^{\ast}|^2/(|U_{e{N}}|^2 + |U_{\mu {N}}|^2)$, for heavy neutral lepton masses in the range of 20 to 1600~GeV. The signal region was defined by requiring two same-sign leptons in the final state, which significantly suppresses SM backgrounds compared to opposite-sign dilepton channels.
CMS has also searched for same-sign dimuon plus dijet signatures in proton-proton collisions at $\sqrt{s} = 8~\mathrm{TeV}$ with an integrated luminosity of $19.7~\mathrm{fb^{-1}}$~\cite{CMS:2015qur}.
This analysis placed constraints on $|U_{\mu {N}}|^2$ for heavy neutral lepton masses between 40 and 500~GeV.

The $W$ bosons produced in the Drell-Yan and VBF processes can also decay into a charged lepton and a neutrino~\cite{CMS:2018iaf}. The CMS Collaboration has searched for events with three charged leptons in the final state using proton-proton collision data at $\sqrt{s} = 13~\mathrm{TeV}$, collected in 2016 with an integrated luminosity of $35.9~\mathrm{fb^{-1}}$. Constraints were placed on the mixing parameters $|U_{e {N}}|^2$ and $|U_{\mu {N}}|^2$ for heavy neutral lepton masses in the range of 1 to 1200~GeV.
The analysis targets final states with three charged leptons in any combination of electrons and muons,
excluding events in which all three leptons have the same electric charge.

Ref.~\cite{CMS:2024xdq} presents an updated analysis based on a larger dataset collected from 2016 to 2018 at
$\sqrt{s} = 13~\mathrm{TeV}$ with an integrated luminosity of $138~\mathrm{fb^{-1}}$, and incorporates improved analysis techniques.
This study provides more precise constraints on the mixing parameters $|U_{eN}|^2\), \(|U_{\mu N}|^2$, and $|U_{\tau N}|^2$
for heavy neutral lepton masses ranging from 10~GeV to 1500~GeV.
The inclusion of $\tau$-flavor mixing is enabled by advancements in $\tau$ lepton reconstruction and identification.
In addition, the analysis considers both scenarios in which $N$ is a Majorana or a Dirac fermion.

Using the same underlying mechanism as the ATLAS same-sign $W^+W^+$ scattering analysis, CMS has conducted a search for the corresponding signal in proton-proton collisions at $\sqrt{s} = 13~\mathrm{TeV}$,
based on the full Run 2 dataset with an integrated luminosity of $138~\mathrm{fb^{-1}}$. This analysis places constraints on $|U_{\mu {N}}|^2$ for heavy neutrino masses in the range of 50~GeV to 25~TeV~\cite{CMS:2022hvh}.

\paragraph{L3}

The constraints reported by the L3 Collaboration are based on the following assumptions~\cite{L3:2001zfe}:
each generation of SM neutrinos mixes exclusively with a corresponding heavy isosinglet neutrino.
The analysis neglects mixing among the light neutrinos, as well as any mixing between the heavy neutrinos or with additional isodoublet states.

In the electron-positron annihilation process $e^{+}e^{-} \to {N}\nu$,
the total cross section is dominated by the $t$-channel exchange of a $W$ boson.
Due to LEP's limited sensitivity to the production of ${N}_\mu$ and ${N}_\tau$ via this channel,
the analysis focuses exclusively on the production of $\mathbb{N}_e$ in association with a $\nu_e$.
The heavy neutrino ${N}$ subsequently decays via ${N} \to W\ell$ or ${N} \to Z\nu$,
with the $W\ell$ channel being dominant, for ${N}$ masses near $W$ and $Z$ masses
where phase space suppression reduces the branching ratio of the $Z\nu$ channel.
The primary signal process is therefore
$e^{+}e^{-} \to {N}_e \nu_e \to \nu_e\, e\, W \to \nu_e\, e\, q'\bar{q}$,
resulting in a final state  characterized by a single isolated electron and hadronic jets.
Based on this signature, the L3 Collaboration performed a search using $e^+e^-$ collision data collected at LEP with
$\sqrt{s} \approx 200~\mathrm{GeV}$,
and placed constraints on the mixing parameter $|U_{e{N}}|^2$ for heavy neutrino masses in the range of $80-205$~GeV.

\paragraph{Future collider searches}

Future collider experiments, such as the LHeC and ILC~\cite{Das:2018usr,Yang:2023ice},
present promising opportunities to probe heavy Majorana neutrinos through a variety of production channels.
These include electron-positron annihilation processes such as $e^{-}e^{+}\to N_1 \bar{\nu}$,
electron-proton scattering processes $eq \to N_1 \bar{q}^{\prime}$,
and electron-photon interactions $e^{-} \gamma \to N W^{-}$.
Notably, utilizing photons generated via proton bremsstrahlung leads to cleaner experimental signatures and significantly reduced Standard Model backgrounds compared to electron bremsstrahlung~\cite{Yang:2023ice}.
These future facilities have the potential to significantly constrain the parameter space of heavy neutrinos with masses at the TeV scale.
In addition, the proposed $\mu$TRISTAN collider offers a unique platform to explore heavy neutrinos via electron-muon collisions, specifically through the process  $e^{-} \mu^{+} \to N_1  \bar{\nu}$~\cite{Das:2024kyk}.
For heavy neutrino masses in the range of $100-300$~GeV,
this setup is projected to probe the mixing angles $|U_{e N}|^2$ and $|U_{\mu N}|^2$
with sensitivities up to two orders of magnitude beyond current bounds from electroweak precision data.

\subsection{Basic PSO setup}

We employ the PSO algorithm to determine the input values of all seesaw parameters. The optimization is performed by requiring the neutrino observables predicted by the seesaw mechanism to reproduce all relevant experimental data discussed above. To this end, we define five fitting functions that quantify the deviations of the neutrino mixing angles, the sum of neutrino masses, and the two neutrino mass-squared differences from their corresponding experimental values. An additional fitting condition is imposed to ensure that the non-unitarity constraint is satisfied.
For all seesaw cases considered in this work, we find that the non-unitarity bound is always more restrictive than the collider constraints on right-handed neutrinos, which we have checked carefully. The observed Dirac CP phase can also be incorporated into the search for viable seesaw parameters, as explained in Appendix~\ref{App:PMNS}. All benchmark seesaw models used in this work are summarized in Appendix~\ref{App:SSBM}.

In the original type-I seesaw framework, consisting of three SM neutrinos and three heavy Majorana neutrinos,
the full $6\times 6$ neutrino mass matrix $\mathbf{M}$ is diagonalized as
\begin{equation}
\mathbf{D} = \mathbf{U}^T \mathbf{M} \mathbf{U}\,,
\end{equation}
where $\mathbf{M}$ is the neutrino mass matrix in the flavor basis, $\mathbf{D}$ is the diagonal mass matrix whose entries correspond to the three light neutrinos and three heavy neutrinos in the mass basis, and $\mathbf{U}$ is the full $6\times 6$ unitary mixing matrix.
For given sets of seesaw parameters, one can diagonalize $\mathbf{M}$ and extract the corresponding neutrino masses and mixing parameters,
which must be consistent with all relevant experimental constraints.

The purpose of employing the PSO algorithm is to identify parameter sets that reproduce the observed neutrino masses and mixing pattern. For each point in parameter space, we calculate the neutrino observables predicted by the model and quantify their deviations from experimental data through a single scalar fitness function, denoted by $F_{\rm fit}$. The optimization problem is thus reduced to finding parameter sets that minimize $F_{\rm fit}$. In the conventional seesaw setup considered here, the fitness function depends on 12 free parameters (17 for Case-SC),
as discussed in Section~\ref{sec:SSP}.
The PSO algorithm is then used to efficiently explore this 12(17)-dimensional parameter space
and identify parameter sets that provide a good fit to the experimental data.

More specifically, at each iteration step $t$, the algorithm generates $n$ candidate parameter sets,
referred to as \emph{particles}. The position of the $i$th particle in parameter space is denoted by
\begin{equation}
p_{t,i} = \left(y_{11}^{\,t,i},\, y_{12}^{\,t,i},\, \ldots \right),
\end{equation}
where the components represent the model parameters to be optimized.
Each particle therefore corresponds to a specific realization of the seesaw parameter set
and hence to a definite value of the fitness function.

Among all particles at iteration step $t$, the one with the smallest fitness value defines the \emph{global-best} position, denoted by $G_t^{\rm best}$.
In addition, for each particle $i$, the best position encountered along its own trajectory up to step $t$ is recorded and denoted by $P_{t,i}^{\rm best}$, which is referred to as the \emph{personal-best} position.
The subsequent evolution of each particle is then determined jointly by its own search history and the collective information of the swarm.

The update rule for the particle velocities and positions is given by
\begin{align}
D_{t,i}^{G} &= \left(G_{t}^{\rm best} - p_{t,i}\right)L^{G} r^{G}, \\
D_{t,i}^{P} &= \left(P_{t,i}^{\rm best} - p_{t,i}\right)L^{P} r^{P}, \\
D_{t,i}^{I} &= v_{t-1,i} L^{I}, \\
v_{t,i} &= D_{t,i}^{P} + D_{t,i}^{G} + D_{t,i}^{I}, \\
p_{t+1,i} &= p_{t,i} + v_{t,i},
\end{align}
where $v_{t,i}$ is the velocity of particle $i$ at step $t$, $L^{P}$ and $L^{G}$ are the learning coefficients associated
with the \emph{personal-best} and \emph{global-best} directions, respectively, and $L^{I}$ is the inertia weight, which quantifies the contribution of the particle's previous velocity to the current update and thus controls the balance between global exploration and local convergence.
The random numbers $r^{P}$ and $r^{G}$ are independently drawn from the interval $(0,1)$ at each update step,
thereby introducing stochasticity into the search and enlarging the coverage of parameter space.
In this work, we adopt
\begin{equation}
L^{P} = 0.4\,, \qquad L^{G} = 0.4\,, \qquad L^{I} = 0.8\,.
\end{equation}

Through successive iterations, both the \emph{global-best} and \emph{personal-best} positions are continuously updated,
and the swarm progressively converges toward regions of parameter space with smaller fitness values.
If the model admits parameter sets compatible with current neutrino data,
the PSO algorithm is able to identify solutions with sufficiently small fitness values.
In our analysis, the final fitness values obtained for all benchmarks are evaluated according to Eq.~\eqref{eq:FitF},
showing that the corresponding \emph{global-best} solutions, namely the resulting seesaw parameter sets,
successfully reproduce the experimentally allowed low-energy neutrino masses and mixing parameters to good accuracy.

The seesaw parameter sets generated in this work by PSO, summarized in Appendix~\ref{App:SSBM}, are not intended to be general in the sense of providing the most representative benchmarks of the full seesaw parameter space.
Rather, their generality is reflected in the parametric relation between the heavy Majorana neutrino masses and the corresponding seesaw couplings.
As discussed in Section~\ref{sec:SSP}, and further illustrated in the one-generation example presented in Appendix~\ref{App:SSdetail}, once the mass of the heavy Majorana neutrino is specified, the corresponding Yukawa coupling can be estimated from Eq.~\eqref{eq:SeesawY}. Although the full seesaw mixing depends on the detailed structure of the $3\times 3$ Yukawa matrix, this relation still provides a useful order-of-magnitude estimate for the Yukawa couplings.

This correspondence between the Majorana masses and Yukawa couplings leads to typical Yukawa scales for a given heavy neutrino mass range.
For electroweak scale right-handed neutrinos with masses of a few hundred GeV,
which are the main focus of this work, one typically finds
\begin{itemize}
  \item in Case-RS, a characteristic \emph{tiny} seesaw Yukawa coupling of order $\mathcal{O}(10^{-7})$;
  \item in Case-SC, a characteristic \emph{small} seesaw Yukawa coupling of order $\mathcal{O}(10^{-3})$;
  \item in Case-SS, characteristic seesaw Yukawa couplings $y^\nu_{\alpha 2}, y^\nu_{\alpha 3}$ of order $\mathcal{O}(10^{-7})$,
while $y^\nu_{\alpha 1}\ll 10^{-7}$ is \emph{ultraweak}.
\end{itemize}

Hence, for right-handed neutrinos whose couplings to the SM lie within these typical ranges
and reproduce the observed neutrino masses through the seesaw mechanism,
the resulting right-handed neutrino portal dark matter evolution is expected to be qualitatively representative.

\subsection{Fitness functions}

The fitting functions are introduced to quantify how closely the low-energy neutrino observables,
obtained from the generated UV seesaw parameters, reproduce the corresponding experimental data.

Assuming normal mass ordering, we define the following five fitting functions:
\begin{enumerate}
  \item \textbf{Neutrino mixing angle fit:}
  Based on the allowed ranges listed in Table~\ref{Tab:NuEX}, the $3\sigma$ allowed intervals for the three neutrino mixing angles are
  \begin{equation}
  \Theta^{\rm ex}_{i=1,2,3} \equiv (\theta_{12}\,, \theta_{23}\,, \theta_{13})
  = (31.63^{\circ} - 35.95^{\circ}\,,\,
  41.3^{\circ} - 49.9^{\circ}\,,\,
  8.19^{\circ} - 8.89^{\circ})\,.
  \end{equation}
  We denote the upper, central, and lower values of $\Theta^{\rm ex}_{i=1,2,3}$ by $\Theta_i^{\rm max}$, $\Theta_i^{\rm mid}$, and $\Theta_i^{\rm min}$, respectively. For each set of seesaw parameters, we compute the corresponding neutrino mixing angles and define the fitting function as
  \begin{equation}
  F_{\rm fit}^\theta \equiv \max_{i=1,2,3}\left\{
  \frac{|\Theta_i - \Theta_i^{\rm mid}| - (\Theta_i^{\rm max} - \Theta_i^{\rm min})/2}
  {(\Theta_i^{\rm max} - \Theta_i^{\rm min})/2}\,, 0
  \right\}\,,
  \end{equation}
  In this definition, only $\Theta_i$ lying outside the experimentally allowed ranges contribute positively, while all negative values are set to zero.
  Hence, $F_{\rm fit}^{\theta}$ measures the largest relative deviation of the mixing angles from their allowed intervals,
  and vanishes if all three angles lie within the experimental bounds.
\item \textbf{Cosmological constraint on the neutrino mass sum:}
  All light neutrino masses in the mass eigenbasis are converted to units of eV, and the corresponding fitting function is defined as
  \begin{equation}
  F_{\rm fit}^{\rm Sum} \equiv
  \max\left\{
  \frac{\sum_\alpha m_{\nu_\alpha} - 0.12}{0.12}\,,\, 0
  \right\}.
  \end{equation}
\item \textbf{Mass-squared difference fit:}
  The fitting functions for the solar and atmospheric mass-squared differences are defined as
  \begin{align}
  F_{\rm fit}^{\Delta m_{21}^2} &\equiv
  \frac{| (m_{\nu_2}^2 - m_{\nu_1}^2) - \Delta m_{21}^2 |}{\Delta m_{21}^2}
  + \frac{| (m_{\nu_2}^2 - m_{\nu_1}^2) - \Delta m_{21}^2 |}{m_{\nu_2}^2 - m_{\nu_1}^2}\,, \\
  F_{\rm fit}^{\Delta m_{31}^2} &\equiv
  \frac{| (m_{\nu_3}^2 - m_{\nu_1}^2) - \Delta m_{31}^2 |}{\Delta m_{31}^2}
  + \frac{| (m_{\nu_3}^2 - m_{\nu_1}^2) - \Delta m_{31}^2 |}{m_{\nu_3}^2 - m_{\nu_1}^2}\,.
  \end{align}
  The second terms in $F_{\rm fit}^{\Delta m_{21}^2}$ and $F_{\rm fit}^{\Delta m_{31}^2}$ penalize configurations
  in which the predicted mass-squared differences approach zero during the PSO search,
  thereby preventing unphysical solutions from yielding artificially small fitting errors.
  \item \textbf{Non-unitarity constraint:}
  The non-unitarity constraint is imposed as
  \begin{equation}
  \eta < \tilde{\eta} =
  \begin{pmatrix}
  1.3\times10^{-3} & 1.2\times10^{-5} & 9\times10^{-4} \\
  1.2\times10^{-5} & 1.1\times10^{-5} & 5.7\times10^{-5} \\
  9\times10^{-4} & 5.7\times10^{-5} & 1\times10^{-3}
  \end{pmatrix},
  \end{equation}
  where the matrix $\eta$ was introduced in Section~\ref{App:NU}. We then define the corresponding fitting function as
  \begin{equation}
  F_{\rm fit}^{\eta} \equiv
  \max_{i,j=1,2,3}
  \left\{
  \frac{\eta_{ij} - \tilde{\eta}_{ij}}{\tilde{\eta}_{ij}}\,,\, 0
  \right\},
  \end{equation}
  which again measures the largest relative deviation from the allowed non-unitarity bounds.
\end{enumerate}

The total fitness function is defined as
\begin{equation}
F_{\rm fit}^{\rm total} \equiv
\exp\!\left(F_{\rm fit}^{\theta}\right)
+ \exp\!\left(F_{\rm fit}^{\rm Sum}\right)
+ \exp\!\left(F_{\rm fit}^{\Delta m_{21}^{2}}\right)
+ \exp\!\left(F_{\rm fit}^{\Delta m_{31}^{2}}\right)
+ \exp\!\left(F_{\rm fit}^{\eta}\right)
- 5 < 10^{-4}\,.
\label{eq:FitF}
\end{equation}

For sufficiently good parameter sets, the total fitness approaches zero and is typically much smaller than unity. In this work, all obtained parameter sets satisfy $F_{\rm fit}^{\rm total} < 10^{-4}$, demonstrating that they provide an excellent fit to the relevant neutrino observables.

\subsection{PSO versus CI}\label{sec:PSOCI}

In this subsection we compare the PSO and CI methods in searching for the seesaw parameters.
The CI parameterization is based on the approximate seesaw relation in Eq.~\eqref{eq:blockdiag},
with the minus sign absorbed into the fields,
\begin{equation}
\mathbf{M}_\nu \equiv \mathbf{M}_D \mathbf{M}_N^{-1} \mathbf{M}_D^T\,,
\end{equation}
which holds in the seesaw limit, where the characteristic scale of $\mathbf{M}_N$ is much larger than that of $\mathbf{M}_D$.
The CI parameterization then takes the form\footnote{
Assume that $U$ diagonalizes $\mathbf{M}_\nu$, namely,
\begin{equation}
\mathbf{M}_{\nu}=U^T \big(\mathbf{M}_D \mathbf{M}_N^{-1} \mathbf{M}_D^T\big) U\,.
\end{equation}
Since $\mathbf{M}_{\nu}$ is diagonal, one may multiply both sides by $\mathbf{M}_{\nu}^{-1/2}$ to obtain
\begin{equation}
\mathbf{1}
= \sqrt{\mathbf{M}_{\nu}^{-1}} U^T \mathbf{M}_D \mathbf{M}_N^{-1} \mathbf{M}_D^T U \sqrt{\mathbf{M}_{\nu}^{-1}}
\equiv R R^T\,,
\label{eq:CIRmatrix}
\end{equation}
where we define the rotation matrix $R \equiv \sqrt{\mathbf{M}_{\nu}^{-1}} U^T \mathbf{M}_D \sqrt{\mathbf{M}_N^{-1}}$
and from which the Yukawa matrix $\mathbf{M}_D$ can be expressed in the CI form.}
\begin{equation}
\mathbf{M}_{D} = U^\ast \sqrt{\mathbf{M}_{\nu}}\, R\, \sqrt{\mathbf{M}_N}\,,
\end{equation}
where $U$ is set equal to the PMNS matrix determined from neutrino oscillation data,
$\mathbf{M}_{\nu}$ and $\mathbf{M}_N$ are the diagonal mass matrices of the light and heavy neutrinos, respectively,
and $R$ is the complex orthogonal matrix defined in Eq.~\eqref{eq:CIRmatrix}, generated from random complex angles in the CI formalism.
Therefore, once one specifies the heavy Majorana masses $\mathbf{M}_N$, the light neutrino masses consistent with the mass-squared difference constraints and the cosmological bound on the neutrino mass sum, the PMNS matrix, and a randomly generated complex orthogonal matrix $R$, the CI parameterization can be used to generate a corresponding $3\times 3$ seesaw Yukawa matrix that reproduces the light neutrino oscillation data.

Unlike the CI parameterization,
which reconstructs the Dirac mass matrix analytically from low-energy neutrino data within the approximate seesaw framework,
our PSO-based approach scans the underlying seesaw parameters directly and tests them through exact diagonalization of the full mass matrix.
This makes the method more flexible in incorporating additional model constraints and more readily applicable to generalized seesaw frameworks.
The main advantages of the PSO over the CI parameterization are summarized as follows:
\begin{enumerate}
  \item \textbf{Exact treatment of the full mass matrix:} In the PSO approach, one diagonalizes the full $6\times 6$ neutrino mass matrix directly.
  The fit is therefore based on the exact numerical diagonalization of the full system,
  rather than on the approximate block-diagonal seesaw formula alone.
  As a result, if the active--heavy neutrino mixing is not extremely small, as in Case-SC,
  or if one wishes to maintain a fully exact treatment throughout, PSO is more general than the CI parameterization.

  \item \textbf{Greater flexibility in parameter scanning and constraint implementation:}
  In the CI parameterization, the scan is typically performed over the heavy neutrino masses $M_i$ and the complex angles in the $R$ matrix.
  This makes the procedure highly efficient, since the scan is automatically restricted to points compatible with low-energy neutrino data.
  However, additional experimental constraints, especially the non-unitarity bounds, are not automatically guaranteed to be satisfied.

  By contrast, PSO can scan directly over the original model parameters,
  such as Yukawa couplings, mass matrix elements, and if necessary, other model-dependent parameters.
  It is therefore more flexible, as it explores a larger parameter space.
  In particular, PSO is advantageous when one wishes to impose neutrino data constraints simultaneously with additional requirements,
  such as heavy neutrino mass conditions, active--sterile mixing bounds, non-unitarity constraints,
  leptogenesis-related conditions, dark matter or collider constraints, as well as texture assumptions or other model-specific relations.

  \item \textbf{Easier generalization to non-minimal seesaw frameworks:}
  Although we do not explore such extensions in detail in the present work,
  the CI parameterization becomes less straightforward in generalized seesaw scenarios,
  such as the inverse seesaw, linear seesaw, extended seesaw, models with additional sterile states,
  setups with special textures or extra symmetries,
  or cases in which one prefers exact diagonalization over the approximate seesaw formula.

  By contrast, PSO does not rely strongly on a particular analytic parameterization.
  As long as one can construct the mass matrix, diagonalize it, compute the relevant observables,
  and define an appropriate fitness function, PSO can be applied.
  In this sense, PSO can be generalized much more easily to non-minimal seesaw setups.
\end{enumerate}

When the active--heavy neutrino mixings are sufficiently small such that CI parametrization remains reliable, a hybrid CI $\to$ PSO procedure could in principle improve the efficiency of the numerical scan. In this strategy, the CI parameterization can first be used to construct seesaw parameter sets consistent with neutrino oscillation data in the standard seesaw limit, while PSO can subsequently be applied to impose additional experimental constraints and model-dependent requirements.

\section{Right-handed neutrino portal dark matter}\label{Sec:RHNP}

The right-handed neutrino portal refers to scenarios in which right-handed Majorana neutrinos act as portal mediators to the dark sector.
The minimal right-handed neutrino portal dark sector consists of a dark fermion $\chi$ and a dark complex scalar $\phi$.\footnote{
The term ``minimal'' in the right-handed neutrino portal dark sector
refers to a framework in which the observed neutrino masses and mixing pattern are explained, while the dark sector couples to the SM exclusively through the right-handed neutrino portal
and reproduces the observed dark matter relic density.}
The stability of the two particles can be protected by either a dark $U(1)$ symmetry or a dark $\mathbb{Z}_2$ symmetry.
If there is a dark $U(1)$ gauge symmetry, in this work we assume the dark $U(1)$
symmetry is already broken at some high scale,
and the $U(1)$ gauge boson is very massive and thus not contributing to the dark sector evolution.

The general right-handed neutrino portal Lagrangian is written as
\begin{equation}
\mathcal{L}_{{\rm RHNP}}=y_{\alpha}^{x}\,\overline{\mathbb{N}_{\alpha}^{0}}\chi\phi
+y_{\alpha}^{x*}\,\overline{\chi}\phi^{\dagger}\mathbb{N}_{\alpha}^{0}\,,
\end{equation}
where the Yukawa coupling $y_{\alpha}^{x}$ can be real or complex, $\mathbb{N}_{\alpha}^{0}$
are the right-handed Majorana fermions  in the original flavor basis of the Type-I seesaw interactions.
The three Majorana fermions mix with the three SM neutrinos through the seesaw mechanism,
and after mass diagonalization the physical neutrino states are linear combinations of the six flavor eigenstates,
c.f., Eq.~(\ref{eq:VTran}).
Thus after the seesaw mixing, the right-handed portal interactions are modified to be (the $h.c.$ terms are shown explicitly)
\begin{align}
\mathcal{L}_{{\rm RHNP}} & =y_{\alpha}^{x}\sum_{i=1}^{6}U_{\alpha+3, i}\overline{\mathbb{V}_{i}}\,\chi\phi
+y_{\alpha}^{x*}\overline{\chi}\phi^{\dagger}\sum_{i=1}^{6}U_{\alpha+3, i}\mathbb{V}_{i}\nonumber \\
 & =\sum_{i=1}^{6}y_{\alpha}^{x}U_{\alpha+3, i}\,\overline{\mathbb{V}_{i}}\,\chi\phi
 +\sum_{i=1}^{6}y_{\alpha}^{x*}U_{\alpha+3, i}\,\overline{\chi}\phi^{\dagger}\mathbb{V}_{i}\nonumber \\
 & \equiv\sum_{i=1}^{6}\big[y_{i}^{x\prime}\,\overline{\mathbb{V}_{i}}\,\chi\phi+y_{i}^{x\prime\,*}\,\overline{\chi}\phi^{\dagger}\mathbb{V}_{i}\big]\,,
 \label{eq:DSN}
\end{align}
where $y_{i}^{x\prime}$ are the hidden sector coupling constants after the seesaw mixing.

As discussed in Section~\ref{Sec:Seesaw}, the right-handed Majorana fermions can
be regarded as  either purely neutral Majorana fermions
or as originally $(B-L)$-charged Dirac fermions $N_{\alpha}$ that become Majorana fermions after the $B-L$ symmetry breaking.
In the latter viewpoint, the dark sector particles $\chi$ or $\phi$ may also carry $B-L$ quantum number
to balance the $B-L$ charge originally carried by $N_{\alpha}$.
Although at low energies the interactions in Eq.~(\ref{eq:DSN}) already violate $B-L$,
any probe of right-handed neutrino portal dark matter is therefore blind to the underlying $B-L$ symmetry, if it exists.

\begin{figure}
	\centering
	\includegraphics[scale=0.4]{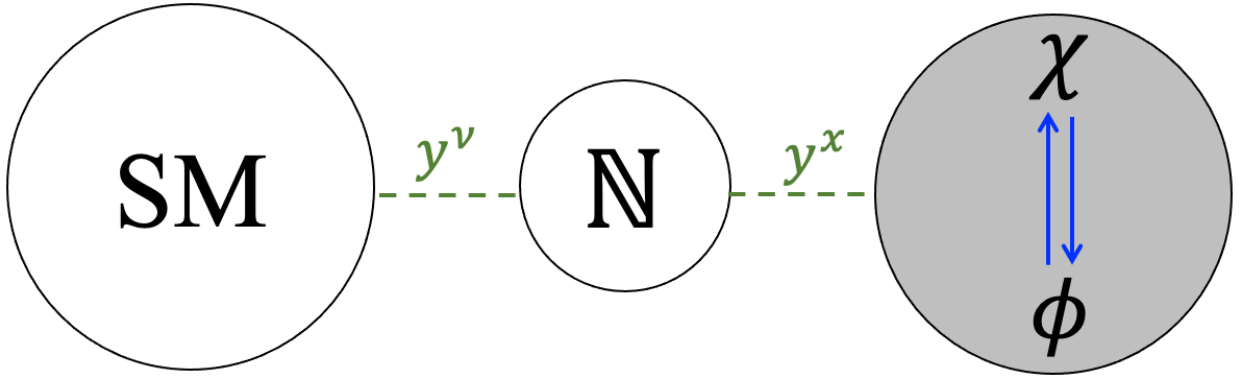}
    \caption{Schematic illustration of the Type-I seesaw right-handed neutrino portal dark matter scenario.
    The right-handed neutrinos generate the active neutrino masses through the seesaw mechanism.
    Once their masses are specified, the corresponding Yukawa couplings $y^\nu$ are determined by fitting the neutrino data.
    After seesaw mixing, the dark sector particles $\chi$ and $\phi$ communicate with the SM sector through the right-handed neutrinos.
    The dark sector Yukawa coupling $y_x$ is fixed by requiring the observed dark matter relic abundance to be reproduced.}
	\label{Fig:RHNP}
\end{figure}

In this work we focus on scenarios where the evolutions of both $\chi$ and $\phi$
are determined purely by the right-handed neutrino portal interactions,\footnote{Especially in freeze-in scenarios, the masses of the dark $U(1)$ gauge boson and of the $U(1)_{B-L}$ gauge boson (if present) should be larger than the reheating temperature. Otherwise, these $U(1)$ gauge bosons can be copiously produced, bringing all dark sector particles into equilibrium with the SM and thereby violating the freeze-in assumption.}
as illustrated schematically in Fig.~\ref{Fig:RHNP}.

The existence of the interaction terms $\sim \overline{\chi}\phi^{\dagger}\mathbb{V}_{i}$ (the discussion for $h.c.$ terms is similar),
where $\mathbb{V}_{i=1,2,3}$ are mass eigenstates of light neutrinos,
allows for a decay of either $\chi \to \overline{\phi} +\mathbb{V}_{i}$ or $\overline{\phi} \to \chi +\mathbb{V}_{i}$
depending on the mass hierarchy of $\chi,\phi$.
In this work we always assume $m_\chi<m_\phi$ and thus $\chi$ is dark matter.
$\phi$ fully decays before BBN in the freeze-out scenario.
In the freeze-in scenario, the couplings between dark sector particles and SM fields are naturally suppressed.
It is therefore possible that the dark coupling $y^{x}$ is extremely small,
in which case the decays $\overline{\phi} \to \chi + \mathbb{V}_i$
have lifetimes exceeding the age of the Universe, and both $\chi$ and $\phi$ can serve as dark matter candidates.

The electroweak right-handed neutrino couplings to the SM particles
can be determined from the three different types of seesaw mixings, as discussed in Section~\ref{Sec:RHN}:
\begin{enumerate}
  \item For Case-RS, the Yukawa couplings of electroweak right-handed neutrinos to $LH$ are of the order $10^{-7}$.
Consequently the hidden sector equilibrates with the SM only at late times ($\lesssim\mathcal{O}({\rm TeV})$),
but before the dark matter freeze-out.
  \item For Case-SC, the relatively large Yukawa couplings keep the hidden sector in thermal equilibrium with the bath at early times,
  after which dark matter undergoes freeze-out.
  \item For Case-SS, we assume that only the lightest $\mathbb{N}_{1}$, which couples ultraweakly to the SM,
  acts as the portal to the dark sector. As a result, the dark sector particles are produced solely via freeze-in.
\end{enumerate}

With the assistance of right-handed neutrinos as a portal connecting
hidden and visible sectors, the abundance of dark particles can be efficiently
depleted through freeze-out,
or generated through freeze-in.
All relevant interactions are summarized as follows, where we use $i,j=1-3$ to label final state particles unless otherwise specified:
\begin{enumerate}
	\item Particles $\chi,\overline{\chi},\phi,\overline{\phi}$ annihilate
	into light neutrinos and SM particles, mediated by portal particles $\mathbb{N}_{i}$:
	$\chi\overline{\chi}\to\mathbb{V}_{i}\mathbb{V}_{j}$, $\phi\overline{\phi}\to\mathbb{V}_{i}\mathbb{V}_{j}$,
	$\chi\phi\,(\overline{\chi\phi})\to h\mathbb{V}_{i},Z\mathbb{V}_{i},W^{\pm}\ell_{i}^{\mp}$.
    Dark sector three-point processes $\phi\leftrightarrow\overline{\chi}\mathbb{V}_{i}$
	($\overline{\phi}\leftrightarrow\chi\,\mathbb{V}_{i}$)
	if $m_{\phi}>(m_{\chi}+m_{i})$.
	\item Particles $\chi,\overline{\chi},\phi,\overline{\phi}$ transform
	into portal particles $\mathbb{N}_{i}$, followed by $\mathbb{N}_{i}$
	decaying into SM particles: $\chi\overline{\chi}\leftrightarrow\mathbb{V}_{i}\mathbb{N}_{j},\mathbb{N}_{i}\mathbb{N}_{j}$,
	$\phi\overline{\phi}\leftrightarrow\mathbb{V}_{i}\mathbb{N}_{j},\mathbb{N}_{i}\mathbb{N}_{j}$,
	$\chi\phi\,(\overline{\chi\phi})\leftrightarrow h\mathbb{N}_{i},Z\mathbb{N}_{i}$,
	and three-point channels $\chi\phi\,(\overline{\chi\phi})\leftrightarrow\mathbb{N}_{i}$
	if $M_{i}>(m_{\chi}+m_{\phi})$, $\phi\leftrightarrow\overline{\chi}\mathbb{N}_{i}$
	($\overline{\phi}\leftrightarrow\chi\mathbb{N}_{i}$) if
	$m_{\phi}>(m_{\chi}+M_{i})$.
	\item Decay channels of right-handed neutrinos: $\mathbb{N}_{i}\leftrightarrow h\mathbb{V}_{j},Z\mathbb{V}_{j},W^{\pm}\ell_{i}^{\mp}$,
	and $\mathbb{N}_{i}\leftrightarrow\chi\phi\,(\overline{\chi\phi})$
	if $M_{i}>(m_{\chi}+m_{\phi})$.
	\item Interactions within the hidden sector: $\chi\chi\leftrightarrow\overline{\phi\phi}$ ($\overline{\chi\chi}\leftrightarrow\phi\phi$), $\chi\overline{\chi}\leftrightarrow\phi\overline{\phi}$.
\end{enumerate}

The couplings related to the right-handed neutrino portal are summarized as follows:
\begin{align}
W_{\mu}^{+}\overline{\nu_{\alpha L}}\gamma^{\mu}\ell_{\alpha L}:\qquad
& \frac{g_{2}}{\sqrt{2}}\sum_{i=1}^{6}U_{\alpha i}W_{\mu}^{+}\overline{\mathbb{V}_{i}}\gamma^{\mu}P_{L}\ell_{\alpha L}\,,\\
W_{\mu}^{-}\overline{\ell_{\alpha L}}\gamma^{\mu}\nu_{\alpha L}:\qquad
& \frac{g_{2}}{\sqrt{2}}\sum_{i=1}^{6}U_{\alpha i}W_{\mu}^{-}\overline{\ell_{\alpha L}}\gamma^{\mu}P_{L}\mathbb{V}_{i}\,,\\
Z_{\mu}\overline{\nu_{\alpha L}}\gamma^{\mu}\nu_{\alpha L}:\qquad
& \frac{g_{2}}{2\cos\theta_{W}}\sum_{i=1}^{6}\sum_{j=1}^{6}U_{\alpha i}U_{\alpha j}Z_{\mu}\overline{\mathbb{V}_{i}}\gamma^{\mu}P_{L}\mathbb{V}_{j}\,,\\
\overline{L_{\alpha}}\widetilde{H}P_{R}\mathbb{N}_{\beta}^{0}+h.c.:\qquad &
-\frac{1}{\sqrt{2}}y_{\alpha \beta}^{\nu}h\,
 \Big(\sum_{i=1}^6 \sum_{j=1}^6 U_{\alpha i} U_{\beta+3, j} \overline{\mathbb{V}_{i}}\,P_{R} \mathbb{V}_{j}
 + \sum_{i=1}^6 \sum_{j=1}^6 U_{\alpha i} U_{\beta+3, j} \overline{\mathbb{V}_{j}} \,P_{L} \mathbb{V}_{i} \Big)\,,\\
%
\chi\phi\overline{\mathbb{N}_{\alpha}^{0}}+h.c.:\qquad
& \sum_{i=1}^{6}y_{\alpha}^{x}U_{\alpha+3, i}\,\overline{\mathbb{V}_{i}}\,\chi\phi
+\sum_{i=1}^{6}y_{\alpha}^{x\ast}U_{\alpha+3, i}\,\overline{\chi}\phi^{\dagger}\mathbb{V}_{i}\,.
\end{align}

In this framework, the most relevant experimental probes arise from the electroweak scale right-handed neutrinos rather than from conventional dark matter direct detection. Since the hidden sector communicates with the SM almost exclusively through the neutrino portal, DM--nucleon scattering is naturally suppressed. The main laboratory signatures are therefore heavy neutral lepton searches at LEP, the LHC, and future colliders, including same-sign dileptons, trileptons, same-sign $W^+ W^+$ scattering, displaced vertices, and other lepton number violating final states induced by active--sterile mixing. These probes are complemented by electroweak precision observables and non-unitarity constraints on the light neutrino mixing matrix. In addition, when kinematically allowed, the portal interaction opens decays such as
$\mathbb{N} \to \chi \phi$, which can modify heavy neutrino branching fractions and may lead to missing energy signatures.
Cosmological requirements, including the observed relic abundance, neutrino data, and the requirement that $\phi$
decay before BBN or, in ultraweak freeze-in regimes, the presence of long-lived dark sector states,
provide further important constraints on the viable parameter space.
The right-handed neutrino portal scenario
therefore requires complementary probes from collider searches, neutrino sector precision tests, and cosmological observations,
and the benchmark models explored here provide useful guidance for future phenomenological investigations.

\section{Benchmarks and evolution of the dark particles}\label{Sec:Pheno}

In this section, we study representative benchmarks of the right-handed neutrino portal dark sector
and analyze the cosmological evolution of the dark particles in detail.
Using the viable seesaw parameter sets obtained in Section~\ref{Sec:PSO} and collected in Appendix~\ref{App:SSBM},
which reproduce the observed light neutrino masses and mixings and satisfy current constraints,
we determine benchmark scenarios for Cases RS, SC, and SS,
and then solve the full coupled Boltzmann equations for the dark fermion $\chi$, the complex scalar $\phi$,
and the relevant right-handed neutrinos.
We consider both freeze-out and freeze-in production of dark matter,
and show how the thermal history depends on the size of the portal couplings and on the strength of internal dark sector interactions.
For the freeze-in case, when the dark sector reaches internal thermal equilibrium,
its evolution must be described by a full coupled treatment,
including a dark sector temperature, in order to obtain the correct relic abundance.

\subsection{Boltzmann equations}\label{sec:BE}

We solve the full set of Boltzmann equations,
including all relevant interactions,
to track the evolution of the dark sector particles $\chi,\phi$ as well as the heavy right-handed neutrinos.
The calculation of scattering processes involving Majorana fermions is summarized in Appendix~\ref{App:SAwM},
while the relevant Feynman diagrams and collision terms are collected in Appendix~\ref{App:CT}.

As discussed in Section~\ref{Sec:RHNP}, under the assumption $m_\chi < m_\phi$, $\chi$ is the dark matter candidate.
The field $\phi$ either decays completely before BBN or, in the freeze-in scenario, can also serve as a dark matter candidate.
The yield of $\chi$ can achieve the observed relic abundance through either freeze-out or freeze-in processes.

\paragraph{Boltzmann equations for freeze-out and freeze-in}

The evolution of all dark sector particles is governed by the coupled
Boltzmann equations for comoving number densities $Y_{\chi},Y_{\phi},Y_{\mathbb{N}_i}$.
Relevant collision terms $\mathcal{C}_\chi, \mathcal{C}_\phi, \mathcal{C}_{\mathbb{N}_{i}}$ are summarized in
Appendix~\ref{App:CT}.

For the {\bf freeze-out evolution,}
\begin{align}
	\frac{\text{d}Y_{\chi}}{\text{d}T} & =-\frac{s}{T\overline{H}}\,\mathcal{C}_{\chi}\,,\label{eq:FOBEchi}\\
	\frac{\text{d}Y_{\phi}}{\text{d}T} & =-\frac{s}{T\overline{H}}\,\mathcal{C}_{\phi}\,,\label{eq:FOBEphi}\\
	\frac{\text{d}Y_{\mathbb{N}_{i}}}{\text{d}T} & =-\frac{s}{T\overline{H}}\,\mathcal{C}_{\mathbb{N}_{i}}\,,\label{eq:FOBENi}
\end{align}
where the entropy density is given by $s=\frac{2\pi^{2}}{45}h_{\text{eff}}T^{3}$, and
\begin{equation} \overline{H}=\frac{H}{1+\frac{1}{3}\frac{T}{h_{\text{eff}}}\frac{\text{d}h_{\text{eff}}}{\text{d}T}}=\sqrt{\frac{\pi^{2}g_{\text{eff}}}{90}}\frac{T^{2}/M_{\text{Pl}}}{1+\frac{1}{3}\frac{T}{h_{\text{eff}}}\frac{\text{d}h_{\text{eff}}}{\text{d}T}}\,.
\end{equation}

For the {\bf freeze-in evolution}, in the presence of sufficiently strong hidden sector interactions
to maintain internal thermal equilibrium, the hidden sector may be characterized by a temperature $T_h$,
defined through its energy density~\cite{Chu:2011be,Hambye:2019dwd}.
The coupled Boltzmann equations for $Y_{\chi}$, $Y_{\phi}$, $Y_{\mathbb{N}_i}$ and the temperature ratio
$\eta \equiv T_v/T_h$ are~\cite{Aboubrahim:2020lnr}
\begin{align}
	\frac{\text{d}Y_{\chi}}{\text{d}T} & =-\frac{s}{H}\frac{{\rm d}\rho_{h}/{\rm d}T_{h}}{4\rho_{h}-j_{h}/H}\,\mathcal{C}_{\chi}\,,\label{eq:FIBEchi}\\
	\frac{\text{d}Y_{\phi}}{\text{d}T} & =-\frac{s}{H}\frac{{\rm d}\rho_{h}/{\rm d}T_{h}}{4\rho_{h}-j_{h}/H}\,\mathcal{C}_{\phi}\,,\label{eq:FIBEphi}\\
	\frac{\text{d}Y_{\mathbb{N}_{i}}}{\text{d}T} & =-\frac{s}{H}\frac{{\rm d}\rho_{h}/{\rm d}T_{h}}{4\rho_{h}-j_{h}/H}\,\mathcal{C}_{\mathbb{N}_{i}}\,,\\
	\frac{{\rm d}\eta}{{\rm d}T_{h}} & =-\frac{A_{v}}{B_{v}}+\frac{\rho_{v}+j_{h}/(4H)}{\rho_{h}-j_{h}/(4H)}\frac{\frac{{\rm d}\rho_{h}}{{\rm d}T_{h}}}{B_{v}}\,,\label{eq:FIBENi}
\end{align}
where $A_{v}$ and $B_{v}$ can be found in Appendix~\ref{App:MultiT},
and the energy transfer density is given by
\begin{align}
	j_{h} & =\sum_{i,j=1}^{3}\bigl[2\,\theta(m_{\phi}-m_{\chi}-m_{i})\bigl(-Y_{\phi}sJ_{\phi\to\overline{\chi}\mathbb{V}_{i}}+Y_{\chi}Y_{\mathbb{V}_{i}}s^{2}J_{\overline{\chi}\mathbb{V}_{i}\to\phi}\bigr)\nonumber \\
	& +\theta(M_{i}-m_{W}-m_{\ell_{j}})\bigl(-Y_{\mathbb{N}_{i}}sJ_{\mathbb{N}_{i}\to W\ell_{j}}+Y_{W}Y_{\ell_{j}}s^{2}J_{W\ell_{j}\to\mathbb{N}_{i}}\bigr)\nonumber \\
	& +2\,\theta(M_{i}-m_{h}-m_{j})\bigl(-Y_{\mathbb{N}_{i}}sJ_{\mathbb{N}_{i}\to h\mathbb{V}_{j}}+Y_{h}Y_{\mathbb{V}_{j}}s^{2}J_{h\mathbb{V}_{j}\to\mathbb{N}_{i}}\bigr)+(h\leftrightarrow Z)\nonumber \\
	& +2\bigl(Y_{h}Y_{\mathbb{V}_{i}}s^{2}J_{h\mathbb{V}_{i}\to\chi\phi}-Y_{\chi}Y_{\phi}s^{2}J_{\chi\phi\to h\mathbb{V}_{i}}\bigr)+2(h\leftrightarrow Z)\nonumber \\
	& +4\bigl(Y_{W}Y_{\ell_{i}}s^{2}J_{W\ell_{i}\to\chi\phi}-Y_{\chi}Y_{\phi}s^{2}J_{\chi\phi\to W\ell_{i}}\bigr)\bigr]\,.
\end{align}
The evolution of dark particles $\chi,\phi$ as well as the right-handed neutrinos $\mathbb{N}_{i=1,2,3}$
is computed using the coupled Boltzmann equations above, and the results are presented in the following subsection.

If hidden sector interactions, such as $\chi\overline{\chi} \leftrightarrow \phi\overline{\phi}$,
are too weak to establish internal equilibrium,
their effects on the dark matter relic density can be safely neglected.
In this regime, the freeze-in production of dark matter can be computed using a two-step procedure:
(1)~first determine the freeze-in production of $\chi$ and $\phi$ from SM particles,
assuming that $\phi$ does not interact with $\chi$;
(2)~if $\phi$ can decay into $\chi$, then include its contribution
by calculating the decay of $\phi$ after the freeze-in production has completed.

However, this treatment is not appropriate when the hidden sector possesses sufficiently strong internal interactions
and can thus maintain thermal equilibrium.
In such cases, the resulting error can be substantial, as shown by explicit comparison.

\subsection{Evolution of the benchmarks}

Using the PSO algorithm described in Section~\ref{Sec:PSO},
we determine seesaw parameter sets for all three cases (Cases RS, SC, and SS),
and identify five representative benchmarks that reproduce the observed neutrino data,
as listed in Appendix~\ref{App:SSBM}.

\subsubsection{Freeze-out evolution, Cases RS, SC}

\begin{figure}
	\centering
	\includegraphics[scale=0.4]{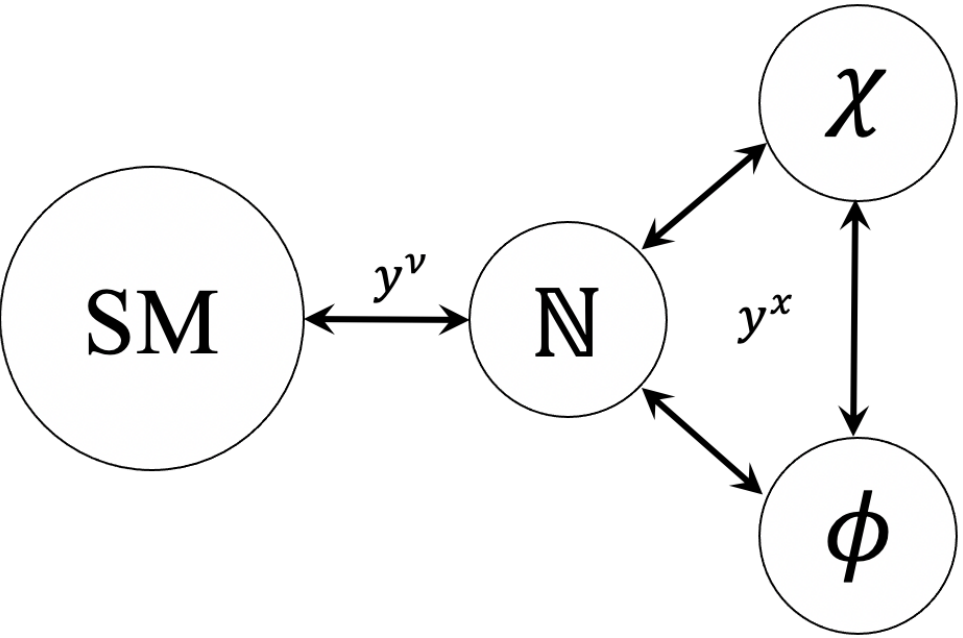}
	\caption{A schematic illustration of the dark sector evolution via freeze-out.
    At high temperatures, the dark particles $\chi$ and $\phi$ remain in thermal
    equilibrium with the SM plasma through the right-handed neutrino portal.
    The dominant freeze-out channels are
    $\chi\overline{\chi}, \phi\overline{\phi} \to \mathbb{N}\mathbb{N}$,
    and $\mathbb{N}$ subsequently decays into SM particles.}
	\label{Fig:FO}
\end{figure}

The freeze-out scenario is illustrated schematically in Fig.~\ref{Fig:FO}, where the dark sector
particles $\chi$ and $\phi$ maintain thermal equilibrium with the SM sector via the right-handed neutrinos.
The dark particles $\chi$ and $\phi$ annihilate efficiently into pairs of right-handed neutrinos,
which subsequently decay into SM particles.
In this paper we assume that $\phi$ is heavier than the dark matter candidate $\chi$,
and that $\phi$ eventually decays completely before BBN.
For the freeze-out scenario,
we investigate five representative benchmarks for Cases RS and SC,
as presented in Table~\ref{Tab:FOmodels}.
Parameters are chosen to satisfy all current experimental constraints, shown in Section~\ref{sec:ex},
as well as the observed abundance of the dark matter.

\begin{table}[h!]
	\setlength{\tabcolsep}{8pt}
    \begin{center}
    \begin{tabular}{|c|c|c|c|c|c|c|c|}
\hline
Freeze-out & Benchmark & $m_{\chi}$ & $m_{\phi}$ & $y_{1,2,3}^{x}$ & $\Omega_{\chi}h^{2}$ & $\Omega_{\chi}^\prime h^{2}$ & $\tau_\phi$ [sec] \tabularnewline
\hline
Model $a$ & \multirow{2}{*}{RS} & $230$ & $420$ & $0.454$ & $0.12$ & $0.106$ & $7.38\times 10^{-13}$ \tabularnewline
\cline{1-1}\cline{3-8}
Model $b$ &  & $230$ & $445$ & $0.485$ & $0.12$  & $9.34\times 10^{-4}$ &  $4.74\times 10^{-23}$ \tabularnewline
\hline
Model $c$ & \multirow{3}{*}{SC} & $230$ & $420$ & $0.573$ & $0.12$  & $4.84\times 10^{-2}$ & $ 4.49\times 10^{-20}$ \tabularnewline
\cline{1-1}\cline{3-8}
Model $d$ &  & $230$ & $445$ & $ 0.600 $ & $0.12$  & $8.42\times 10^{-4}$ & $3.50\times 10^{-23}$ \tabularnewline
\cline{1-1}\cline{3-8}
Model $e$ &  & $98$ & $110$ & $0.17$ & $0.12$ & $-$ & $2.24\times 10^{-17}$ \tabularnewline
\hline
     \end{tabular}
     \end{center}
	 \caption{Freeze-out benchmarks we discuss in this work for Cases RS and SC. All masses are in units of GeV. The lifetimes of $\phi$ ($\tau_\phi$) and right-handed neutrinos $\mathbb{N}_{1,2,3}$ for each model are ensured to be less than one second. The dark fermion $\chi$ serves as the dark matter candidate with a full occupation of the observed dark matter relic density. For simplicity, we set the values of coupling $y_{1,2,3}^{x}$ to be the same. The lifetimes of $\phi$ in seconds are listed in the last column for the five models.
For Models $a-d$, we also calculate the evolution of $\chi$ alone under the assumption that all other species remain in equilibrium throughout,
and summarize the corresponding relic density in the $\Omega_{\chi}' h^{2}$ column.
This is intended to illustrate the importance of the internal dark sector interactions, as will be discussed in detail at the end of this section.}
	\label{Tab:FOmodels}
\end{table}

In the dark sector, we restrict ourselves to a minimal field content consisting only of $\chi$ and $\phi$.
Their cosmological evolution is assumed to proceed exclusively through the right-handed neutrino portal,
without any additional dark interactions or mediators, such as a $U(1)$ dark photon.
The benchmarks considered here are therefore representative of the minimal right-handed neutrino portal dark sector.
Specifically, we focus on the case in which the lighter fermion $\chi$ serves as the dark matter candidate with a mass around the electroweak scale,
and the freeze-out process $\chi\overline{\chi}\to \mathbb{N}\mathbb{N}$ can efficiently deplete its abundance.
We also distinguish two qualitatively different possibilities for the scalar $\phi$,
depending on whether the decay $\phi\to \mathbb{N}\chi$ is kinematically open.
In addition, within Case-SC there exists a special regime in which both $\chi$ and $\phi$ are lighter than $\mathbb{N}_1$.
In this case the channels $\chi\overline{\chi},\,\phi\overline{\phi}\to \mathbb{N}_i\mathbb{N}_j$ are suppressed,
and the coannihilation process $\chi\phi\to LH$ instead becomes the dominant mechanism for depleting the dark sector abundance.

\paragraph{Case-RS freeze-out evolution}

\begin{figure}
	\centering
	\subfigure[Model $a$]{
		\includegraphics[scale=0.26]{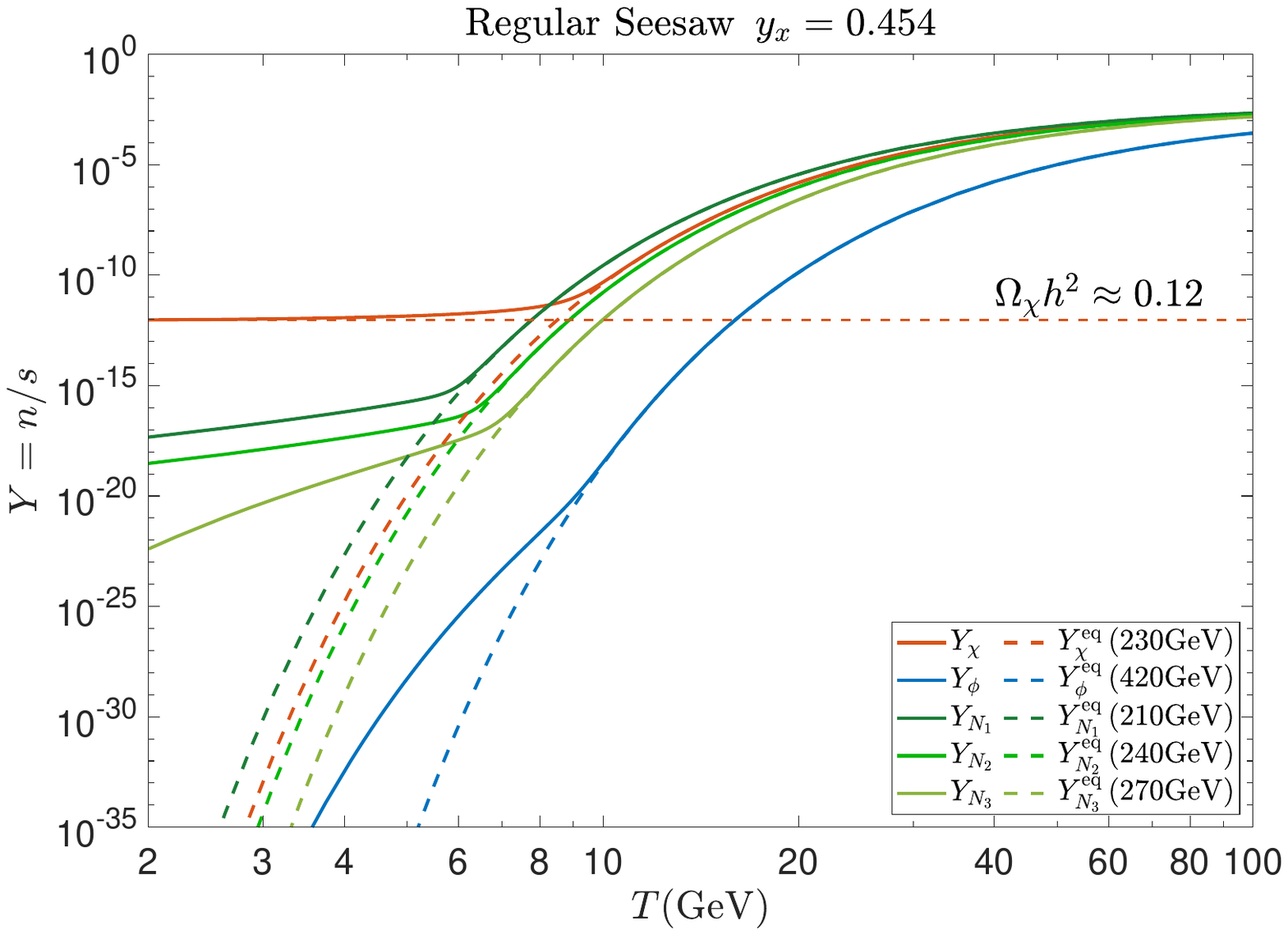} }\quad
	\subfigure[Model $b$]{
		\includegraphics[scale=0.26]{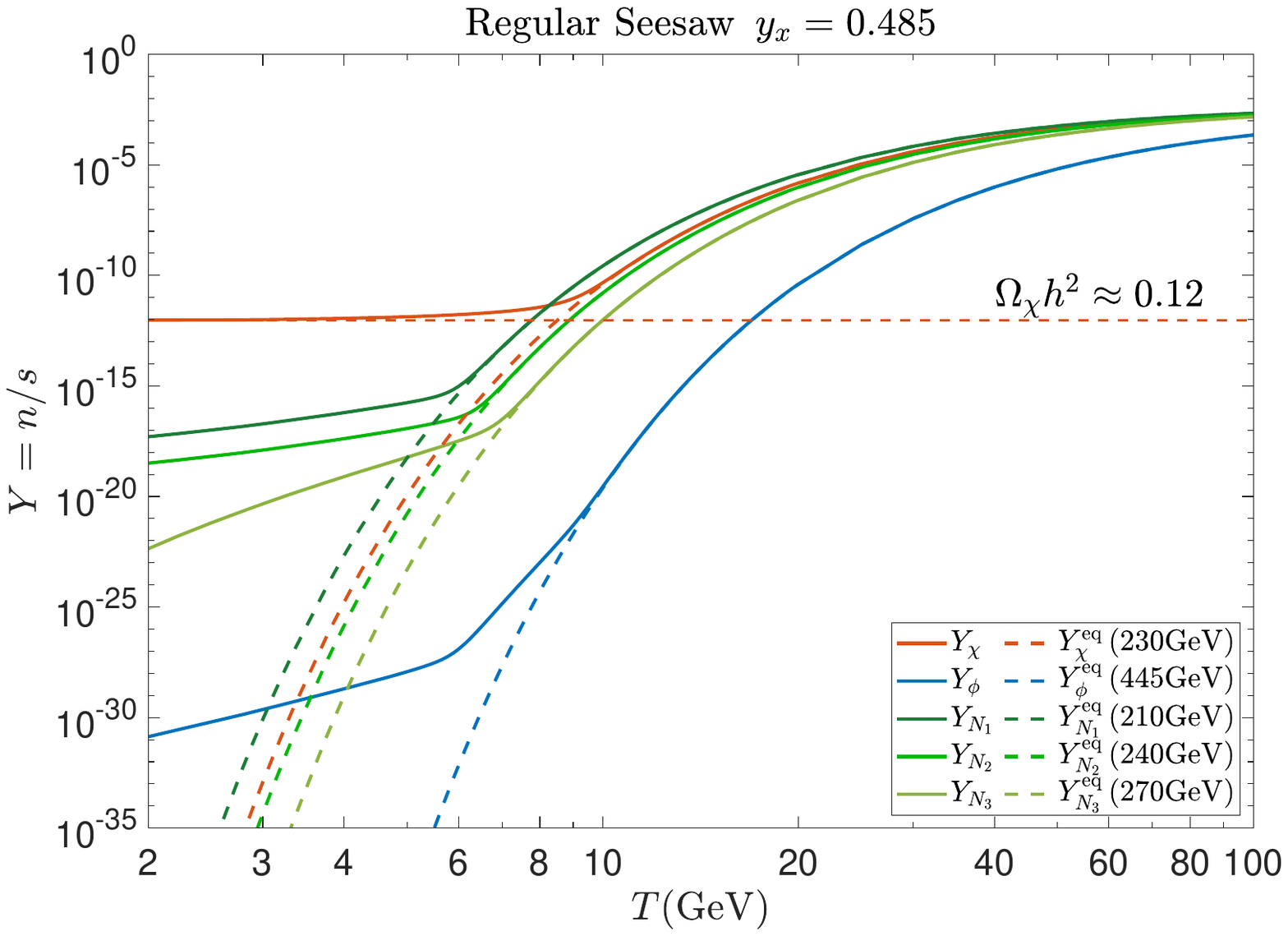} }\\
	\subfigure[Model $c$]{
	\includegraphics[scale=0.26]{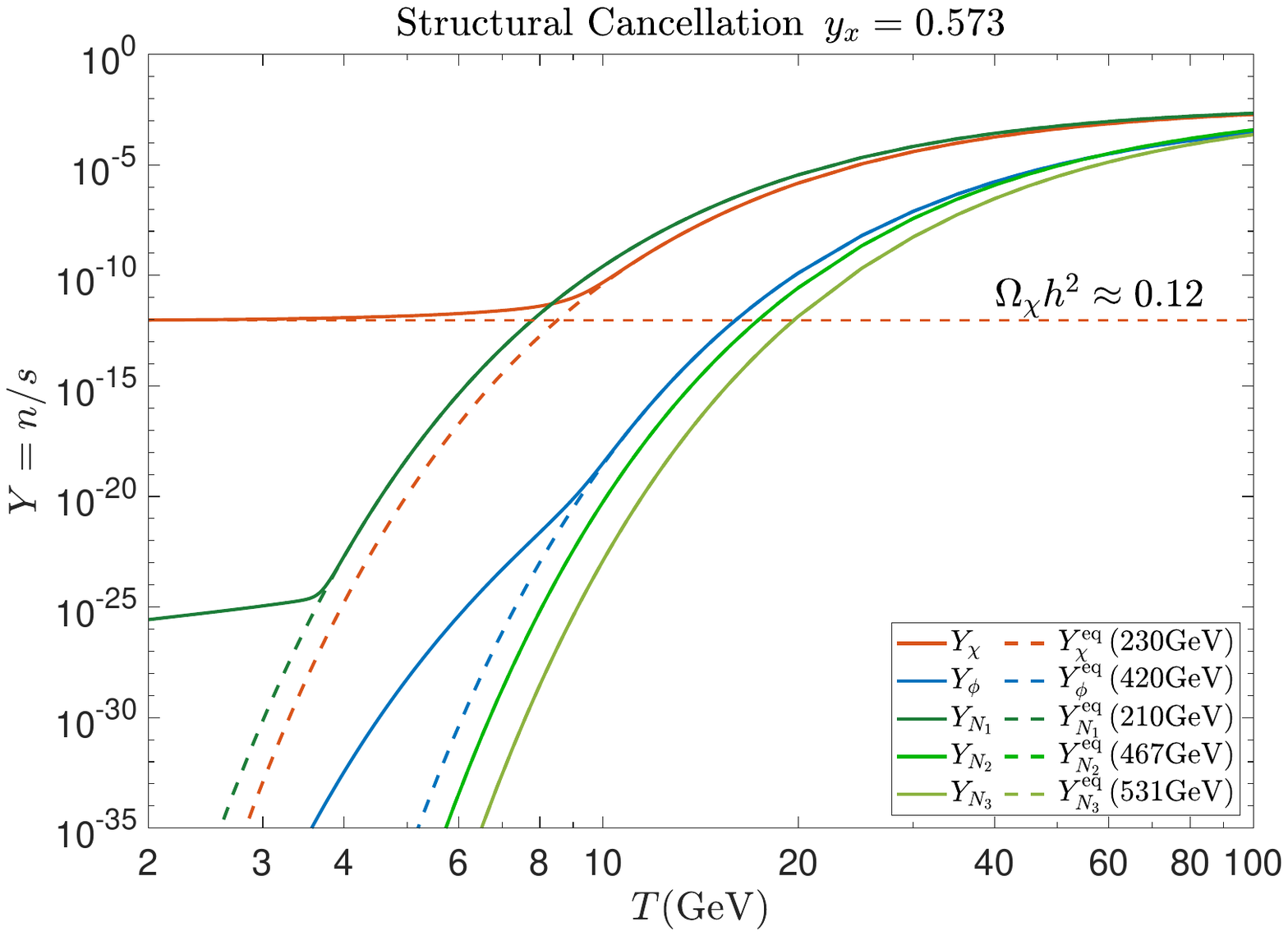} }\quad	
	\addtocounter{subfigure}{+1}
	\subfigure[Model $e$]{
	\includegraphics[scale=0.26]{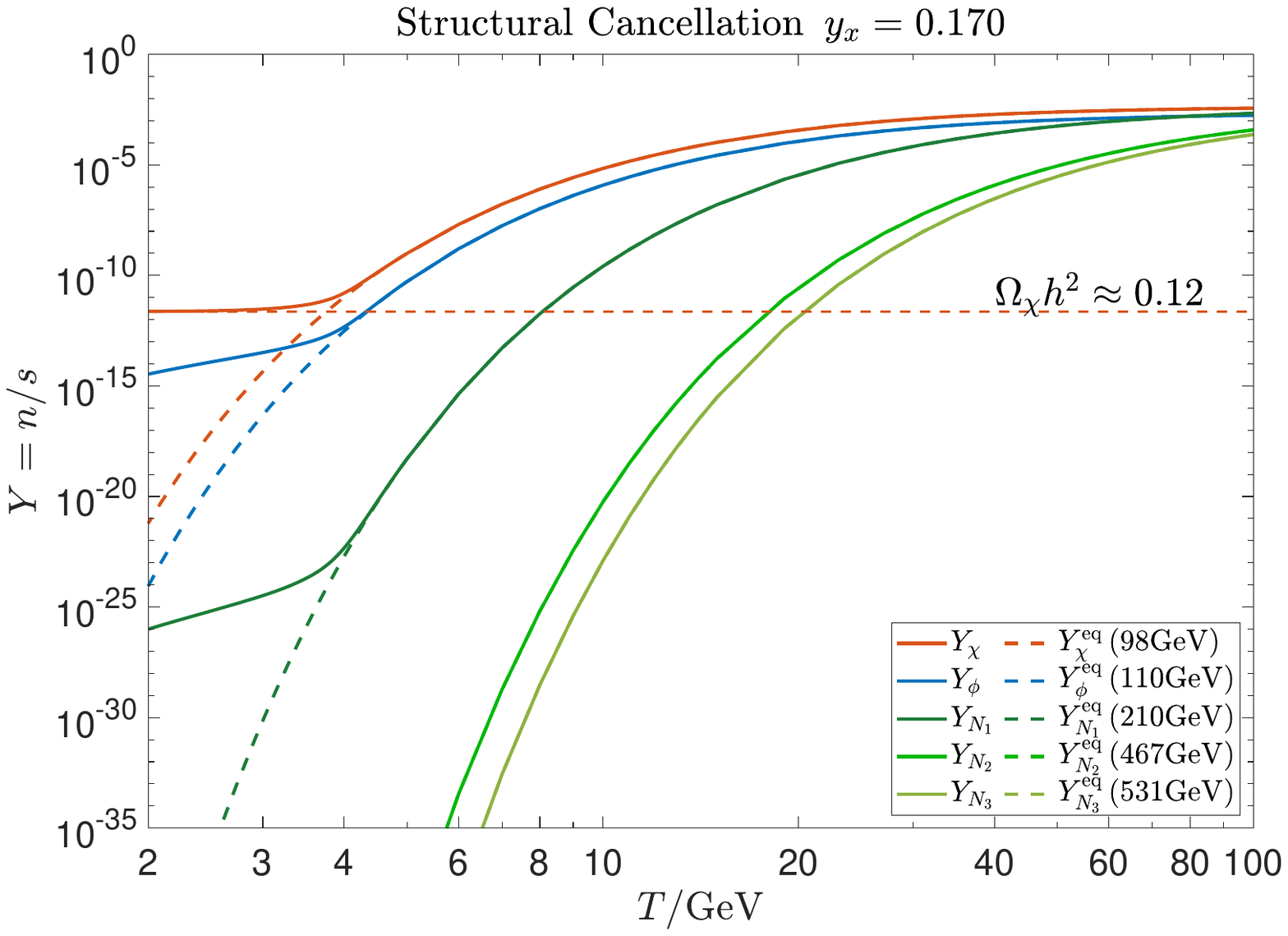} }
	\caption{Dark sector freeze-out evolution for Models~$a,b$ (Benchmark-RS) and Models~$c,e$ (Benchmark-SC). For each model, the figure shows the evolution of the comoving number densities of all dark particles $\chi,\phi$, and $\mathbb{N}_{i=1,2,3}$, where the dark fermion $\chi$ serves as the dark matter candidate. The red horizontal dashed line indicates the observed dark matter relic density with dark matter mass equal to $m_{\chi}$. For Benchmark-RS, the difference between two models is as follows. In Model $a$, $\phi$ freezes out via $\phi\overline{\phi}\to\mathbb{N}_{i}\mathbb{N}_{j}$ and annihilates into $\chi$ through $\phi\phi\to\overline{\chi\chi}$, $\phi\overline{\phi}\to\chi\overline{\chi}$. At later times, all remaining $\phi$ decay through $\phi\to\overline{\chi}\mathbb{V}_{i=1,2,3}$ before BBN. While in Model $b$, the mass hierarchy $m_{\phi}>(m_{\chi}+M_{1})$ allows the internal decay channel $\phi\to\overline{\chi}\mathbb{N}_{1}$. For Benchmark-SC,
	the portal neutrinos $\mathbb{N}_{i=1,2,3}$ remain tightly coupled to the thermal bath, leading to a delayed freeze-out.
In Model~$e$, the dominant depletion channel for the dark particles is the resonant coannihilation process $\chi \phi \to L H$.}
	\label{Fig:Y_RS_SC}
\end{figure}

For Model~$a$ taking the Benchmark-RS seesaw parameters, the evolution is shown in Fig.~\ref{Fig:Y_RS_SC} (a). Owing to the mass hierarchy $m_{\chi}>M_{1}$, the dominant process depleting dark matter $\chi$ is the secluded channel $\chi\overline{\chi}\to\mathbb{N}_{1}\mathbb{N}_{1}$.
In this case the couplings between the hidden and visible sectors are relatively weak, so the portal neutrinos $\mathbb{N}_{i=1,2,3}$ decrease through interactions
$\mathbb{N}_{i}\mathbb{N}_{j}\to\chi\overline{\chi}$ ($i,j=1,2,3$) for $M_i + M_j > 2m_\chi$,
which simultaneously act as a production channel for $\chi$. At the same time, $\mathbb{N}_{i}$ decay to SM particles via $\mathbb{N}_{i}\to W^{\pm}\ell_{j}^{\mp}$, with subdominant channels $\mathbb{N}_{i}\to h\mathbb{V}_{j},Z\mathbb{V}_{j}$ being suppressed by the tiny Yukawa couplings or neutrino mixing. All right-handed neutrinos eventually decrease to zero. For the particle $\phi$, the dominant interactions
are $\phi\phi\to\overline{\chi\chi}$, $\phi\overline{\phi}\to\chi\overline{\chi}$, and $\phi\overline{\phi}\to\mathbb{N}_{i}\mathbb{N}_{j}$.
Dark particle annihilation through $\chi \phi \to L H$ plays only a subdominant role.
At later times, all remaining $\phi$ decay through $\phi\to\overline{\chi}\mathbb{V}_{i=1,2,3}$ before BBN.

The evolution for Model $b$ is shown in Fig.~\ref{Fig:Y_RS_SC} (b). In this case, the scalar mass $m_{\phi}$ is raised such that $m_{\phi}>(m_{\chi}+M_{1})$, which opens an efficient decay channel $\phi\to\overline{\chi}\mathbb{N}_{1}$ together with
its inverse process. 
We will discuss the importance of this process at the end of this subsection together with the other benchmark points.
The portal neutrinos can deplete via internal interactions $\mathbb{N}_{i}\mathbb{N}_{j}\to\chi\overline{\chi}$,
and can decay into SM particles via $\mathbb{N}_{i}\to W^{\pm}\ell_{j}^{\mp},h\mathbb{V}_{j},Z\mathbb{V}_{j}$,
similar to Model $a$.
With a suitable choice of parameters, the abundance of the dark fermion $\chi$
eventually reaches the observed dark matter relic density,
while other dark particles are depleted to negligible levels before BBN.

\paragraph{Case-SC freeze-out evolution}


For Case-SC, the Yukawa couplings $y^\nu_{ij}$ and the neutrino mixing elements $|U_{ij}|$, presented in Appendix~\ref{App:SSBM},
ensure that the two sectors are initially in thermal equilibrium.
The evolution of Model~$c$ taking the Benchmark-SC seesaw parameters is plotted in Fig.~\ref{Fig:Y_RS_SC} (c),
where the portal neutrinos $\mathbb{N}_{i}$ remain closely coupled to the thermal bath.
The evolution is similar to Model~$a$, in which $\chi$ predominantly annihilates through $\chi\overline{\chi}\to\mathbb{N}_{1}\mathbb{N}_{1}$, and processes $\mathbb{N}_{i}\mathbb{N}_{j}\to\chi\overline{\chi}$,
$\phi\phi\to\overline{\chi\chi}$, and $\phi\overline{\phi}\to\chi\overline{\chi}$ proceed simultaneously.
$\phi$ can annihilate via $\phi\phi\to\overline{\chi\chi}$, $\phi\overline{\phi}\to\chi\overline{\chi}$, $\phi\overline{\phi}\to\mathbb{N}_{i}\mathbb{N}_{j}$,
and after $\phi$ freezes out, it also decays completely via $\phi \to \overline{\chi}\mathbb{V}_{i}$ before BBN.
The coannihilation $\chi \phi \to L H$ is suppressed.

Model~$e$ provides another representative scenario in which both $\chi$ and $\phi$ are lighter than the heavy Majorana neutrinos,
i.e., $m_{\chi}, m_{\phi} < M_i$.
In this case, the annihilation channels $\chi\overline{\chi}\to \mathbb{N}\mathbb{N}$ and $\phi\overline{\phi}\to \mathbb{N}\mathbb{N}$
are kinematically suppressed.
As a result, the effective depletion process for the dark particles is the coannihilation channel $\chi\phi\to LH$,
mediated by heavy Majorana neutrinos in the $s$-channel.
Given the sizes of the couplings $y^\nu$ and $y^x$,
this process can reproduce the observed relic density only in the resonant regime,
where $m_{\chi}+m_{\phi}\simeq M_i$.
Other annihilation channels, such as $\chi\overline{\chi}\to \mathbb{V}_i\mathbb{V}_j$, $\chi\phi\to h\mathbb{V}_i,\, Z\mathbb{V}_i,\, W\ell_i$, are all suppressed.
After the coannihilation of $\chi$ and $\phi$ into the SM sector, $\phi$ subsequently decays into $\chi$ and light neutrinos and becomes negligible before BBN, while $\chi$ attains the observed dark matter relic abundance.
In Model~$e$, we choose the resonant condition $m_{\chi}+m_{\phi}\simeq M_1$. Other representative benchmarks can also be obtained through the $\mathbb{N}_2$ or $\mathbb{N}_3$ resonance, with $m_{\chi}+m_{\phi}\simeq M_2$ or $M_3$, respectively.\footnote{
The resonance enhancement through the coannihilation channel
\begin{equation}
\chi + \phi \xrightarrow[y_x  \qquad \quad y_\nu]{\mathbb{N}}  L + H
\end{equation}
differs from the well-studied resonant annihilation in a $U(1)$ portal scenario.
In the present case, the cross section near the resonance can be schematically written as
\begin{equation}
\sigma_{\mathbb{N}_1} \sim \frac{\mathcal{O}(y_x^2 y_\nu^2)}{\Gamma_1^2 M_1^2}
\sim \frac{\mathcal{O}(y_x^2 y_\nu^2)}{\mathcal{O}(y_x^4) M_1^2}\,,
\end{equation}
where $\Gamma_1$ is the decay width of $\mathbb{N}_1$ and approximately scales as $y_x^2$.
By contrast, in the $U(1)$ resonant case one typically has
\begin{equation}
\sigma_{Z^\prime} \sim \frac{\mathcal{O}(g_x^4)}{\Gamma^{\prime 2} M^{\prime 2}}
\sim \frac{\mathcal{O}(g_x^4)}{\mathcal{O}(g_x^4) M^{\prime 2}}\,,
\end{equation}
where $\Gamma^{\prime}$ and $M^{\prime}$ denote the decay width and mass of the $U(1)$ gauge boson $Z^\prime$, respectively.

Therefore, in the present setup, the resonant enhancement arises from a nontrivial interplay
between the numerator and denominator of the cross section.
In particular, increasing $y_x$ does not necessarily lead to a larger cross section:
if $y_x$ becomes too large, the corresponding increase in the decay width can instead significantly suppress the resonant contribution.
By contrast, in the $U(1)$ portal case, the cross section increases monotonically with $g_x$.}

\paragraph{Impact of dark sector dynamics on the dark matter evolution}

One of the most important findings of the above analysis is that dark sector dynamics can have a substantial impact on the dark matter evolution.
As shown in Fig.~\ref{Fig:Y_RS_SC}, Models $a,b,c$ exhibit qualitatively similar behavior.
In particular, until $\chi$ freezes out, the abundances of $\phi$ and the mediators $\mathbb{N}_{i=1,2,3}$ remain at their equilibrium values.
Only after $\chi$ departs from equilibrium do $\phi$ and $\mathbb{N}_i$ also begin to deviate from equilibrium.

This may give the impression that the evolution of $\phi$ and $\mathbb{N}_i$ has no impact on the evolution of  the dark matter $\chi$,
and therefore no effect on the $\chi$ relic density.
This is, however, not the case.
To demonstrate the importance of this effect,
we also compute the evolution of $\chi$ alone, assuming that all other species remain in equilibrium \emph{throughout.}
This approximation yields a relic abundance significantly smaller than that obtained from the full coupled Boltzmann system,
as shown in the $\Omega_{\chi}' h^{2}$ column of Table~\ref{Tab:FOmodels}.
The reason is that forcing $\phi$ and $\mathbb{N}_i$ to remain in equilibrium effectively keeps $\chi$ in active interactions for a longer time,
delays freeze-out, and thereby further suppresses the final relic abundance.

Comparing Models $a$ and $b$, the process $\phi \leftrightarrow \chi \mathbb{N}_1$ is forbidden in Model $a$,
while it is kinematically allowed in Model $b$.
Assuming that $\phi$ and $\mathbb{N}_i$ remain in equilibrium,
this process in Model $b$ keeps $\chi$ active in the thermal bath for a longer time and thus delays its freeze-out.
Consequently, the relic abundance in Model $b$ is further suppressed compared with that in Model $a$, where this process is forbidden.
A similar pattern is also observed for Models $c$ and $d$,
where the process $\phi \leftrightarrow \chi \mathbb{N}_1$ is forbidden in Model $c$ but allowed in Model $d$.

Comparing Models $a$ and $c$, although the process $\phi \leftrightarrow \chi \mathbb{N}_1$ is forbidden in both models,
$\phi$ can still decay into $\chi$ and light neutrinos via the channel $\phi \leftrightarrow \chi \mathbb{V}_{i=1,2,3}$.
Since the active--heavy neutrino mixings are relatively larger in Model $c$ under Case-SC than in Model $a$ under Case-RS,
this channel is more effective in Model $c$.
Consequently, assuming that all other species remain in equilibrium,
the relic abundance obtained by evolving $\chi$ alone is further suppressed in Model $c$ than in Model $a$.

The departures from equilibrium exhibited by other species after dark matter freeze-out,
as shown in Fig.~\ref{Fig:Y_RS_SC}, are therefore of essential importance.
They encode the coupled evolution of the dark sector particles
and show that the full set of relevant Boltzmann equations must be solved in order to obtain the correct dark matter evolution and relic density.
Although the evolution curves of the dark sector particles may appear similar from Models $a-d$,
the underlying physics, as well as the detailed evolution, can differ significantly.

Model~$e$ represents a distinct regime in which both $\chi$ and $\phi$ are lighter than the heavy Majorana neutrinos,
and thus the channels $\chi \overline{\chi}\to \mathbb{N}\mathbb{N}$ and $\phi \overline{\phi}\to \mathbb{N}\mathbb{N}$ become ineffective.
The relic abundance is then determined mainly by the resonant coannihilation process $\chi\phi\to LH$ with $m_{\chi}+m_{\phi}\simeq M_i$, followed by the decay of $\phi$ into $\chi$ and light neutrinos before BBN.
This case further demonstrates that, in the right-handed neutrino portal scenario,
dark matter depletion can proceed directly into SM particles,
rather than primarily through secluded annihilation into right-handed neutrinos.

With seesaw parameters that consistently reproduce the observed light neutrino masses and mixings,
the common features of right-handed neutrino portal dark matter in the freeze-out scenario can be summarized as follows:
\begin{itemize}
\item The dominant depletion channel of the dark matter particle $\chi$ is generally $\chi\overline{\chi}\to \mathbb{N}\mathbb{N}$, as long as $m_\chi$ is not significantly smaller than $M_1$.
    Otherwise, the depletion of $\chi$ is governed mainly by the resonant coannihilation process $\chi\phi\to LH$.
\item Although the evolution plots show that the abundances of $\phi$ and $\mathbb{N}_{i=1,2,3}$ remain close to their equilibrium values before $\chi$ freeze-out, their effect on the dark matter evolution cannot be neglected by simply setting them equal to their equilibrium abundances.
\end{itemize}

\subsubsection{Freeze-in evolution, Cases RS, SC, SS}

\begin{figure}
	\centering
		\includegraphics[scale=0.4]{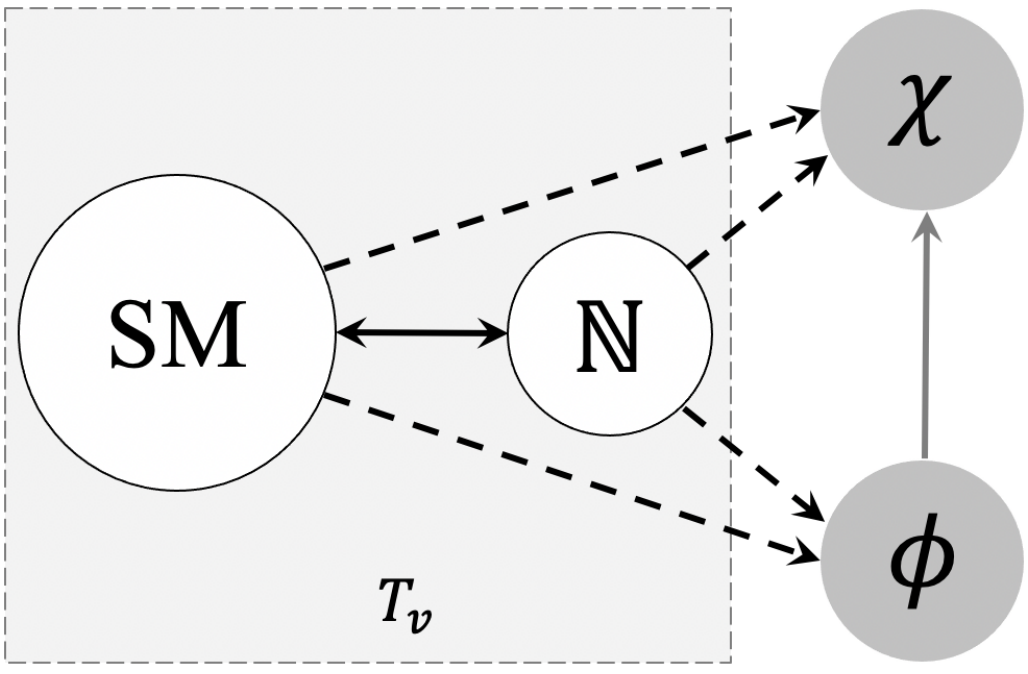}
	\caption{For Cases RS and SC, the right-handed neutrinos reach thermal equilibrium
    during the main freeze-in epoch of the dark particles and can thus be treated
    as part of the visible sector, sharing the visible sector temperature.
    A consistent model further requires that $\phi$ decay completely before BBN,
    so the corresponding coupling $y^{x}$ is not large enough to maintain internal thermal equilibrium within the dark sector.
    Hence, the dark particles are produced via freeze-in through the processes
    $\mathbb{N}\mathbb{N}\to \chi\overline{\chi},\phi\overline{\phi}$ and $L H \to \chi \phi$.}
	\label{Fig:FI1}
\end{figure}

\begin{figure}
	\centering
		\includegraphics[scale=0.45]{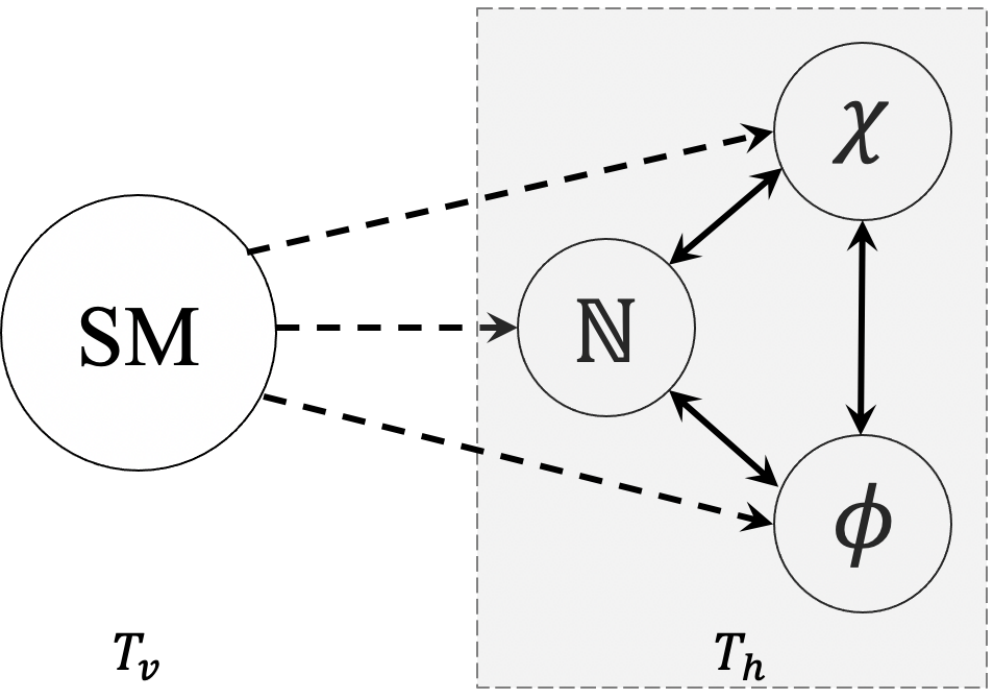}\qquad
		\includegraphics[scale=0.45]{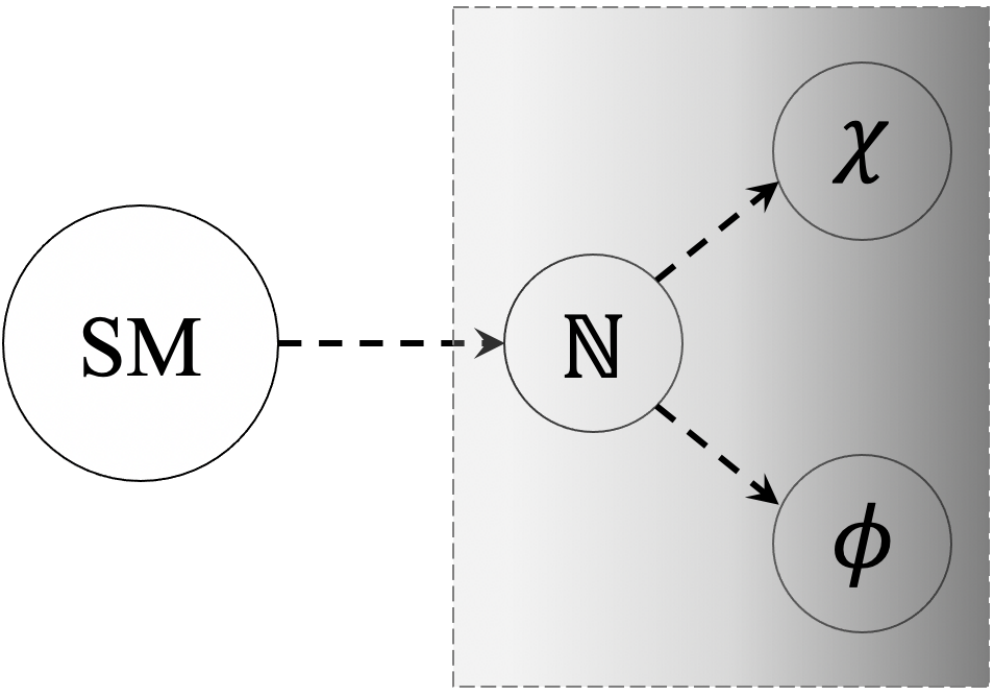}
	\caption{For Case-SS, the portal right-handed neutrino $\mathbb{N}_1$ couples ultraweakly to the SM
    and therefore never reaches thermal equilibrium with it.
    Depending on the value of $y^{x}$, the dark sector can either reach internal equilibrium (left panel),
    in which case one can define a hidden sector temperature $T_h$ (Models $d,e$),
    or remain out of internal equilibrium (right panel), in which case both $\chi$ and $\phi$ are dark matter candidates (Model $f$).}
	\label{Fig:FI2}
\end{figure}

The freeze-in scenario is illustrated schematically in Figs.~\ref{Fig:FI1} and~\ref{Fig:FI2}.
There are two distinct possibilities, depending on whether
the right-handed neutrinos can reach thermal equilibrium with the SM sector.

Seen from Eq.~(\ref{eq:DSN}), after including the seesaw mixing,
$\phi$ can decay into $\chi$ and a light neutrino via
$\phi \to \overline{\chi} +\mathbb{V}$,
with an effective coupling constant
\begin{equation}
  y^{x\prime} \sim y^{x} U_{\mathbb{N}\mathbb{V}} \,,
\end{equation}
where $U_{\mathbb{N}\mathbb{V}}$ is the seesaw mixing between light and heavy
neutrinos in the mass eigenstate basis, as defined in Eqs.~(\ref{eq:LNu}) and~(\ref{eq:HNu}).

For $\chi$ and $\phi$ masses of order a few hundred GeV,
the scalar $\phi$ decays before BBN if $y^{x\prime} \gtrsim 10^{-12}$,
whereas its lifetime exceeds the age of the Universe for $y^{x\prime} \lesssim 10^{-22}$.

\paragraph{Case-RS freeze-in}
In Case-RS, the electroweak right-handed neutrinos have Yukawa couplings of order $y^{\nu}\sim 10^{-7}$,
resulting in a seesaw mixing $U_{\mathbb{N}\mathbb{V}}\sim 10^{-8} - 10^{-7}$.
These electroweak right-handed neutrinos reach thermal equilibrium with the SM particles below the TeV scale.
There are two cases:
\begin{enumerate}
  \item Both $\chi,\phi$ are DM.
  The requirement of
  $y^{x\prime} \lesssim 10^{-22}$ gives rise to $y^{x} \lesssim 10^{-14}$.
  \begin{itemize}
  \item For $M_{N}>m_{\chi}+m_{\phi}$, the dominant freeze-in  process is $\mathbb{N}\to\chi+\phi$.
  Given $y^{x} \lesssim 10^{-14}$, the freeze-in production is not sufficient. {\color{red}\ding{55}}
  \item For $M_{N}<m_{\chi}+m_{\phi}$, the freeze-in processes are $\mathbb{N}\mathbb{N}\to\chi\overline\chi$
  and $\mathbb{N}\mathbb{N}\to\phi\overline\phi$.
  The four-point freeze-in production is less efficient compared to the decaying process in the previous case.
  The contribution from the other freeze-in process $LH\to\chi\phi$ is even more suppressed. {\color{red}\ding{55}}
  \end{itemize}
  \item Only $\chi$ is DM. In this case $\phi$ decays before BBN,
  thus $y^{x\prime} \gtrsim 10^{-12}$ requires $y^{x}\gtrsim 10^{-7}$:
  \begin{itemize}
  \item For $M_{N}>m_{\chi}+m_{\phi}$, the dominant freeze-in process is $\mathbb{N}\to\chi+\phi$.
  Given that $y_{x}\gtrsim 10^{-7}$, the freeze-in production is too large,
  leading to an overproduction of dark matter. {\color{red}\ding{55}}
  \item For $M_{N}<m_{\chi}+m_{\phi}$, the freeze-in processes are $\mathbb{N}\mathbb{N}\to\chi\overline\chi$, $\mathbb{N}\mathbb{N}\to\phi\overline\phi$, and $LH\to\chi\phi$. 
  Freeze-in models $a,b$ for Benchmark RS are shown in Table~\ref{Tab:FImodels}. {\color{Green}\ding{51}}
\end{itemize}
\end{enumerate}

\paragraph{Case-SC freeze-in}
In Case-SC, the electroweak right-handed neutrinos have Yukawa couplings of order $y^{\nu}\sim 10^{-3}$,
resulting in a seesaw mixing $U_{\mathbb{N}\mathbb{V}}\sim 10^{-4} - 10^{-3}$.
These electroweak right-handed neutrinos remain in thermal equilibrium with the SM particles.
There are also two cases:
\begin{enumerate}
  \item Both $\chi,\phi$ are DM.
  The requirement $y^{x\prime} \lesssim 10^{-22}$ implies $y^{x} \lesssim 10^{-18}$,
  which is far too small to generate a significant freeze-in abundance. {\color{red}\ding{55}}
  \item Only $\chi$ is DM.
  In this case  $y^{x\prime} \gtrsim 10^{-12}$ then requires $y^{x} \gtrsim 10^{-9}$.
  \begin{itemize}
  \item For $M_{N} > m_{\chi} + m_{\phi}$, the dominant freeze-in process is $\mathbb{N} \to \chi + \phi$,
  which leads to an overproduction of dark matter via freeze-in. {\color{red}\ding{55}}
  \item For $M_{N}<m_{\chi}+m_{\phi}$, the freeze-in processes are $\mathbb{N}\mathbb{N}\to\chi\overline\chi$, $\mathbb{N}\mathbb{N}\to\phi\overline\phi$, and $LH\to\chi\phi$.
  Freeze-in model $c$ for Benchmark SC is shown in Table~\ref{Tab:FImodels}. {\color{Green}\ding{51}}
  \end{itemize}
\end{enumerate}

The freeze-in production of dark matter for Cases RS and SC is illustrated schematically in Fig.~\ref{Fig:FI1}.
During the main freeze-in epoch ($T \sim m_{\rm DM}$), the right-handed neutrinos are in thermal equilibrium with the SM sector.
However, as discussed above, the dark sector coupling $y^x$ is not large enough,
so the internal dark sector interactions are too weak to establish thermal equilibrium within the dark sector.

\paragraph{Case-SS freeze-in}

In Case-SS, one of the electroweak right-handed neutrinos (taken to be $\mathbb{N}_1$)
has a Yukawa coupling much smaller than the typical seesaw coupling, $y^{\nu} \ll 10^{-7}$,
resulting in the seesaw mixing $U_{\mathbb{N}_1\mathbb{V}}\lesssim 10^{-10}$ for $\mathbb{N}_1$.
Because of this feeble interaction, $\mathbb{N}_1$ never reaches thermal equilibrium with the SM bath.
Consequently, dark matter can only be produced via freeze-in through the $\mathbb{N}_1$ portal.\footnote{
In Case-SS, we assume that the heavier right-handed neutrinos $\mathbb{N}_{2,3}$ are responsible for generating the observed neutrino masses and mixing pattern. Accordingly, the corresponding Yukawa couplings $y_{i2}, y_{i3}$ are taken to lie in the range estimated by Eq.~\eqref{eq:SeesawY}.
For electroweak scale $\mathbb{N}_{2,3}$, however, the $\mathbb{N}_{2,3}$ portal to dark sector may lead to evolution of the dark particles similar to that in Case-RS, as discussed above, if it is required to reproduce the observed relic density.
To provide a qualitatively different and representative benchmark scenario, we therefore assume in Case-SS that the dark portal proceeds only through the electroweak scale state $\mathbb{N}_{1}$, whose Yukawa coupling to the SM is much smaller than the usual seesaw Yukawa scale, namely well below $\mathcal{O}(10^{-7})$, and thus it can naturally realize freeze-in production.}
There are also two cases:
\begin{enumerate}
  \item Both $\chi$ and $\phi$ are dark matter.
  The requirement $y_1^{x\prime} \lesssim 10^{-22}$ implies
  $y_1^{x} \lesssim 10^{-22}/y^{\nu}$, which typically yields
  $y_1^{x} \lesssim 10^{-10}$. Such a tiny coupling prevents the
  dark sector from reaching thermal equilibrium. In this case, the
  observed dark matter relic abundance can only be obtained if $M_{N} > m_{\chi} + m_{\phi}$. {\color{Green}\ding{51}}
  \item In this case only $\chi$ constitutes dark matter. The coupling $y_{1}^{x}$ must be
  sufficiently large so that  $y_1^{x\prime}\sim y_{1}^{x}U_{\mathbb{N}_1\mathbb{V}} \sim 10^{-12}$,
  which implies $y_{1}^{x} \gtrsim 0.1$.
  Depending on the precise values of $y_{1}^{x}$ and dark particle masses,
  the dark sector may or may not reach internal thermal equilibrium.
  In the former case one can consistently define a hidden sector temperature,
  whereas in the latter case the dark sector remains out of equilibrium
  throughout the freeze-in process. {\color{Green}\ding{51}}
\end{enumerate}

\paragraph{Benchmark analysis}
The freeze-in production of dark matter for Case-SS is illustrated in Fig.~\ref{Fig:FI2}.
In this case, the portal right-handed neutrino never reaches thermal equilibrium with the SM.
Depending on the size of the dark sector coupling $y^x$, the dark sector
may either attain internal thermal equilibrium or remain out of equilibrium throughout its evolution.

\begin{table}
\begin{center}
\begin{tabular}{|c|c|c|c|c|c|c|c|c|}
\hline
Freeze-in  & Benchmark  & $m_{\chi}$  & $m_{\phi}$  & $y_{1}^{x}$  & $y_{2,3}^{x}$  & $\Omega_{\chi}h^{2}$  & $\Omega_{\phi}h^{2}$  & $\tau_{\phi}$ {[}sec{]} \tabularnewline
\hline
Model $a$  & \multirow{2}{*}{RS} & $230$  & $370$  & \multicolumn{2}{c|}{$2.2\times10^{-6}$ } & $0.12$  & $0$  & $4.64\times10^{-2}$ \tabularnewline
\cline{1-1}\cline{3-9}
Model $b$  &  & $110$  & $330$  & \multicolumn{2}{c|}{$2.7\times10^{-6}$ } & $0.12$  & $0$  & $4.95\times10^{-13}$ \tabularnewline
\hline
Model $c$  & SC  & $160$  & $380$  & \multicolumn{2}{c|}{ $1.5\times 10^{-9}$ } & $0.12$  & $0$  & $2.01\times10^{-6}$ \tabularnewline
\hline
Model $d$  & SS1  & $440$  & $450$  & $0.29$  & $-$  & $0.12$  & $0$  & $9.00\times10^{-2}$ \tabularnewline
\hline
Model $e$  & SS2  & $120$  & $200$  & $0.90$  & $-$  & $0.12$  & $0$  & $0.92$\tabularnewline
\hline
Model $f$  & SS3  & $90$  & $100$  & $1.7\times10^{-12}$  & $-$  & $0.057$  & $0.063$  & $1.89\times10^{19}$ \tabularnewline
\hline
\end{tabular}
\par\end{center}
\caption{Freeze-in benchmarks for Cases RS, SC, and SS, covering all viable configurations.
  All masses are given in GeV. In Models $a-d$ where only $\chi$ is dark matter, the dark fermion
  $\chi$ accounts for the entire observed dark matter relic abundance, and the lifetimes of
  $\phi$ and the right-handed neutrinos $\mathbb{N}_{1,2,3}$ are shorter than 1~sec, so that
  they decay completely before BBN for Models $a-e$. In model~$f$ where both $\chi$ and $\phi$ contribute to
  dark matter, the lifetime of $\phi$ exceeds the age of the Universe. In Case-SS we assume
  that only $\mathbb{N}_1$, which couples ultraweakly to the SM, serves as the portal to the
  dark sector. The difference between benchmarks SS1, SS2 and SS3 lies in the seesaw Yukawa
  coupling $y^\nu_1$, taken to be of order $10^{-11}$ and $10^{-13}$, respectively, which
  in turn fixes the corresponding values of $y_{1}^{x}$.}
\label{Tab:FImodels}
\end{table}

\begin{figure}
	\centering
	\addtocounter{subfigure}{+3}
		\includegraphics[scale=0.28]{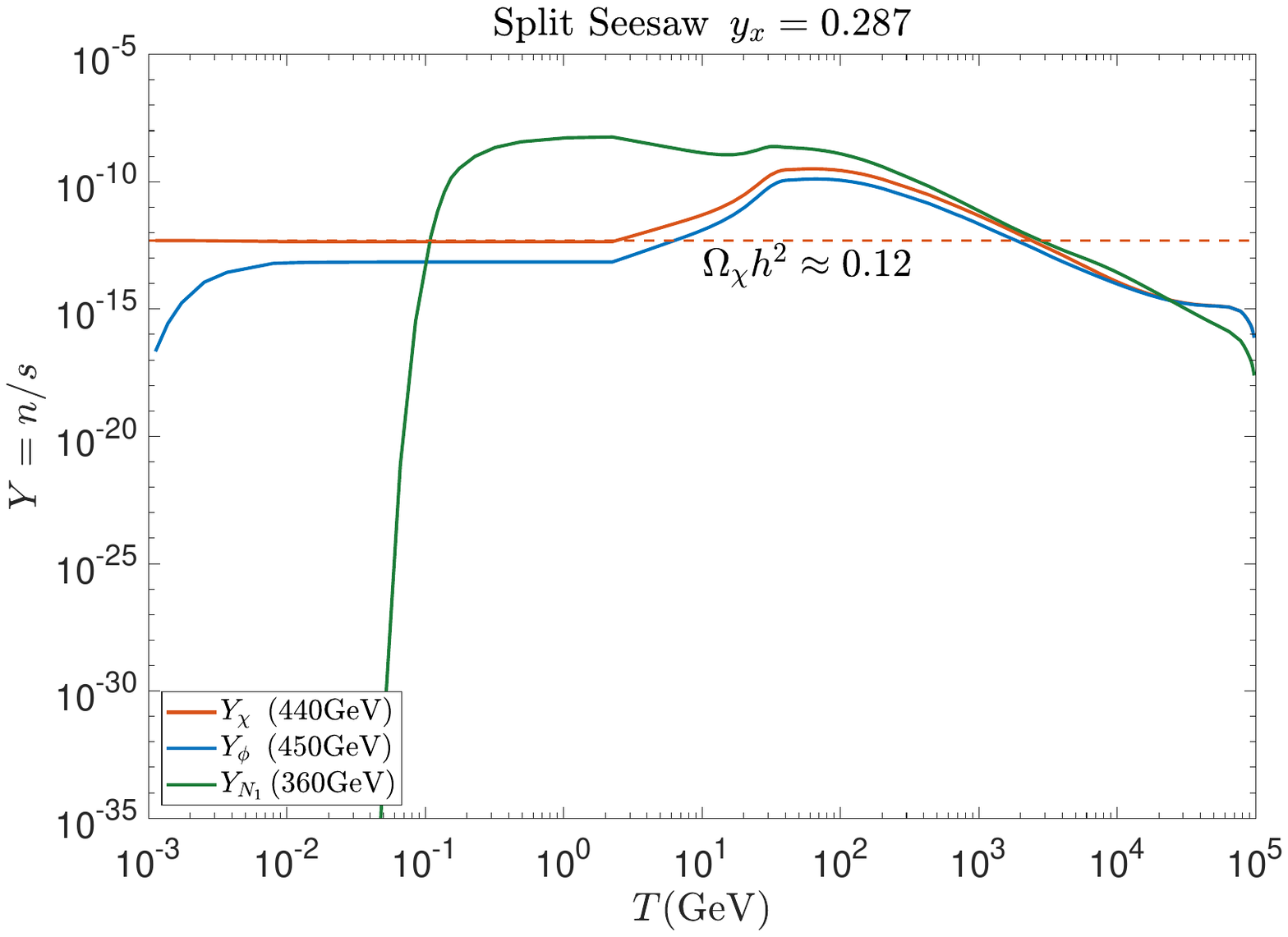}
		\includegraphics[scale=0.28]{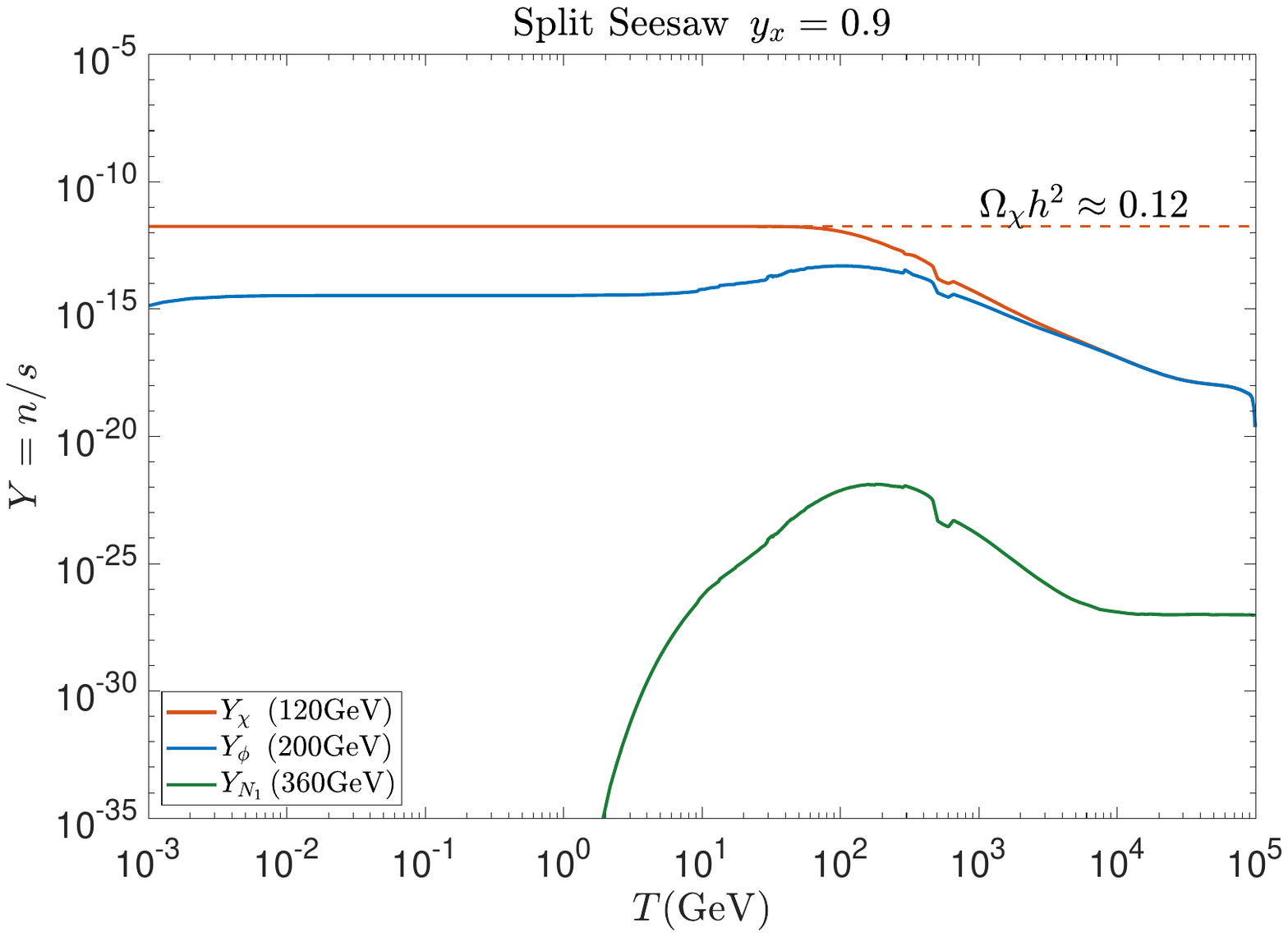}\\
	\subfigure[Model $d$, Benchmark-SS1]{
		\includegraphics[scale=0.28]{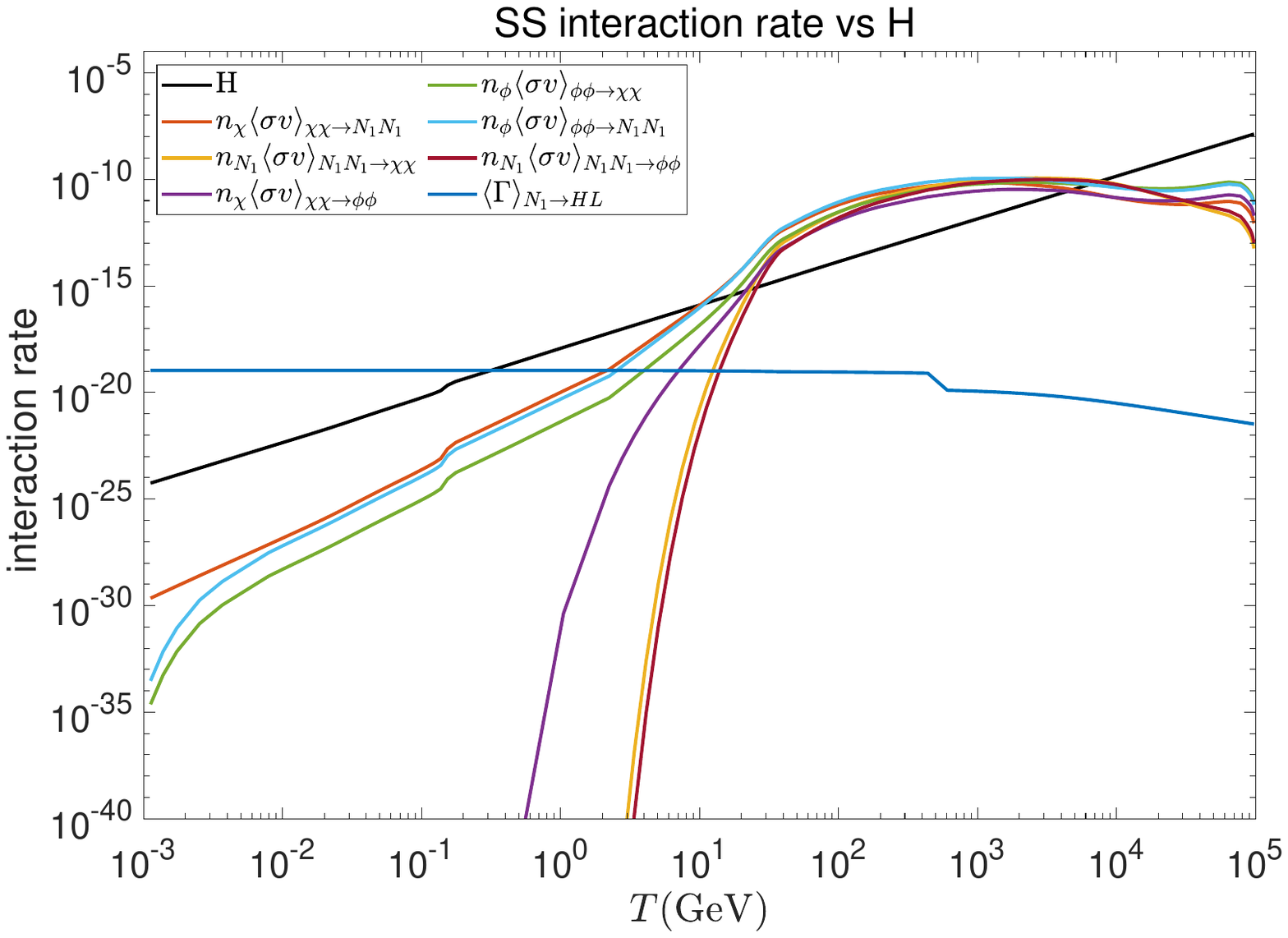}}\quad
	\subfigure[Model $e$, Benchmark-SS2]{
		\includegraphics[scale=0.28]{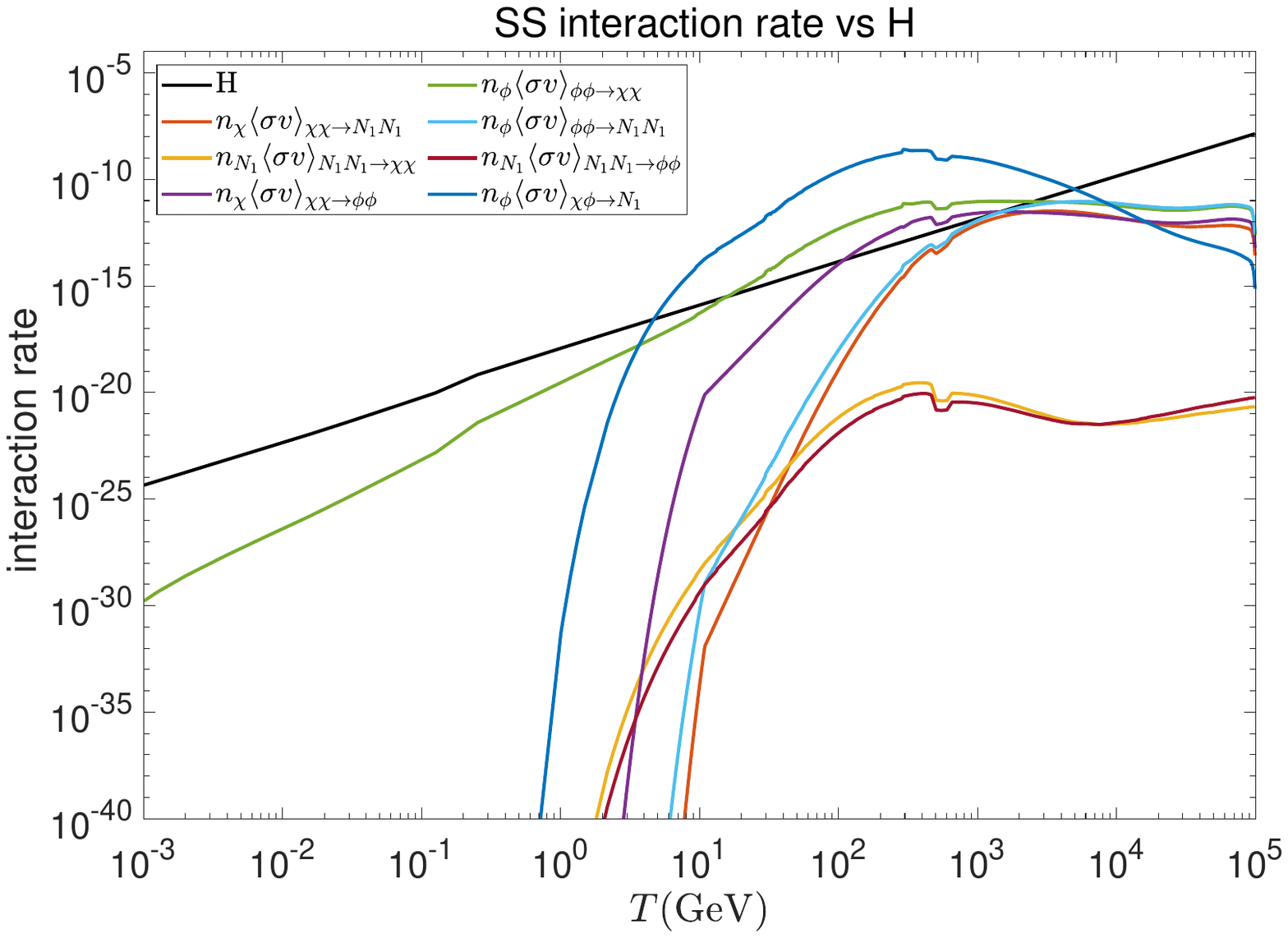}}
	\caption{The freeze-in evolution for Model~$d$ (Benchmark-SS1) and Model~$e$ (Benchmark-SS2) in terms of the visible sector temperature $T$ (GeV). The top panels show the evolution of the comoving number densities of dark particles $\chi,\phi$, and $\mathbb{N}_{1}$, where the red horizontal dashed line denotes the observed dark matter relic density for a dark matter mass $m_{\chi}$. In the lower figure, we trace the dark sector interaction rates compared to the Hubble rate. In both cases, the dark particles are produced via freeze-in processes $L H\to \chi\phi$ and $LH\to\mathbb{N}_{1}$, and the dark sector reaches internal thermal equilibrium once the relevant interaction rates exceed the Hubble. For Case-SS2, the decay channel $\mathbb{N}_{1}\to\chi\phi$ is allowed because $M_1 > (m_{\chi}+m_{\phi})$. All remaining $\phi$ and $\mathbb{N}_{1}$ decays completely with lifetimes shorter than 1 second.}
	\label{Fig:Y_SS12}
\end{figure}

For the various cases discussed above, we compute the evolution of the dark sector
particles and determine the resulting dark matter relic abundance, as summarized in Table~\ref{Tab:FImodels}.
Models~$d,e$ are particularly interesting, since in these two cases the dark sector
reaches internal thermal equilibrium; the corresponding evolutions are shown in Fig.~\ref{Fig:Y_SS12}.

In Model $d$ taking the Benchmark-SS1 seesaw parameters, dark particles $\chi,\phi$, and $\mathbb{N}_{1}$ are produced via freeze-in through the processes $L H\to \chi\phi$ and $LH\to\mathbb{N}_{1}$. Particles $\chi$ and $\phi$ interact through $\phi\phi\leftrightarrow\overline{\chi\chi}$ ($\overline{\phi\phi}\leftrightarrow\chi\chi$), and annihilate into $\mathbb{N}_{1}$ via $\chi\overline{\chi}\to\mathbb{N}_{1}\mathbb{N}_{1}$, $\phi\overline{\phi}\to\mathbb{N}_{1}\mathbb{N}_{1}$, causing $Y_\chi$ and $Y_\phi$ to decrease while $Y_{\mathbb{N}_{1}}$ increases. Once the thermally averaged decay rate of $\mathbb{N}_{1}$ exceeds the Hubble expansion rate, the decay $\mathbb{N}_{1}\to L H$ becomes active and $Y_{\mathbb{N}_{1}}$ rapidly drops to zero. Since $m_{\phi}>m_{\chi}$, the dark fermion $\chi$ accounts for the observed dark matter relic density, while $\phi$ eventually decays away at lower temperatures before BBN.

For this model, it is useful to compare the full calculation with a simplified treatment.
As discussed in~\cite{Abdelrahim:2025fiz}, dark sector internal interactions can qualitatively affect the thermal history,
and the dark sector evolution needs to be described by coupled Boltzmann equations for the number densities and total energy density,
thereby capturing cosmological histories such as freeze-in, double freeze-in, and dark sector equilibration with a separate dark temperature.

In our analysis, we solve the full coupled evolution of the freeze-in dark matter $\chi$ together with other dark sector particles $\phi$ and $\mathbb{N}_1$ produced during the evolution, including all relevant internal dark sector interactions, and obtain $\Omega_\chi h^2 = 0.12$.
For comparison,
a simplified approach computes the freeze-in production of $\chi$, $\phi$, and $\mathbb{N}_1$ separately,
neglects internal dark sector interactions during the production epoch,
and includes the late decays of $\phi$ and $\mathbb{N}_1$ only afterward in the final $\chi$ relic abundance.
In the present setup, however, this approximation does not provide an accurate description.
In particular, since the right-handed neutrino $\mathbb{N}_1$ cannot decay into dark particles,
the contribution associated with $\mathbb{N}_1 \to \chi \phi$ is absent,
and the resulting relic density is $\Omega_\chi h^2 = 0.004$,
differing from the full result by more than $95\%$.
This comparison shows that, for the benchmark considered here,
a full coupled treatment of the dark sector evolution is essential for obtaining the correct relic abundance.

In Model $e$ taking the Benchmark-SS2 seesaw parameters, since $M_1 > m_{\chi}+m_{\phi}$, the three-point decay within the dark sector $\mathbb{N}_{1}\to\chi\phi$ is kinematically allowed. Dark particles $\chi$ and $\phi$ then accumulate both by the freeze-in process $LH\to\chi\phi$ and by $LH\to\mathbb{N}_{1}$ followed by $\mathbb{N}_{1}\to\chi\phi$. When the interaction rate of processes $\overline{\chi\chi}\to\phi\phi$ falls below the Hubble rate at $T_{v}\simeq10^{2}$ GeV, this channel becomes inefficient, marking the onset of the dark freeze-out during which the number density of $\phi$ starts to decrease. The dark freeze-out ends once the reverse process $\phi\phi\to\overline{\chi\chi}$ becomes inactive, with its rate dropping below the Hubble scale. Afterwards, the abundance of $\phi$ gradually levels off and remains constant until $\phi$ eventually decays at low temperatures, with a lifetime shorter than 1 second. During the dark freeze-out, the dark matter $\chi$ continues to grow and ultimately attains the observed relic density. In this case, the eventual decrease of $Y_{\mathbb{N}_{1}}$ is driven by several effects, including $\mathbb{N}_{1}\to\chi\phi,\,\mathbb{N}_{1}\mathbb{N}_{1}\to\chi\overline{\chi}$, and $\mathbb{N}_{1}\mathbb{N}_{1}\to\phi\overline{\phi}$.

The dark matter particle $\chi$ ultimately reaches the observed relic abundance $\Omega_\chi h^2 = 0.12$,
once the full evolution of all dark sector particles is taken into account.
Unlike in the previous case, the right-handed neutrino is heavier than $m_\chi + m_\phi$,
so the decay of $\mathbb{N}_1$ makes a strong contribution to the dark matter freeze-in.
The approximate procedure,
where the freeze-in production of $\chi$, $\phi$, and $\mathbb{N}_1$ is computed separately
and the contributions from the late decays of $\phi$ and $\mathbb{N}_1$ are added to the dark matter relic density after freeze-in,
leads to a dark matter relic density $\Omega_\chi h^2 = 0.08$, corresponding to a $30\%$ error.


In Model $f$ taking the Benchmark-SS3 seesaw parameters, the tiny coupling $y_{1}^{x} \lesssim 10^{-10}$ prevents the dark sector from reaching internal thermal equilibrium. The decays $\phi\to \overline{\chi}+\mathbb{V}_{i=1,2,3}$ have lifetimes exceeding the age of the Universe. Both $\chi$ and $\phi$ are dark matter components that can only be produced via freeze-in through the $\mathbb{N}_1$ portal, and together they account for the entire dark matter relic density.

\paragraph{Summary of the freeze-in results}
With seesaw parameters that consistently reproduce the observed light neutrino masses and mixings, the common features of right-handed neutrino portal dark matter in the freeze-in scenario can be summarized as follows:
\begin{itemize}
\item The freeze-in evolution falls into two qualitatively distinct classes, depending on whether the portal right-handed neutrinos thermalize with the SM sector. In Cases RS and SC, the right-handed neutrinos are in thermal equilibrium with the SM bath during the main freeze-in epoch and can therefore be treated as part of the visible sector. A viable freeze-in benchmark then requires $\phi$ to decay before BBN, while the dark sector coupling $y_x$ remains too small to establish internal thermal equilibrium within the dark sector. In these cases, the viable benchmarks have only $\chi$ as dark matter and are mainly populated through the processes $\mathbb{N}\mathbb{N}\to\chi\overline{\chi}$, $\mathbb{N}\mathbb{N}\to\phi\overline{\phi}$, and $LH\to\chi\phi$, whereas the regime $M_{\mathbb{N}}>m_\chi+m_\phi$ generally leads to overproduction through $\mathbb{N}_1\to\chi\phi$.

\item In Case-SS, by contrast, the portal state $\mathbb{N}_1$ never reaches thermal equilibrium with the SM bath, so the dark sector is produced entirely through the ultraweak $\mathbb{N}_1$ portal. If the dark sector coupling is sufficiently large, the dark sector can still attain internal thermal equilibrium and subsequently undergo a dark freeze-out, even though its overall population is generated by freeze-in; and this is realized in
    Models $d,e$. If the dark sector coupling is much smaller, internal equilibrium is never established, and both $\chi$ and $\phi$ can survive as dark matter components, as in Model $f$.

\item Dark sector dynamics are therefore essential in the freeze-in regime. When the hidden sector interactions are too weak to establish internal equilibrium, a simplified freeze-in treatment may be adequate. However, once the dark sector reaches internal equilibrium, one must solve the full coupled Boltzmann equations, including the hidden sector temperature when applicable, in order to obtain the correct relic abundance. In particular, neglecting dark sector interactions and treating $\chi$, $\phi$, and $N_1$ as independent freeze-in components can misestimate the final relic abundance by about $30\%$ when the three-point decay $\mathbb{N}_1\to\chi\phi$ is kinematically allowed, and by more than $95\%$ when such three-point dark sector processes are absent.
\end{itemize}

\subsubsection{Summary of the electroweak right-handed neutrino portal scenario}

Finally, we summarize in Table~\ref{Tab:RHNP} the general results for the minimal right-handed neutrino portal dark matter scenario
illustrated in Fig.~\ref{Fig:RHNP}.
The seesaw Yukawa couplings $y^{\nu}$ are determined by fitting neutrino data within the Type-I seesaw mechanism,
with magnitudes ranging from ultraweak to tiny and small values in Cases SS, RS, and SC, respectively.
The dark Yukawa coupling $y^x$ is fixed by requiring the observed dark matter relic abundance to be reproduced through either freeze-out or freeze-in.
For Case-SS, $y^{\nu}$ and $y^x$ specifically refer to the couplings $y_1^{\nu}$ and $y_1^x$
associated with the first-generation right-handed neutrino $\mathbb{N}_1$.
In this case, freeze-in production through $\mathbb{N}_1$ is the only viable mechanism for obtaining the observed dark matter relic density.

\begin{table}
\begin{center}
\begin{tabular}{|c|c|c|c|c|}
\hline
Seesaw Type & $y^{\nu}$ & behavior of $\mathbb{N}_{i=1,2,3}$ & $y^{x}$ (FO) & $y^{x}$ (FI)\tabularnewline
\hline
\hline
Case-RS & $\mathcal{O}(10^{-7})$ & EQ with SM below $10^{3}$ GeV & $\mathcal{O}(10^{-1})$ & $\mathcal{O}(10^{-6})$\tabularnewline
\hline
Case-SC & $\mathcal{O}(10^{-3})$ & EQ with SM below $10^{5}$ GeV & $\mathcal{O}(10^{-1})$ & $\mathcal{O}(10^{-9})$\tabularnewline
\hline
Case-SS & $\lesssim10^{-10}$ & $\mathbb{N}_{1}$ never achieves EQ with SM & $-$ & $\mathcal{O}(10^{-1})$ or $\lesssim10^{-12}$\tabularnewline
\hline
\end{tabular}
\end{center}
\caption{Summary of the electroweak right-handed neutrino portal scenario, with the orders of magnitude of $y^{\nu}$ and $y^x$ indicated for various cases studied in this paper. The Yukawa couplings $y^{\nu}$ are determined from neutrino data, while $y^x$ is fixed by requiring dark matter to reproduce the observed relic abundance through either freeze-out or freeze-in.
For Case-SS, $y^{\nu}$ and $y^x$ refer to $y_1^{\nu}$ and $y_1^x$ associated with $\mathbb{N}_1$.
For a given set of couplings $y^{\nu}$, we also indicate whether the right-handed neutrinos are in thermal equilibrium with the SM bath.}
\label{Tab:RHNP}
\end{table}

\section{Conclusion}\label{Sec:Con}

In this work, we have explored electroweak scale implementations of the Type-I seesaw mechanism and their connection to dark matter via a right-handed neutrino portal. We have studied a minimal dark sector coupled to right-handed neutrinos and precisely carried out a detailed analysis of its cosmological evolution.

We began by emphasizing the central role of right-handed neutrinos in neutrino mass model building.
As electroweak singlet fermions, their neutral nature allows Majorana mass terms,
opens the door to leptogenesis, and makes them natural portal fields to hidden sectors.
When their masses lie in electroweak to TeV scale,
it is plausible that the origin of this mass scale is linked to electroweak symmetry breaking, and crucially,
they become experimentally accessible at colliders and in precision neutrino experiments.
In this mass range, constraints from colliders, electroweak precision tests,
and flavor physics are significant but still leave substantial room for viable heavy neutral leptons.

We investigated three representative classes of electroweak scale seesaw models: Cases RS, SC, and SS.
These cases span small, tiny, and ultraweak couplings of the electroweak right-handed neutrinos to the SM sector,
while reproducing the observed light neutrino mass hierarchy and mixing pattern.
A detailed calculation of the seesaw mixing, using the two-component spinor formalism, is presented in Appendix~\ref{App:SSdetail}.
Building on this, all seesaw parameters in our analysis, including the couplings of the right-handed neutrinos to the SM,
are obtained from a consistent seesaw framework using the PSO algorithm,
rather than from rough order-of-magnitude estimates as in part of the existing literature.
We then embedded these seesaw constructions into a minimal dark sector containing a fermion $\chi$ and a complex scalar $\phi$,
which communicate with the SM solely through renormalizable Yukawa interactions with the right-handed neutrinos.

We solved the full set of Boltzmann equations, including all relevant interactions involving light and heavy neutrinos after seesaw mixing,
and thereby determined the complete evolution of the dark particles.
We find that there is a wide region of viable parameter space reproducing the observed dark matter relic abundance,
demonstrating that electroweak right-handed neutrino portal dark matter provides a robust framework
in light of current and future searches.
Among the three representative cases, Case-SC is particularly interesting, as the right-handed neutrinos couple to the SM sector with sizable couplings as large as $\mathcal{O}(10^{-3})$, making this scenario especially promising for collider searches and neutrino experiments.

For the freeze-out benchmarks, we find that the dark matter depletion is typically dominated by $\chi\overline{\chi}\to \mathbb{N}\mathbb{N}$,
unless $m_\chi$ is sufficiently below the portal-neutrino mass scale,
in which case the resonant coannihilation channel $\chi\phi\to LH$ becomes the leading depletion process in Case-SC.
Moreover, despite the fact that $\phi$ and $\mathbb{N}_{i=1,2,3}$ remain close to equilibrium before $\chi$ freeze-out,
their coupled evolution cannot be neglected.
This demonstrates that a full coupled Boltzmann treatment is essential for obtaining the correct relic abundance also in the freeze-out regime.

For ultraweak seesaw couplings, we show that strong internal interactions within the dark sector can have a significant impact on freeze-in predictions. In Case-SS, where the portal right-handed neutrino never thermalizes with the SM,
the dark sector can only be produced via freeze-in.
When the hidden sector couplings are large enough to drive $\chi$ and $\phi$ toward internal equilibrium,
the evolution of the dark sector can no longer be captured by treating each species as an independent freeze-in component.
We explicitly compare the full multi-temperature evolution with a commonly used approximate treatment in which one computes the freeze-in production of $\chi$, $\phi$ and right-handed neutrinos separately, neglecting their mutual interactions, and then allows the unstable particles to decay only after freeze-in.
In representative benchmark points, we find that this simplified procedure underestimates the final dark matter abundance by about $30\%$ when three-point internal interactions such as $\mathbb{N} \leftrightarrow \chi \phi$ are present, or by more than $95\%$ when such interactions are absent. This demonstrates that a careful treatment of dark sector dynamics is essential in right-handed neutrino portal freeze-in models.

Overall, our results show that electroweak scale right-handed neutrino portal dark matter provides a robust and predictive framework linking neutrino masses, the dark matter relic abundance, and the properties of heavy neutral leptons. The three seesaw realizations considered here exhibit qualitatively different portal couplings and thermal histories, yet all admit viable benchmark models that reproduce the observed relic density in agreement with current laboratory and cosmological bounds, as summarized in Table~\ref{Tab:RHNP}.
In this framework, the most relevant experimental and observational probes arise from collider searches for heavy neutral leptons, precision tests of neutrino mixing and non-unitarity, and cosmological measurements, which together provide complementary ways to further constrain the parameter space studied in this work.\\

\noindent\textbf{Acknowledgments: }

This work is
supported in part by the National Natural Science Foundation of China under Grant No. 11935009,
and Tianjin University Self-Innovation Fund Extreme Basic Research Project Grant No. 2025XJ21-0007.

\newpage

\appendix

\section{Notations and useful formulas}\label{App:Conven}

\paragraph{Convention on the charge conjugation}

We use the metric ${\rm diag}\{+,---\}$. The charge conjugation operation is
defined as
\begin{equation}
C:\psi\to\psi^{c}\qquad\psi^{c}\equiv-{\rm i}\gamma^{2}\psi^{*}=-{\rm i}(\bar{\psi}\gamma^{0}\gamma^{2})^{T}=-{\rm i}\gamma^{2}\gamma^{0}\bar{\psi}^{T}\,.
\end{equation}
The gamma matrices are given by
\begin{equation}
\gamma^{\mu}=\left(\begin{array}{cc}
0 & \sigma^{\mu}\\
\bar{\sigma}^{\mu} & 0
\end{array}\right)\,,\qquad\sigma^{\mu}=(\mathbb{\mathbf{1}},\vec{\sigma})\,,\qquad\bar{\sigma}^{\mu}=(\mathbb{\mathbf{1}},-\vec{\sigma})\,.
\end{equation}
$\gamma^{0},\gamma^{2}$ are symmetric matrices, each equal to its
own transpose
\begin{equation}
\gamma^{0}=\left(\begin{array}{cc}
0 & \mathbb{\mathbf{1}}\\
\mathbb{\mathbf{1}} & 0
\end{array}\right)=\left(\begin{array}{cccc}
 &  & 1\\
 &  &  & 1\\
1 & {\rm }\\
{\rm } & 1
\end{array}\right)\,,\qquad\gamma^{2}=\left(\begin{array}{cc}
0 & \sigma_{2}\\
-\sigma_{2} & 0
\end{array}\right)=\left(\begin{array}{cccc}
 &  &  & -{\rm i}\\
 &  & {\rm i}\\
 & {\rm i}\\
-{\rm i}
\end{array}\right)\,.
\end{equation}
Thus a charge conjugation $C:\,\psi\to\psi^{c}$ transforms a Dirac
field to be
\begin{equation}
C:\,\psi=\left(\begin{array}{c}
\psi_{L}\\
\psi_{R}
\end{array}\right)\equiv\left(\begin{array}{c}
\xi\\
\eta
\end{array}\right)\,\longrightarrow\psi^{c}=-{\rm i}\gamma^{2}\psi^{*}=\left(\begin{array}{c}
-{\rm i}\sigma_{2}\psi_{R}^{*}\\
{\rm i}\sigma_{2}\psi_{L}^{*}
\end{array}\right)=\left(\begin{array}{c}
-{\rm i}\sigma_{2}\eta^{*}\\
{\rm i}\sigma_{2}\xi^{*}
\end{array}\right)\,.
\end{equation}
A Majorana fermion $\psi_{M}$ is defined as $\psi_{M}=\psi_{M}^{c}$
and thus a Majorana 4-spinor can be defined by either using a left-handed
2-component spinor $\xi$
\begin{equation}
\psi_{M}=\left(\begin{array}{c}
\xi\\
\xi^{c}
\end{array}\right)=\left(\begin{array}{c}
\xi\\
{\rm i}\sigma_{2}\xi^{*}
\end{array}\right)\,,\label{eq:LHM4spinor}
\end{equation}
or using a right-handed 2-component spinor $\eta$
\begin{equation}
\psi_{M}=\left(\begin{array}{c}
\eta^{c}\\
\eta
\end{array}\right)=\left(\begin{array}{c}
-{\rm i}\sigma_{2}\eta^{*}\\
\eta
\end{array}\right)\,,
\end{equation}
it is therefore convenient to define the charge conjugation for two-component
spinor fields as the following
\begin{align}
 & \text{for\,left-handed\,field}\,\,\xi:\quad\qquad{\rm \xi^{c}\equiv{\rm i}\sigma_{2}\xi^{*}\,,}\\
 & \text{for\,right-handed\,field}\,\,\eta:\ \qquad\eta^{c}\equiv-{\rm i}\sigma_{2}\eta^{*}\,.
\end{align}

\paragraph{Dirac and Majorana mass terms}

For the Dirac fermion
\begin{equation}
\psi=\left(\begin{array}{c}
\psi_{L}\\
\psi_{R}
\end{array}\right)=\left(\begin{array}{c}
\xi\\
\eta
\end{array}\right)\,,\nonumber
\end{equation}
its Dirac mass term can be written in terms of two-component spinors
\begin{equation}
-\mathcal{L}=m_{D}\overline{\psi}\psi=m_{D}(\overline{\psi_{L}}\psi_{R}+\overline{\psi_{R}}\psi_{L})=m_{D}(\xi^{\dagger}\eta+\eta^{\dagger}\xi)\,.
\end{equation}
The key step in the seesaw mechanism is to rewrite the Dirac mass terms
in terms of the charge conjugate fields, to mix with the Majorana mass terms
\begin{equation}
-\mathcal{L}=m_{D}\overline{\psi}\psi=m_{D}(\xi^{c\dagger}\eta^{c}+\eta^{c\dagger}\xi^{c})\,,
\end{equation}
which can be verified by
\begin{align}
\xi^{c\dagger}\eta^{c} & =({\rm i}\sigma_{2}\xi^{*})^{\dagger}(-{\rm i}\sigma_{2}\eta^{*})=-\xi^{T}\eta^{*}\nonumber \\
 & =-\left(\begin{array}{cc}
\xi_{1} & \xi_{2}\end{array}\right)\left(\begin{array}{c}
\eta_{1}^{*}\\
\eta_{2}^{*}
\end{array}\right)=\left(\begin{array}{cc}
\eta_{1}^{*} & \eta_{2}^{*}\end{array}\right)\left(\begin{array}{c}
\xi_{1}\\
\xi_{2}
\end{array}\right)=\eta^{\dagger}\xi\,.
\end{align}

On the other hand, a Majorana mass term is defined as
\begin{equation}
-\mathcal{L}_{M}=\frac{M}{2}\overline{\psi_{M}}\psi_{M}\,,
\end{equation}
for the 4-component Majorana fermion
\begin{equation}
\psi_{M}=\left(\begin{array}{c}
\psi_{M,L}\\
\psi_{M,R}
\end{array}\right)=\left(\begin{array}{c}
\xi\\
\eta
\end{array}\right)\,,\qquad{\rm with}\ \eta=\xi^{c}\,.
\end{equation}
The Majorana mass term can be written in terms of either the left-handed
two-component spinor $\xi$
\begin{equation}
-\mathcal{L}_{M}^{L}=\frac{M}{2}\big(\xi^{c\dagger}\xi+\xi^{\dagger}\xi^{c}\big)=\frac{M}{2}\big[\xi^{T}(-{\rm i}\sigma_{2})\xi+\xi^{\dagger}({\rm i}\sigma_{2})\xi^{*}\big]\,,\label{eq:LHMmass}
\end{equation}
or in terms of the right-handed two-component spinor $\eta$
\begin{equation}
-\mathcal{L}_{M}^{R}=\frac{M}{2}\big(\eta^{c\dagger}\eta+\eta^{\dagger}\eta^{c}\big)=\frac{M}{2}\big[\eta^{T}({\rm i}\sigma_{2})\eta+\eta^{\dagger}(-{\rm i}\sigma_{2})\eta^{*}\big]\,.\label{eq:RHMmass}
\end{equation}

\paragraph{Neutrino masses in the SM}

The SM contains only left-handed neutrinos
and we have ($\alpha=e,\mu,\tau$)
\begin{equation}
V_{{\rm SM}\,\alpha}=\left(\begin{array}{c}
\nu_{\alpha\,L}\\
0
\end{array}\right)\equiv\left(\begin{array}{c}
\nu_{\alpha}\\
0
\end{array}\right)\,,\qquad2~{\rm d.o.f.}.
\end{equation}
Mass terms can be generated by introducing gauge-singlet right-handed neutrinos  
\begin{equation}
\left(\begin{array}{c}
0\\
\nu_{\alpha\,R}
\end{array}\right)\,,\qquad \alpha = 1,2,3\,,
\end{equation}
and we define the right-handed neutrino in the full Majorana 4-spinor
form
\begin{equation}
\mathbb{N}_{\alpha}=\left(\begin{array}{c}
-{\rm i}\sigma_{2}\nu_{\alpha\,R}^{*}\\
\nu_{\alpha\,R}
\end{array}\right)\equiv\left(\begin{array}{c}
-{\rm i}\sigma_{2}N_{\alpha}^{*}\\
N_{\alpha}
\end{array}\right)\,,\qquad2~{\rm d.o.f.},
\end{equation}
where we use $N$ instead of $\nu_{R}$ to avoid possible confusion.
From this point forward, we will use normal capital letters to denote
Dirac 4-spinors; and use capital letters with double lines, such as
$\mathbb{N}$ and $\mathbb{V}$, to denote Majorana 4-spinors.

We note that the use of the subscripts ``$L$'' and ``$R$'' to
denote left-handed and right-handed fields can be confusing. For instance,
``$\nu_{R}$'', typically identified as the right-handed neutrino,
may refer either to the right-handed component of a 4-spinor, which
is 2-component, or to a right-handed Majorana 4-spinor. This ambiguity
becomes particularly problematic when performing charge conjugation
on spinors labeled with ``$L$'' and ``$R$''. We thus define
two-component spinors $\nu_{i},N_{i}$, representing the two-component
left-handed SM neutrinos and right-handed Majorana neutrinos respectively.
Thus in the flavor basis, the full neutrino Dirac 4-spinor reads
\begin{equation}
V_{\alpha}=\left(\begin{array}{c}
\nu_{\alpha\,L}\\
\nu_{\alpha\,R}
\end{array}\right)\equiv\left(\begin{array}{c}
\nu_{\alpha}\\
N_{\alpha}
\end{array}\right)\,,\qquad4~{\rm d.o.f.}.
\end{equation}

\section{Seesaw mechanism for one generation, details}\label{App:SSdetail}

In this Appendix, we present the most detailed derivation of seesaw-induced neutrino mixing
currently available in the literature, including the intermediate steps that are often omitted even in pedagogical textbooks and reviews.
This derivation explicitly determines the couplings of all neutrinos
(three light and three heavy) to SM particles Eqs.~(\ref{eq:SMWp})--(\ref{eq:SMH})
and to dark sector particles Eq.~(\ref{eq:DSN}) after the seesaw mixing.
It also shows explicitly how the SM neutrinos become Majorana particles through the seesaw mechanism.

The full mass term for a one-generation neutrino is written as
\begin{equation}
\mathcal{L}_{{\rm mass}}=-y^{\nu}\overline{L}\widetilde{H}P_{R}\mathbb{N}
+h.c.-\frac{M}{2}\overline{\mathbb{N}}\mathbb{N}\,,
\end{equation}
where $\widetilde{H}={\rm i}\sigma_{2}H$. In terms of 2-component spinors,
\begin{align}
\mathcal{L}_{{\rm mass}}^{{\rm neutrino}} & =-\frac{y^{\nu}v}{\sqrt{2}}(\nu^{\dagger}N
+N^{\dagger}\nu)-\frac{M}{2}\big[N^{T}({\rm i}\sigma_{2})N+(N^{*})^{T}(-{\rm i}\sigma_{2})N^{*}\big]\nonumber \\
 & = -\frac{m_{D}}{2}(\nu^{\dagger}N+N^{\dagger}\nu+N^{c\dagger}\nu^{c}+\nu^{c\dagger}N^{c})
 -\frac{M}{2}(N^{\dagger}N^{c}+N^{c\dagger}N)\,,
\end{align}
where in the second line we define $m_{D}= y^{\nu}v/\sqrt{2}$ and
\begin{equation}
\nu^{c}\equiv+{\rm i}\sigma_{2}\nu^{*}\,,\qquad N^{c}\equiv-{\rm i}\sigma_{2}N^{*}\,.
\end{equation}
Hence we can obtain the mass mixing matrix:
\begin{align}
\mathcal{L}_{{\rm mass}}^{{\rm neutrino}} & =-\frac{1}{2}\left(\begin{array}{cc}
\nu^{c\dagger} & N^{\dagger}\end{array}\right)\left(\begin{array}{cc}
0 & m_{D}\\
m_{D} & M
\end{array}\right)\left(\begin{array}{c}
\nu\\
N^{c}
\end{array}\right)-\frac{1}{2}\left(\begin{array}{cc}
\nu^{\dagger} & N^{c\dagger}\end{array}\right)\left(\begin{array}{cc}
0 & m_{D}\\
m_{D} & M
\end{array}\right)\left(\begin{array}{c}
\nu^{c}\\
N
\end{array}\right)\,,\label{eq:1GMM}
\end{align}
and these two terms are just the hermitian conjugate of each other.
First we analyze the first term:
\begin{align*}
{\rm 1st\,term} & =-\frac{1}{2}\left(\begin{array}{cc}
\nu^{c\dagger} & N^{\dagger}\end{array}\right)\left(\begin{array}{cc}
0 & m_{D}\\
m_{D} & M
\end{array}\right)\left(\begin{array}{c}
\nu\\
N^{c}
\end{array}\right)\\
 & =-\frac{1}{2}\left(\begin{array}{cc}
({\rm i}\sigma_{2}\nu^{*})^{\dagger} & N^{\dagger}\end{array}\right)OO^{T}\left(\begin{array}{cc}
0 & m_{D}\\
m_{D} & M
\end{array}\right)OO^{T}\left(\begin{array}{c}
\nu\\
-{\rm i}\sigma_{2}N^{*}
\end{array}\right)\\
 & =-\frac{1}{2}\left(\begin{array}{cc}
\nu^{T}(-{\rm i}\sigma_{2}) & (N^{*})^{T}({\rm i}\sigma_{2})(-{\rm i}\sigma_{2})\end{array}\right)OO^{T}\left(\begin{array}{cc}
0 & m_{D}\\
m_{D} & M
\end{array}\right)OO^{T}\left(\begin{array}{c}
\nu\\
-{\rm i}\sigma_{2}N^{*}
\end{array}\right)\\
 & =-\frac{1}{2}\left(\begin{array}{cc}
\nu^{T} & (-{\rm i}\sigma_{2}N^{*})^{T}\end{array}\right)(-{\rm i}\sigma_{2})OO^{T}\left(\begin{array}{cc}
0 & m_{D}\\
m_{D} & M
\end{array}\right)OO^{T}\left(\begin{array}{c}
\nu\\
-{\rm i}\sigma_{2}N^{*}
\end{array}\right)\\
 & =-\frac{1}{2}\left(\begin{array}{cc}
\nu^{T} & (-{\rm i}\sigma_{2}N^{*})^{T}\end{array}\right)O(-{\rm i}\sigma_{2})O^{T}\left(\begin{array}{cc}
0 & m_{D}\\
m_{D} & M
\end{array}\right)OO^{T}\left(\begin{array}{c}
\nu\\
-{\rm i}\sigma_{2}N^{*}
\end{array}\right)\\
& =-\frac{1}{2}\left(\begin{array}{cc}
\nu_{\ell}^{T} & \nu_{h}^{T}\end{array}\right)(-{\rm i}\sigma_{2})\left(\begin{array}{cc}
m_{\ell} & 0\\
0 & m_{h}
\end{array}\right)\left(\begin{array}{c}
\nu_{\ell}\\
\nu_{h}
\end{array}\right)\\
 & \Rightarrow-\frac{1}{2}\left(\begin{array}{cc}
\nu_{\ell}^{(L)\,T} & \nu_{h}^{(L)\,T}\end{array}\right)(-{\rm i}\sigma_{2})\left(\begin{array}{cc}
m_{\ell} & 0\\
0 & m_{h}
\end{array}\right)\left(\begin{array}{c}
\nu_{\ell}^{(L)}\\
\nu_{h}^{(L)}
\end{array}\right)\\
 & =-\frac{1}{2}\big[m_{\ell}\nu_{\ell}^{(L)\,T}(-{\rm i}\sigma_{2})\nu_{\ell}^{(L)}+m_{h}\nu_{h}^{(L)\,T}(-{\rm i}\sigma_{2})\nu_{h}^{(L)}\big]\,.
\end{align*}
The $2\times 2$ orthogonal matrix $O$ (in the full three-generation seesaw mixing case, the orthogonal matrix is $6\times 6$),
can pass through $(-{\rm i}\sigma_{2})$, since $O$ operates only on the neutrino flavor basis,
whereas $(-{\rm i}\sigma_{2})$ acts only on the two-component spinors, which can be seen schematically as
\begin{align}
\mathbf{V}^{T}\mathbf{M}\mathbf{V}	&=\mathbf{V}_{I,a}^{T}(-{\rm i}\sigma_{2})_{ab}\mathbf{M}_{IJ}\mathbf{V}_{J,b}
	=\mathbf{V}_{I,a}^{T}(-{\rm i}\sigma_{2})_{ab}O_{IP}\big(O^{T}\mathbf{M}O\big)_{PQ}O_{QJ}^{T}\mathbf{V}_{J,b}\nonumber\\
	&=\mathbf{V}_{I,a}^{T}O_{IP}(-{\rm i}\sigma_{2})_{ab}\big(O^{T}\mathbf{M}O\big)_{PQ}O_{QJ}^{T}\mathbf{V}_{J,b}\,.
\end{align}
The mass eigenbasis is defined by
\begin{equation}
O^{T}\left(\begin{array}{c}
\nu\\
-{\rm i}\sigma_{2}N^{*}
\end{array}\right)\equiv\left(\begin{array}{c}
\nu_{\ell}\\
\nu_{h}
\end{array}\right)\,,
\end{equation}
where the orthogonal matrix $O$ is determined by diagonalizing the
mass matrix. One can see from the last step that $\nu_{\ell}$ and
$\nu_{h}$ are indeed left-handed (component) Majorana fields, c.f.,
Eqs.~(\ref{eq:LHM4spinor}) and~(\ref{eq:LHMmass}). The derivation of the second term is similar
\begin{align*}
{\rm 2nd\,term} & =-\frac{1}{2}\left(\begin{array}{cc}
\nu^{\dagger} & N^{c\dagger}\end{array}\right)\left(\begin{array}{cc}
0 & m_{D}\\
m_{D} & M
\end{array}\right)\left(\begin{array}{c}
\nu^{c}\\
N
\end{array}\right)\\
 & =-\frac{1}{2}\left(\begin{array}{cc}
(\nu^{*})^{T} & [(-{\rm i}\sigma_{2}N^{*})^{*}]^{T}\end{array}\right)\left(\begin{array}{cc}
0 & m_{D}\\
m_{D} & M
\end{array}\right)\left(\begin{array}{c}
{\rm i}\sigma_{2}\nu^{*}\\
N
\end{array}\right)\\
 & =-\frac{1}{2}\left(\begin{array}{cc}
({\rm i}\sigma_{2}\nu^{*})^{T} & N^{T}\end{array}\right)({\rm i}\sigma_{2})OO^{T}\left(\begin{array}{cc}
0 & m_{D}\\
m_{D} & M
\end{array}\right)OO^{T}\left(\begin{array}{c}
{\rm i}\sigma_{2}\nu^{*}\\
N
\end{array}\right)\\
 & =-\frac{1}{2}\left(\begin{array}{cc}
\nu_{\ell}^{T} & \nu_{h}^{T}\end{array}\right)({\rm i}\sigma_{2})\left(\begin{array}{cc}
m_{\ell} & 0\\
0 & m_{h}
\end{array}\right)\left(\begin{array}{c}
\nu_{\ell}\\
\nu_{h}
\end{array}\right)\\
 & \Rightarrow-\frac{1}{2}\left(\begin{array}{cc}
\nu_{\ell}^{(R)\,T} & \nu_{h}^{(R)\,T}\end{array}\right)({\rm i}\sigma_{2})\left(\begin{array}{cc}
m_{\ell} & 0\\
0 & m_{h}
\end{array}\right)\left(\begin{array}{c}
\nu_{\ell}^{(R)}\\
\nu_{h}^{(R)}
\end{array}\right)\\
 & =-\frac{1}{2}\big[m_{\ell}\nu_{\ell}^{(R)\,T}({\rm i}\sigma_{2})\nu_{\ell}^{(R)}+m_{h}\nu_{h}^{(R)\,T}({\rm i}\sigma_{2})\nu_{h}^{(R)}\big]\,,
\end{align*}
which is the Majorana mass term for the right-handed fields.

The diagonalization of the mass matrix is identical for both the first and second terms
\begin{equation}
O^{T}\left(\begin{array}{cc}
0 & m_{D}\\
m_{D} & M
\end{array}\right)O\approx\left(\begin{array}{cc}
1 & \frac{m_{D}}{M}\\
-\frac{m_{D}}{M} & 1
\end{array}\right)\left(\begin{array}{cc}
0 & m_{D}\\
m_{D} & M
\end{array}\right)\left(\begin{array}{cc}
1 & -\frac{m_{D}}{M}\\
\frac{m_{D}}{M} & 1
\end{array}\right)\approx\left(\begin{array}{cc}
m_{\ell} & 0\\
0 & m_{h}
\end{array}\right)\,,
\end{equation}
for $m_D \ll M$, and the two mass eigenvalues are
\begin{equation}
m_{h}\approx M\,,\qquad m_{\ell}\approx\frac{m_{D}^{2}}{M}=\frac{y^{2}v^{2}}{2M}\,.
\end{equation}
For the set of values $y^{\nu}\sim\mathcal{O}(1)$, $M\sim{\rm GUT}\sim10^{14-15}$~GeV
(GUT Majorana right-handed neutrinos), or $y^{\nu}\sim10^{-7}-10^{-6}$,
$M\sim100-1000$~GeV (EW--TeV Majorana right-handed neutrinos), one
roughly gets $m_{\ell}\sim10^{-2}$~eV, which is about the correct
order of the active neutrino masses.

Using the rotation matrix $O$ we obtain the mass eigenbasis
\begin{align}
\mathbb{V}_{\ell} & =\left(\begin{array}{c}
\nu_{\ell}^{(L)}\\
\nu_{\ell}^{(R)}
\end{array}\right)=\left(\begin{array}{c}
\nu+\frac{m_{D}}{M}(-{\rm i}\sigma_{2}N^{*})\\
{\rm i}\sigma_{2}\nu^{*}+\frac{m_{D}}{M}N
\end{array}\right)\,,\\
\mathbb{V}_{h} & =\left(\begin{array}{c}
\nu_{h}^{(L)}\\
\nu_{h}^{(R)}
\end{array}\right)=\left(\begin{array}{c}
-\frac{m_{D}}{M}\nu-{\rm i}\sigma_{2}N^{*}\\
-\frac{m_{D}}{M}({\rm i}\sigma_{2}\nu^{*})+N
\end{array}\right)\,.
\end{align}
One easily verifies
\begin{equation}
\mathbb{V}_{\ell}^{c}=\mathbb{V}_{\ell}\,,\qquad\qquad\mathbb{V}_{h}^{c}=\mathbb{V}_{h}\,,
\end{equation}
and thus they are Majorana fields. Recall the right-handed Majorana
neutrino $\mathbb{N}$ and further define a new Majorana field $\mathbb{V}$
using the original SM left-handed neutrino $\nu$:
\begin{equation}
\mathbb{V}\equiv\left(\begin{array}{c}
\nu\\
{\rm i}\sigma_{2}\nu^{*}
\end{array}\right)\,,\qquad\mathbb{N}=\left(\begin{array}{c}
-{\rm i}\sigma_{2}N^{*}\\
N
\end{array}\right)\,,
\end{equation}
one writes down the final results:
\begin{align}
\mathbb{V}_{\ell} & =\mathbb{V}+\frac{m_{D}}{M}\mathbb{N}\,,\\
\mathbb{V}_{h} & =-\frac{m_{D}}{M}\mathbb{V}+\mathbb{N}\,,
\end{align}
where $\mathbb{V}_{\ell},\mathbb{V}_{h}$ are the mass eigenstates
for the light and heavy Majorana neutrinos,
and are constructed using  the SM neutrino fields $\nu$ (only left-handed, 2 d.o.f.)
and the seesaw right-handed Majorana neutrino $N$ (also 2 d.o.f.).
Especially, $\mathbb{V}_{\ell}$ is supposed to be the (observed) light active neutrino.

On the other hand, one can express the original basis in terms of
the mass eigenbasis
\begin{align}
\mathbb{V} & =\mathbb{V}_{\ell}-\frac{m_{D}}{M}\mathbb{V}_{h}\,,\\
\mathbb{N} & =\frac{m_{D}}{M}\mathbb{V}_{\ell}+\mathbb{V}_{h}\,.
\end{align}
Consequently, all SM interactions involving neutrinos are modified
once the flavor eigenstates are expressed in terms of the corresponding mass eigenstates after the seesaw mixing.

\section{PMNS parameterizations and the Dirac phase}\label{App:PMNS}

The PDG parameterization of $U_{{\rm PMNS}}$ is
\begin{align}
U_{{\rm PMNS}}^{{\rm PDG}} & =\left(\begin{array}{ccc}
1 & 0 & 0\\
0 & c_{23} & s_{23}\\
0 & -s_{23} & c_{23}
\end{array}\right)\left(\begin{array}{ccc}
c_{13} & 0 & s_{13}{\rm e}^{-{\rm i}\delta_{{\rm CP}}}\\
0 & 1 & 0\\
-s_{13}{\rm e}^{{\rm i}\delta_{{\rm CP}}} & 0 & c_{13}
\end{array}\right)\left(\begin{array}{ccc}
c_{12} & s_{12} & 0\\
-s_{12} & c_{12} & 0\\
0 & 0 & 1
\end{array}\right)\left(\begin{array}{ccc}
{\rm e}^{{\rm i}\eta_{1}} & 0 & 0\\
0 & {\rm e}^{{\rm i}\eta_{2}} & 0\\
0 & 0 & 1
\end{array}\right)\nonumber \\
 & =\left(\begin{array}{ccc}
c_{12}c_{13} & s_{12}c_{13} & s_{13}{\rm e}^{-{\rm i}\delta_{{\rm CP}}}\\
-s_{12}c_{23}-c_{12}s_{13}s_{23}{\rm e}^{{\rm i}\delta_{{\rm CP}}} & c_{12}c_{23}-s_{12}s_{13}s_{23}{\rm e}^{{\rm i}\delta_{{\rm CP}}} & c_{13}s_{23}\\
s_{12}s_{23}-c_{12}s_{13}c_{23}{\rm e}^{{\rm i}\delta_{{\rm CP}}} & -c_{12}s_{23}-s_{12}s_{13}c_{23}{\rm e}^{{\rm i}\delta_{{\rm CP}}} & c_{13}c_{23}
\end{array}\right)\left(\begin{array}{ccc}
{\rm e}^{{\rm i}\eta_{1}} & 0 & 0\\
0 & {\rm e}^{{\rm i}\eta_{2}} & 0\\
0 & 0 & 1
\end{array}\right)\,,
\end{align}
where $\theta_{ij}$ are mixing angles and $s_{ij}\equiv\sin\theta_{ij}$,
$c_{ij}\equiv\cos\theta_{ij}$; there is one Dirac phase $\delta_{{\rm CP}}$
and two Majorana phases $\eta_{1,2}$.

The $\nu_{\alpha}\to\nu_{\beta}$ neutrino oscillation probability is theoretically given by:
\begin{equation}
P_{\alpha\beta}=\frac{\mathop{\underset{i}{\sum}}|U_{\alpha i}|^{2}|U_{\beta i}|^{2}+2\underset{i<j}{\sum}[{\rm Re}(U_{\alpha i}U_{\beta j}U_{\alpha j}^{*}U_{\beta i}^{*})\cos\triangle_{ij}-{\rm Im}(U_{\alpha i}U_{\beta j}U_{\alpha j}^{*}U_{\beta i}^{*})\sin\triangle_{ij}]}{(UU^{\dagger})_{\alpha\alpha}(UU^{\dagger})_{\beta\beta}}\,,\label{Eq:NTProb}
\end{equation}
where $\triangle_{ij}=\frac{(m_{i}^{2}-m_{j}^{2})L}{2E}$, and $E$ is beam energy.
The probability $P_{\alpha\beta}$ is determined experimentally and
depends on the ratio $L/E$.
With data from various neutrino experiments,
and the measurements of $P_{ee}, P_{\mu\mu}, P_{\mu e}$,
entries of the PMNS matrix can be determined.

\paragraph{Determine the seesaw parameters}

The Type-I seesaw mass matrix $M$ including Dirac phases,
is a complex symmetric matrix that can be diagonalized by the Autonne--Takagi decomposition:
\begin{equation}
\mathbf{D}=\mathbf{U}^{T}\mathbf{M}\mathbf{U}=\mathbf{U}^{\dagger}\mathbf{M}^{\dagger}\mathbf{U}^{*}\,,
\end{equation}
where $\mathbf{D}$ is non-negative real diagonal matrix containing all
mass eigenvalues, and $\mathbf{U}$ is a $6\times6$ unitary matrix,
which transforms the flavor basis into the mass eigenbasis
\begin{equation}
\mathbf{U}^{\dagger}\left(\begin{array}{cccccc}
\nu_{e} & \nu_{\mu} & \nu_{\tau} & N_{1}^{c} & N_{2}^{c} & N_{3}^{c}\end{array}\right)^{T}=\left(\begin{array}{cccc}
\mathbb{N}_{1}^{(L)} & \mathbb{N}_{2}^{(L)} & \cdots\cdots & \mathbb{N}_{6}^{(L)}\end{array}\right)^{T}\,.
\end{equation}
$\nu_{e},\nu_{\mu},\nu_{\tau}$ are the left-handed SM neutrinos (two-component),
$N_{1,2,3}$ are the right-handed Majorana neutrinos (two-component) introduced by
the Seesaw Mechanism, and $\mathbb{N}_{1-6}^{(L)}$ are the left-handed components of Majorana fields, c.f., Eq.~(\ref{eq:LHM4spinor}),
in the mass eigenbasis.

Since the seesaw mixing is small, the light neutrino mixing can be approximated
by the upper-left $3\times3$ sub-matrix of $\mathbf{U}$, denoted by ${U}$, c.f., Eq.~\eqref{eq:U66},
\begin{equation}
{U}^{\dagger}\left(\begin{array}{c}
\nu_{e}\\
\nu_{\mu}\\
\nu_{\tau}
\end{array}\right)\approx\left(\begin{array}{c}
\mathbb{N}_{1}^{(L)}\\
\mathbb{N}_{2}^{(L)}\\
\mathbb{N}_{3}^{(L)}
\end{array}\right)\,,\qquad {U}^{\dagger}\left(\begin{array}{c}
\nu_{e}^{c}\\
\nu_{\mu}^{c}\\
\nu_{\tau}^{c}
\end{array}\right)\approx\left(\begin{array}{c}
\mathbb{N}_{1}^{(R)}\\
\mathbb{N}_{2}^{(R)}\\
\mathbb{N}_{3}^{(R)}
\end{array}\right)\,,
\end{equation}
and ${U}$ should reproduce the PMNS matrix obtained from the neutrino experimental data,
up to allowed non-unitarity effects.

Mixing angles $\theta_{ij}$ are determined by
\begin{equation}
\theta_{13}=\arcsin|{U}_{13}|\,,\qquad\theta_{23}=\arctan\frac{|{U}_{23}|}{|{U}_{33}|}\,,\qquad\theta_{12}=\arctan\frac{|{U}_{12}|}{|{U}_{11}|}\,,
\end{equation}
and the Dirac phase can be calculated using the entries of $U$, e.g.,
\begin{equation}
\delta_{{\rm CP}}=\arccos\frac{|{U}_{21}|^{2}-s_{12}^{2}c_{23}^{2}-c_{12}^{2}s_{13}^{2}s_{23}^{2}}{2s_{12}s_{13}s_{23}c_{12}c_{23}}\,,
\end{equation}
which will be used to search for suitable seesaw benchmarks.

\section{Scattering processes involving Majorana fermions}\label{App:SAwM}

\paragraph{Three-point scattering}
Considering a Majorana fermion $\lambda$ coupling to a complex scalar
field $\phi$ and a Dirac fermion $\psi$ with the coupling constant
$g$
\begin{equation}
\mathcal{L}=g\phi\bar{\psi}\lambda+g^{*}\phi^{\dagger}\bar{\lambda}\psi\,.
\end{equation}
Consider the case: $m_{\lambda}>m_{\psi}+m_{\phi}$. In this case the Majorana
fermion $\lambda$ can decay to $\psi+\overline{\phi}$ and their anti-particle
pair. First consider the decaying processes $\lambda(q)\to\psi(p)+\overline{\phi}(k)$
with $q=p+k$. The Wick contraction is given by
\begin{equation}
\braket{
		\acontraction{}{{\rm out}|}{g\phi}{\bar{\psi}}  
		\acontraction{{\rm out}|g\phi\bar{\psi}}{\lambda}{}{|{\rm in}}  
{\rm out}|g\phi\bar{\psi}\lambda|{\rm in} } 
=g\bar{u}(p)u_{M}(q)\,,
\end{equation}
where the lower subscript $_M$ indicates the Majorana wave function.

Following the usual Feynman rule, the contraction $\lambda|{\rm in}\rangle$ corresponds to an incoming particle.
The same process can also be interpreted as an incoming antifermion with momentum $-p$ combining with
an incoming $\phi$ with momentum $-k$,
producing an ``antifermion'' with momentum $-q$,
and the corresponding amplitude is written as
\begin{equation}
\braket{ 
		\acontraction{}{{\rm out}|}{g\phi\bar{\psi}}{\lambda}  
		\acontraction[2ex]{{\rm out}|g\phi}{\bar{\psi}}{\lambda}{|{\rm in}}
{{\rm out}|g\phi\bar{\psi}\lambda|{\rm in} }} 
=-g\bar{v}(-p)v_{M}(-q)\,,
\end{equation}
with $\lambda$ outgoing with a momentum $-q$ and $\psi$ incoming
with a momentum $-p$, and thus $\phi$ is also incoming with momentum
$-k$. Here the $\bar{\psi}|{\rm in}\rangle$ contraction gives rise
to an incoming antifermion with momentum $-p$ and $\langle{\rm out}|\lambda$
corresponds to an outgoing antifermion with momentum $-q$.
The extra minus sign comes from taking the $\langle{\rm out}|\lambda$ contraction
by exchanging the order of the two fermionic fields $\psi$ and $\lambda$.

Since the spinor product should come out to be a scalar,
\begin{align}
\bar{v}(-p)v_{M}(-q) & =\big[\bar{v}(-p)v_{M}(-q)\big]^{T}=\big[v^{\dagger}(-p)\gamma^{0}v_{M}(-q)\big]^{T}=v_{M}^{T}(-q)\big(\gamma^{0}\big)^{T}v^{\dagger T}(-p)\nonumber \\
 & =v_{M}^{T}(-q)\bar{v}^{T}(-p)=v_{M}^{T}(-q)C^{T}C\bar{v}^{T}(-p)=\bar{u}_{M}(-q)C^{T}C^{T}u(-p)\nonumber \\
 & =-\bar{u}_{M}(-q)u(-p)\,.
\end{align}
where we have used $C\equiv-{\rm i}\gamma^{2}\gamma^{0}$ and
\begin{equation}
C^{T}C=+1\,,\qquad C^{T}C^{T}=-1\,.
\end{equation}
The charge conjugation of the positive / negative energy wave functions
are
\begin{gather}
u = -{\rm i}\gamma^{2} v^* =-{\rm i}\gamma^{2}\gamma^{0}\bar{v}^{T}=C\bar{v}^{T}\,,\\
u^T = \Big[-{\rm i}\gamma^{2}\gamma^{0}\bar{v}^{T}\Big]^{T}=\bar{v}C^{T}\,,\\
v = -{\rm i}\gamma^{2}u^* = -{\rm i}\gamma^{2}\gamma^{0}\bar{u}^{T}=C\bar{u}^{T}\,,\\
v^T = \Big[-{\rm i}\gamma^{2}\gamma^{0}\bar{u}^{T}\Big]^{T}=\bar{u}C^{T}\,.
\end{gather}
Thus we find
\begin{equation}
\langle{\rm out}|g\phi\bar{\psi}\lambda|{\rm in}\rangle=-g\bar{v}(-p)v_{M}(-q)=g\overline{{u}_{M}}(-q)u(-p)\,.
\end{equation}
The spin-sum gives rise to the same result:
\begin{equation}
g^{2}{\rm Tr}\big[(\slashed p+m_{\psi})(\slashed q+m_{\lambda})\big]=g^{2}{\rm Tr}\big[(-\slashed p+m_{\psi})(-\slashed q+m_{\lambda})\big]=g^{2}{\rm Tr}(+\slashed p\slashed q+m_{\psi}m_{\lambda})\,.
\end{equation}
On the other hand, the $h.c.$ term contributes
\begin{equation}
\langle{\rm out}|g^{*}\phi^{\dagger}\bar{\lambda}\psi|{\rm in}\rangle=-g^{*}\overline{{v}_{M}}(q)v(p)\,.
\end{equation}

\paragraph{Four-point scattering}
To compute the quantum number (fermion number)-violating process such as $LL\to HH$ or $\mathbb{N}\mathbb{N}\to HH$
(Majorana fermion can be the external leg, or the propagator), the common approach is:
\begin{enumerate}
  \item Choose a fermion-flow direction on each fermion line:
  for a Dirac fermion, the fermion-flow aligns with charge flow;
  whereas for a Majorana fermion, the fermion-flow direction is just a convention.
  Starting from an external fermion line,
  draw the fermion-flow arrow, passing through the propagator, and keep drawing until it goes out from another external leg.
  \item For an external leg, compare its physical momentum direction and the fermion-flow direction just drawn.
  For an incoming fermion (antifermion) with its momentum direction opposite to the fermion-flow direction,
  the corresponding Feynman rule is $\bar{v}$ ($u$) instead of $u$ ($\bar{v}$).
  Similarly, for an outgoing fermion (antifermion) with its momentum direction opposite to the fermion-flow direction,
  the corresponding Feynman rule is $v$ ($\bar{u}$) instead of $\bar{u}$ ($v$).
  \item Write down the spinor chain in the amplitude and compute the squared amplitude using the usual techniques.
\end{enumerate}
This fermion-flow formulation also applies to the above three-point scattering computations.

\section{Relevant Feynman diagrams and collision terms}\label{App:CT}

\begin{figure}[h!]
	\centering
	\includegraphics[scale=0.1]{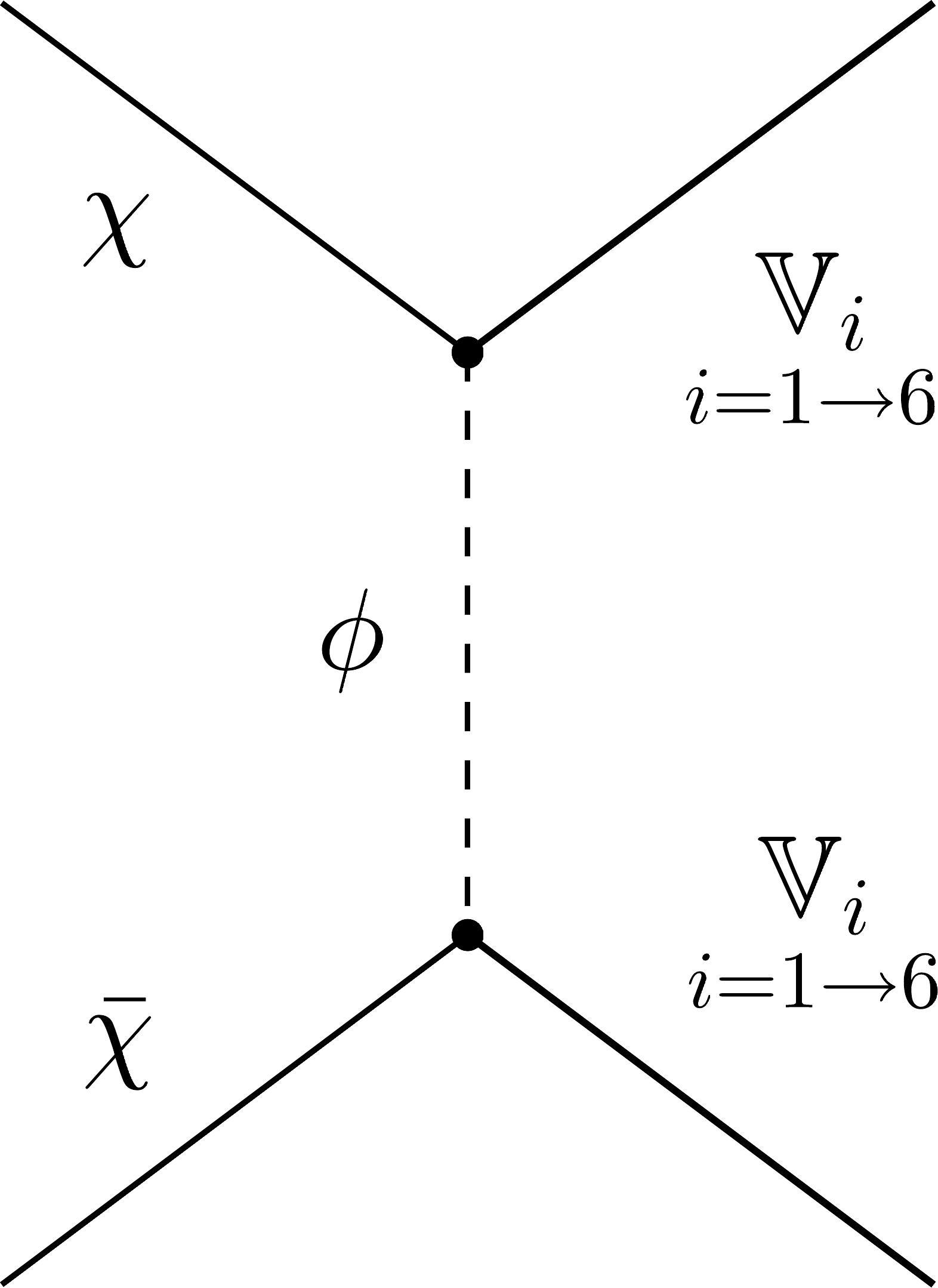}\qquad
	\includegraphics[scale=0.1]{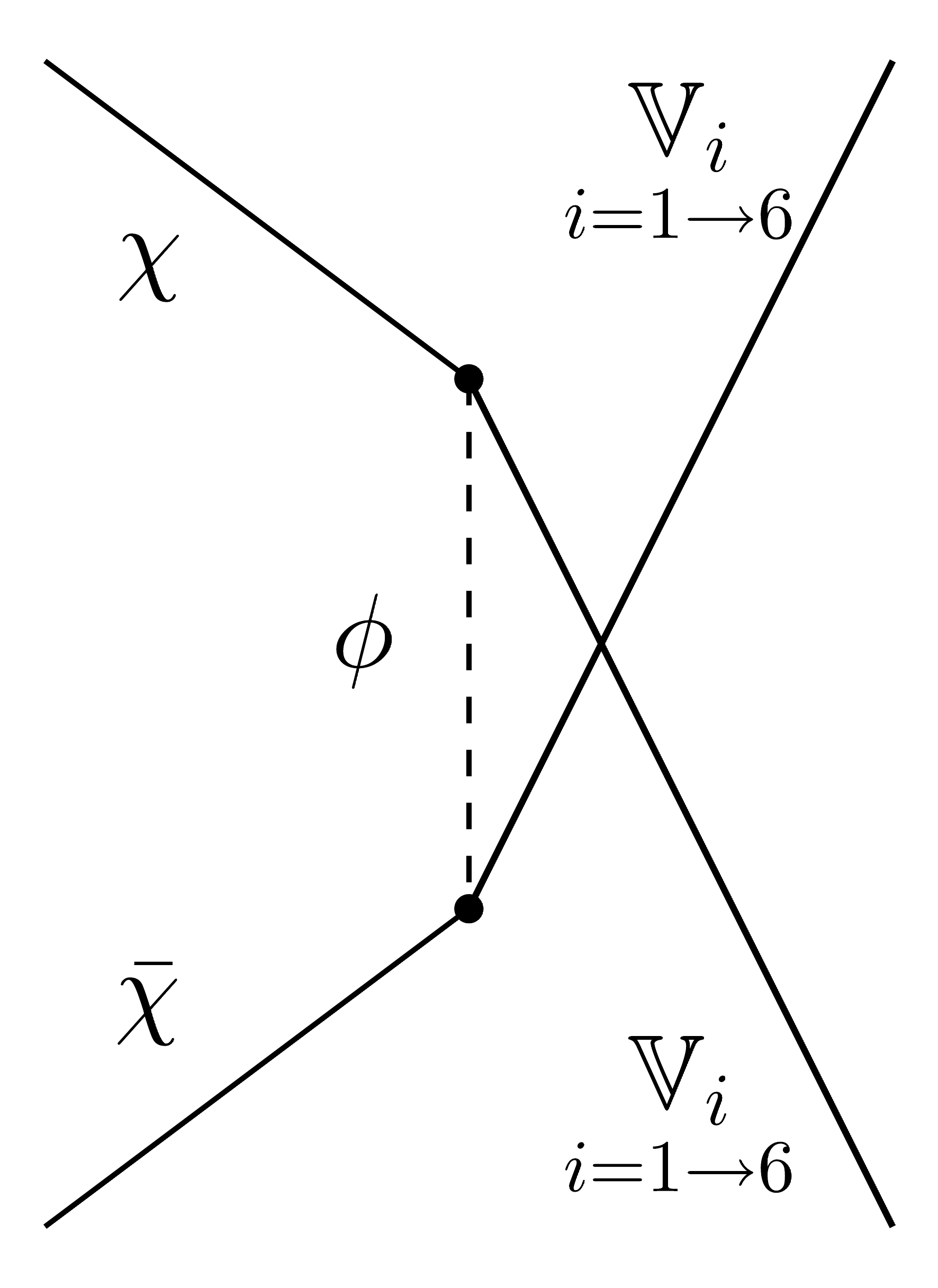}\qquad
	\includegraphics[scale=0.1]{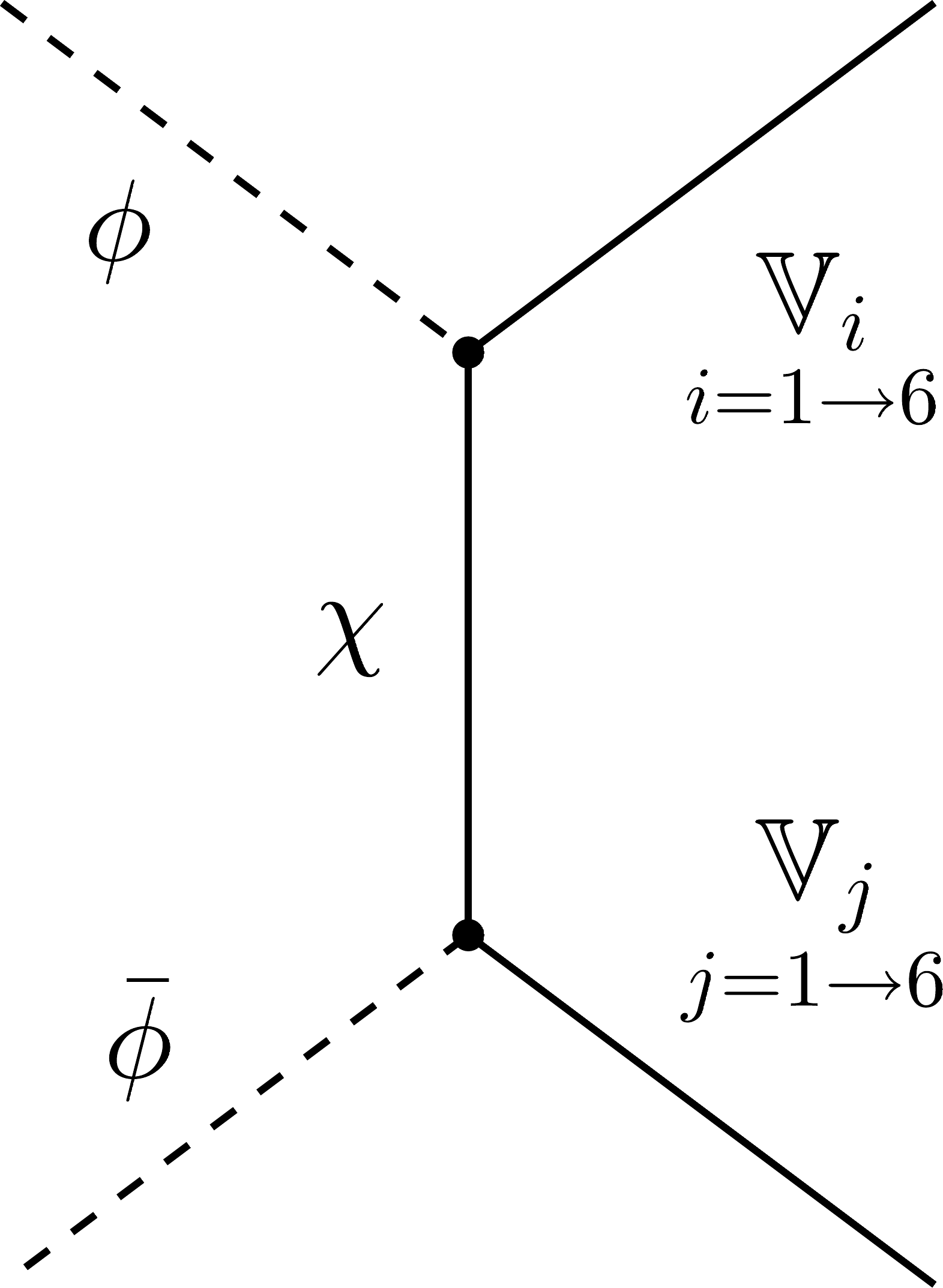}\qquad
	\includegraphics[scale=0.1]{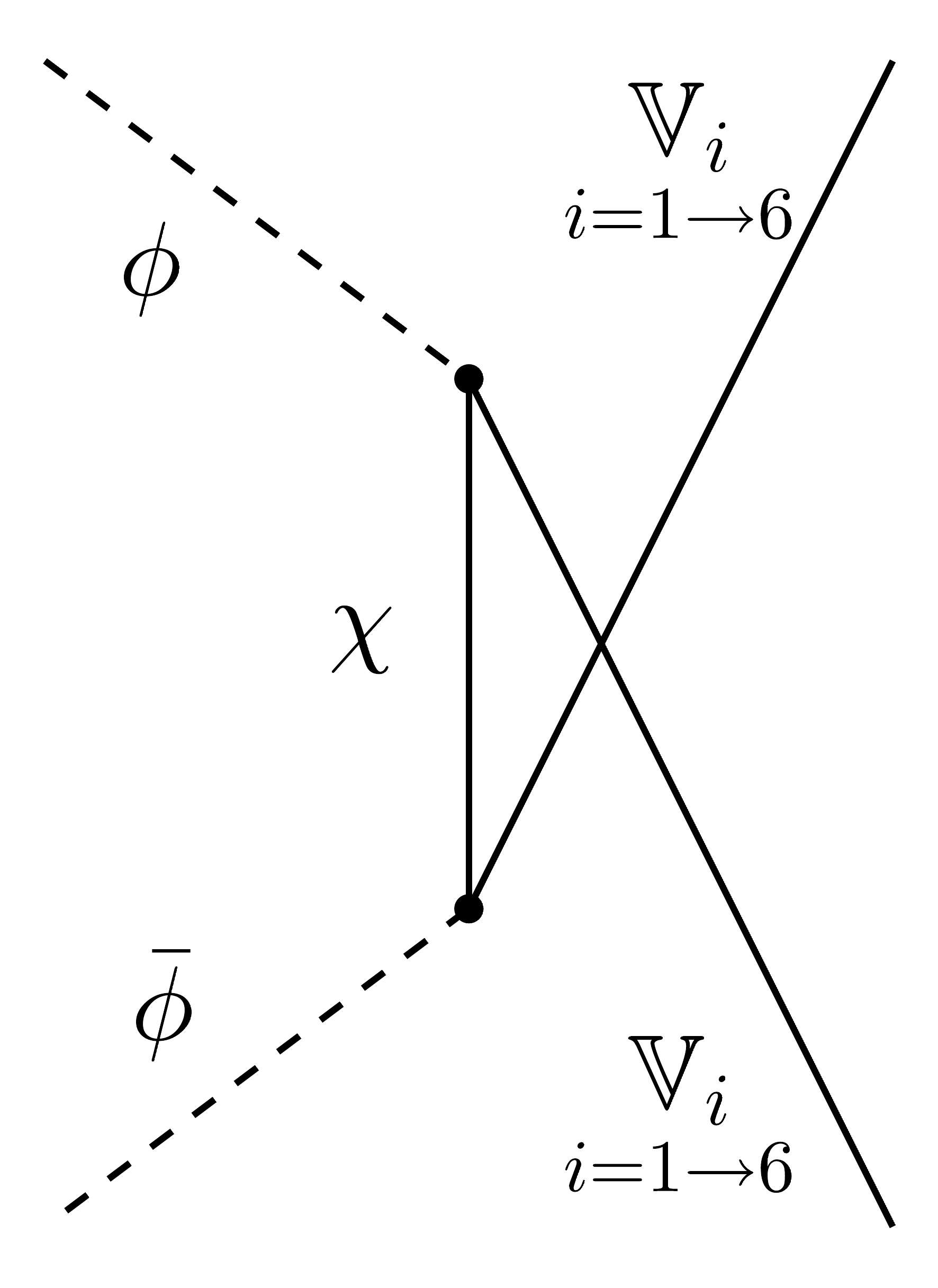}\\~\\~\\
	\includegraphics[scale=0.1]{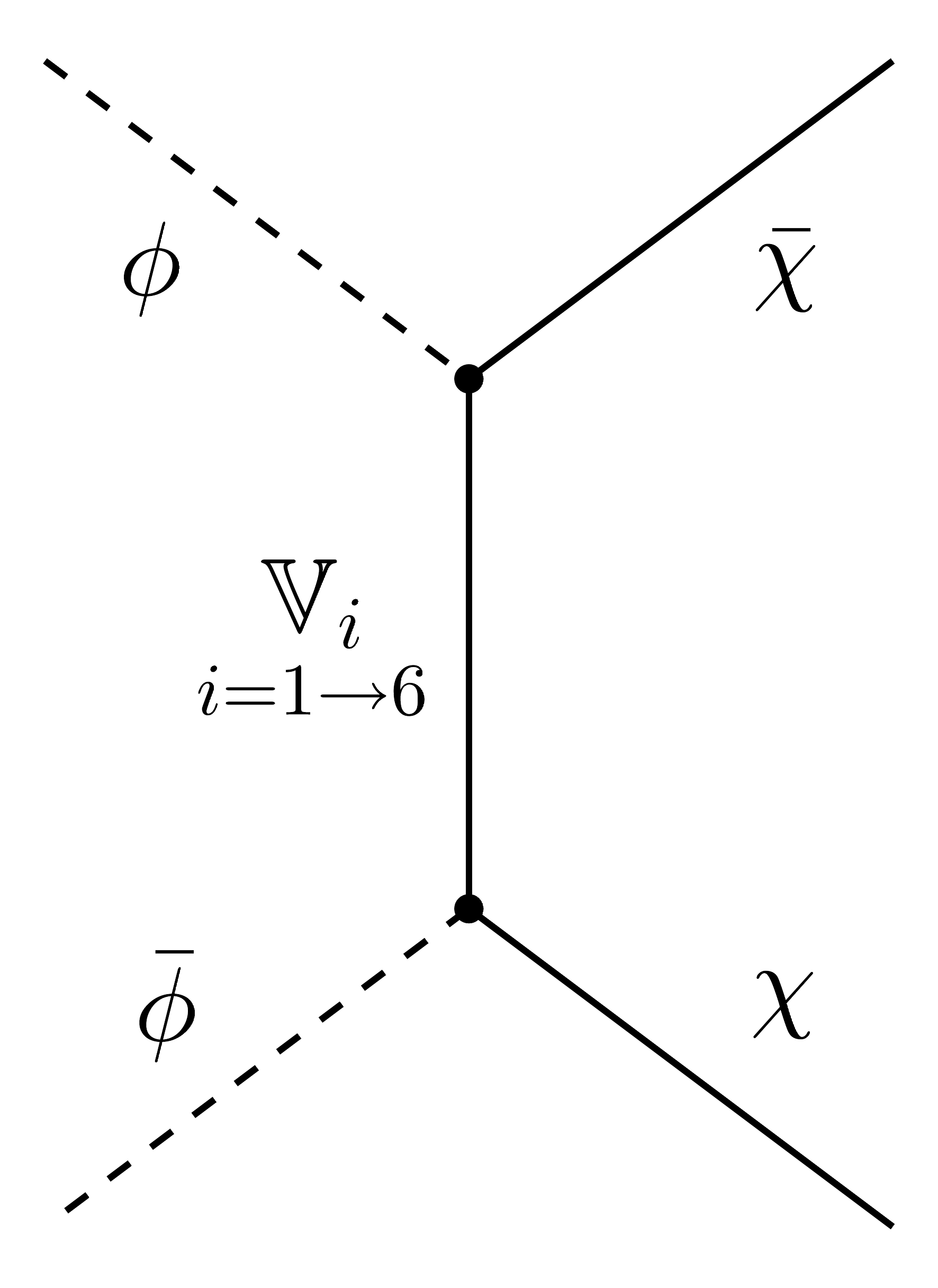}\qquad
	\includegraphics[scale=0.1]{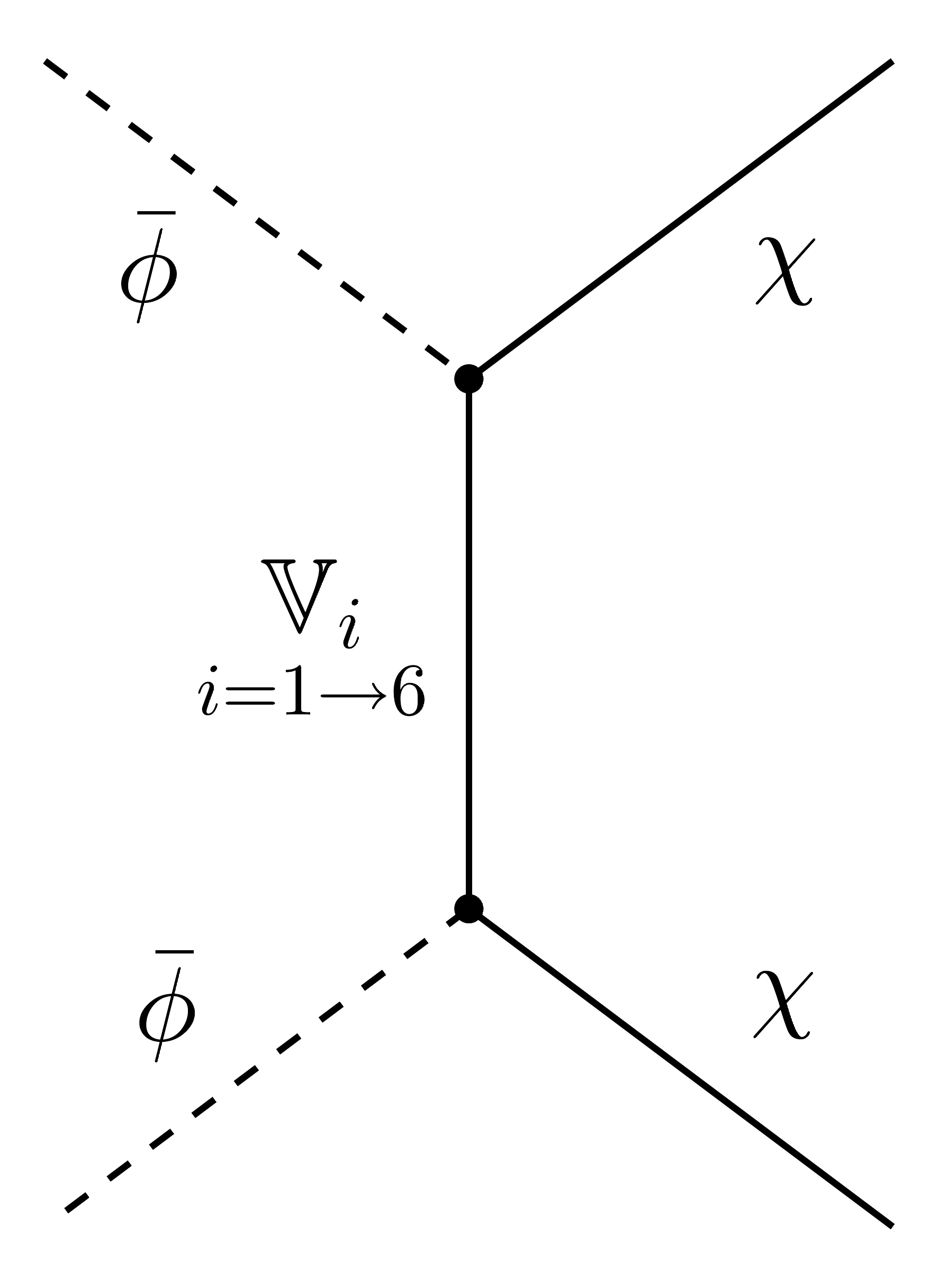}\\~\\~\\
	\includegraphics[scale=0.17]{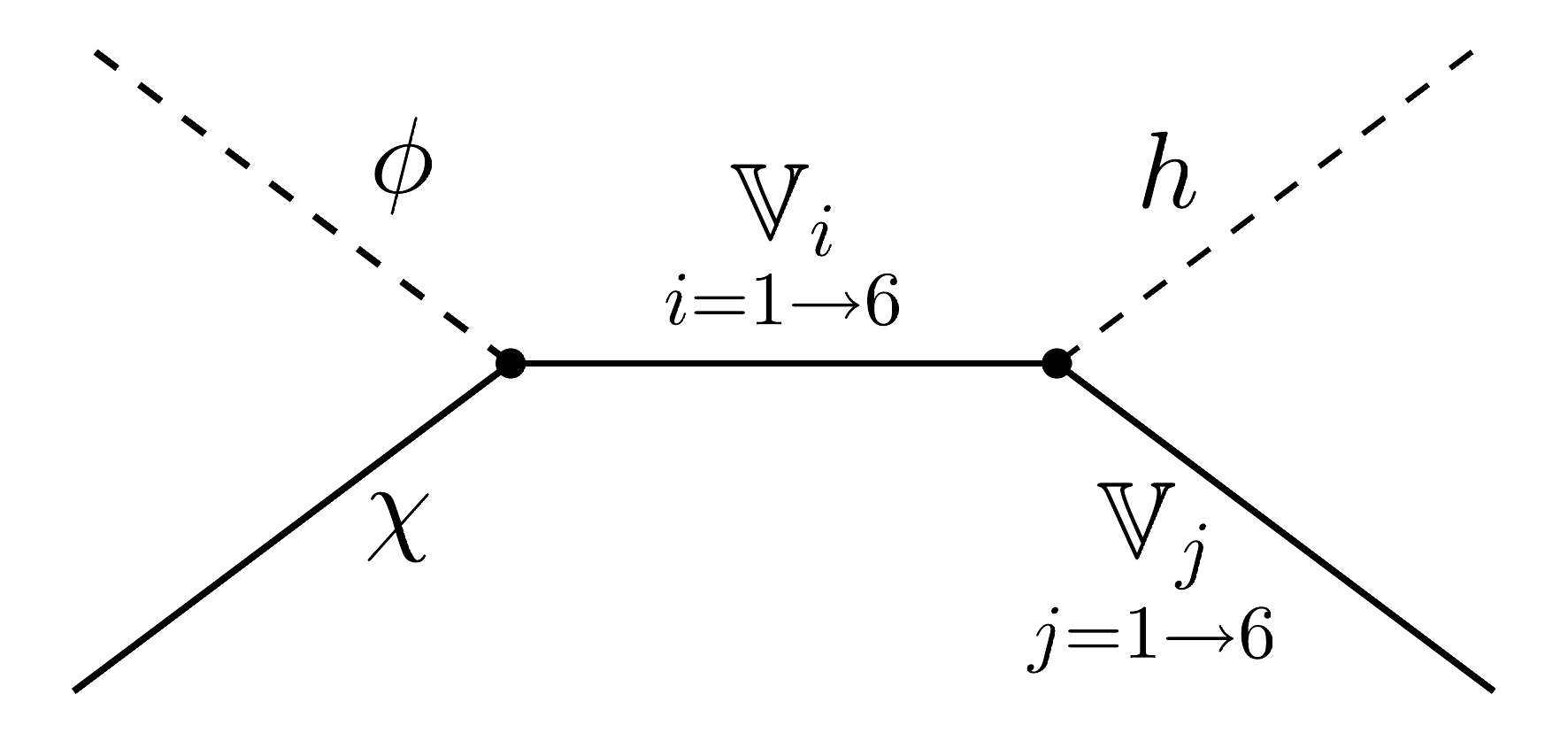}\qquad
	\includegraphics[scale=0.17]{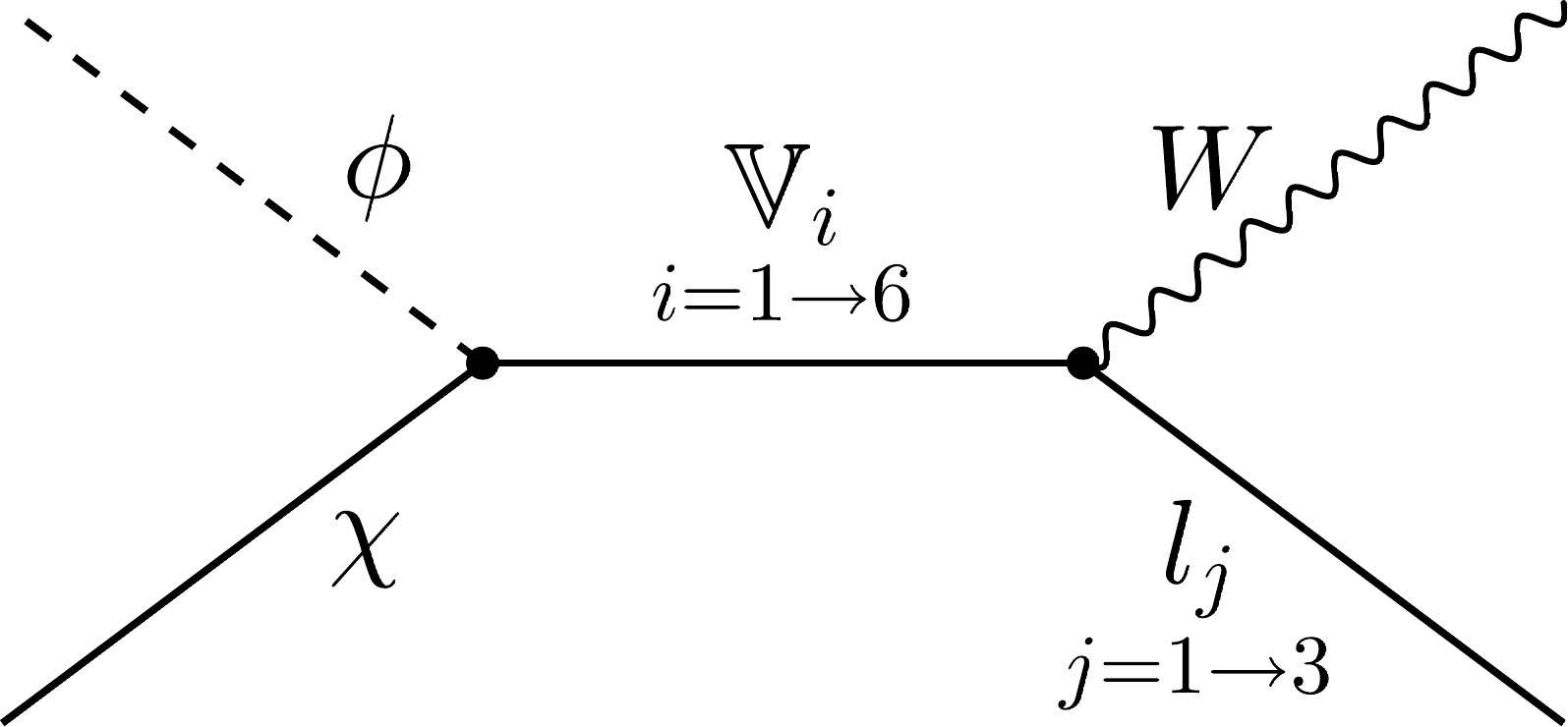}\qquad
	\includegraphics[scale=0.17]{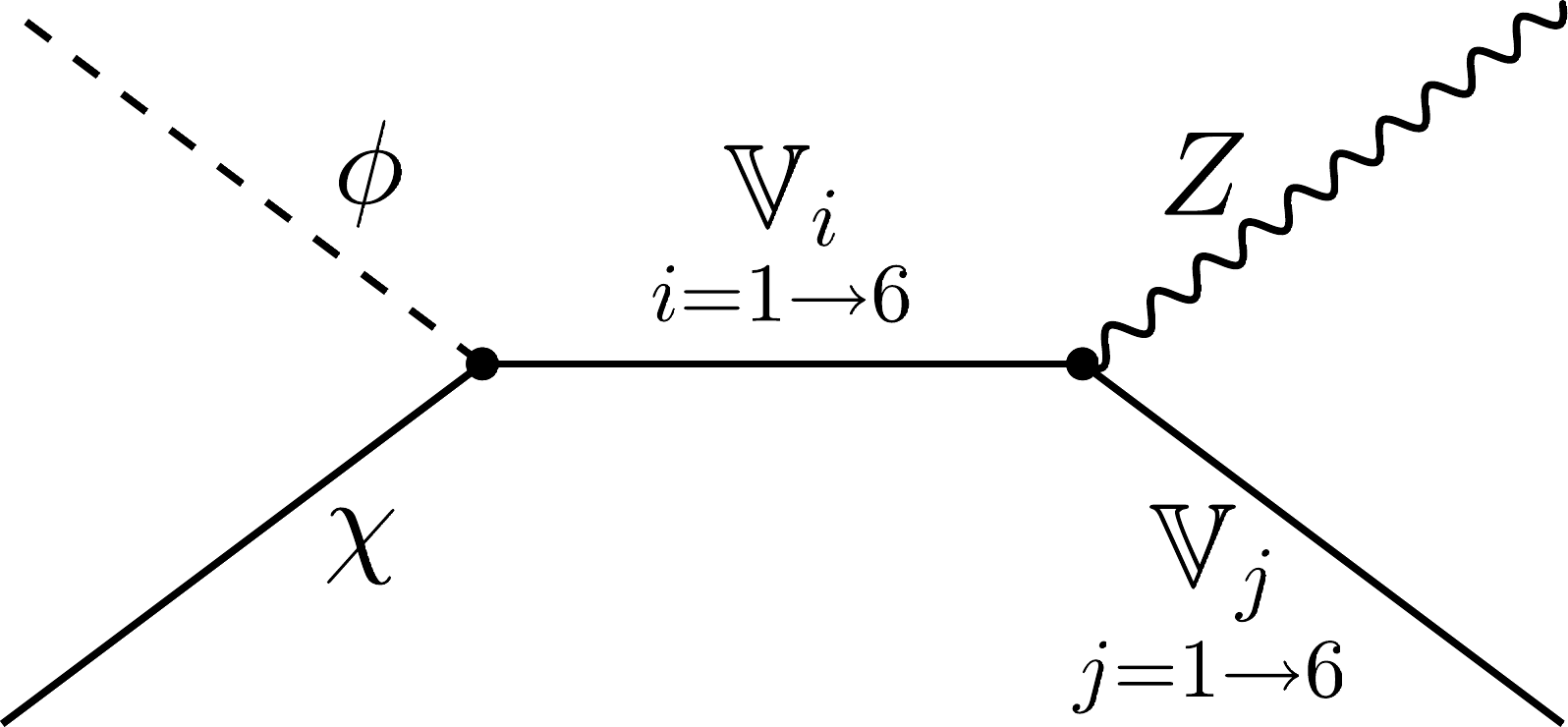}
	\caption{Relevant Feynman diagrams for $2 \to 2$ scattering processes are shown.
    The first row depicts dark particle $\chi,\phi$ annihilation into light or heavy neutrinos after seesaw mixing;
    the inverse processes lead to freeze-in production of the dark sector particles.
    The second row shows internal dark sector interactions mediated by light or heavy neutrinos.
    The third row illustrates processes arising from $\chi \phi \to L H$ after electroweak symmetry breaking.}
	\label{Fig:FD22}
\end{figure}

\begin{figure}[h!]
	\centering
	\includegraphics[scale=0.14]{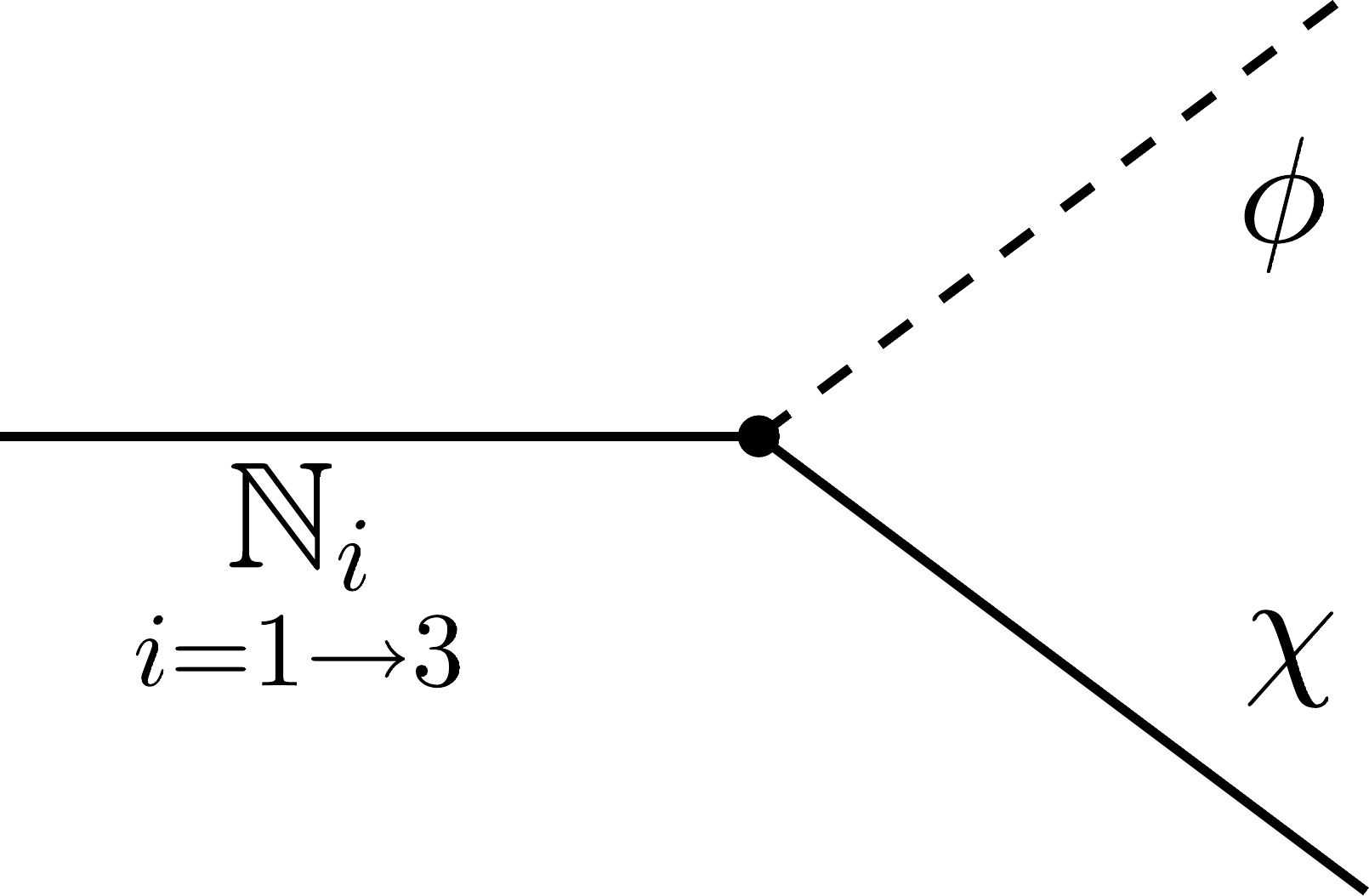}\\~\\
	\includegraphics[scale=0.15]{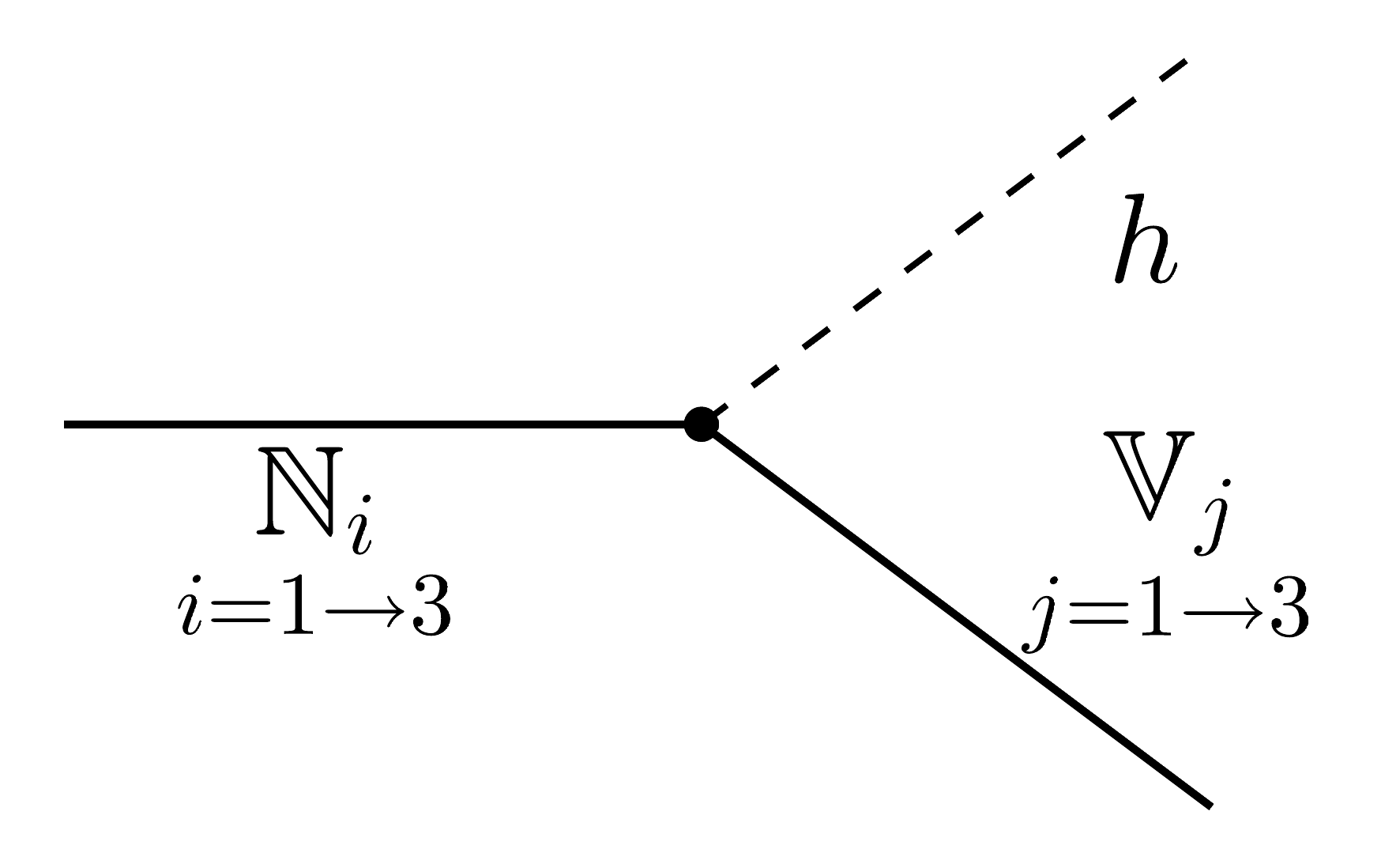}\qquad
	\includegraphics[scale=0.15]{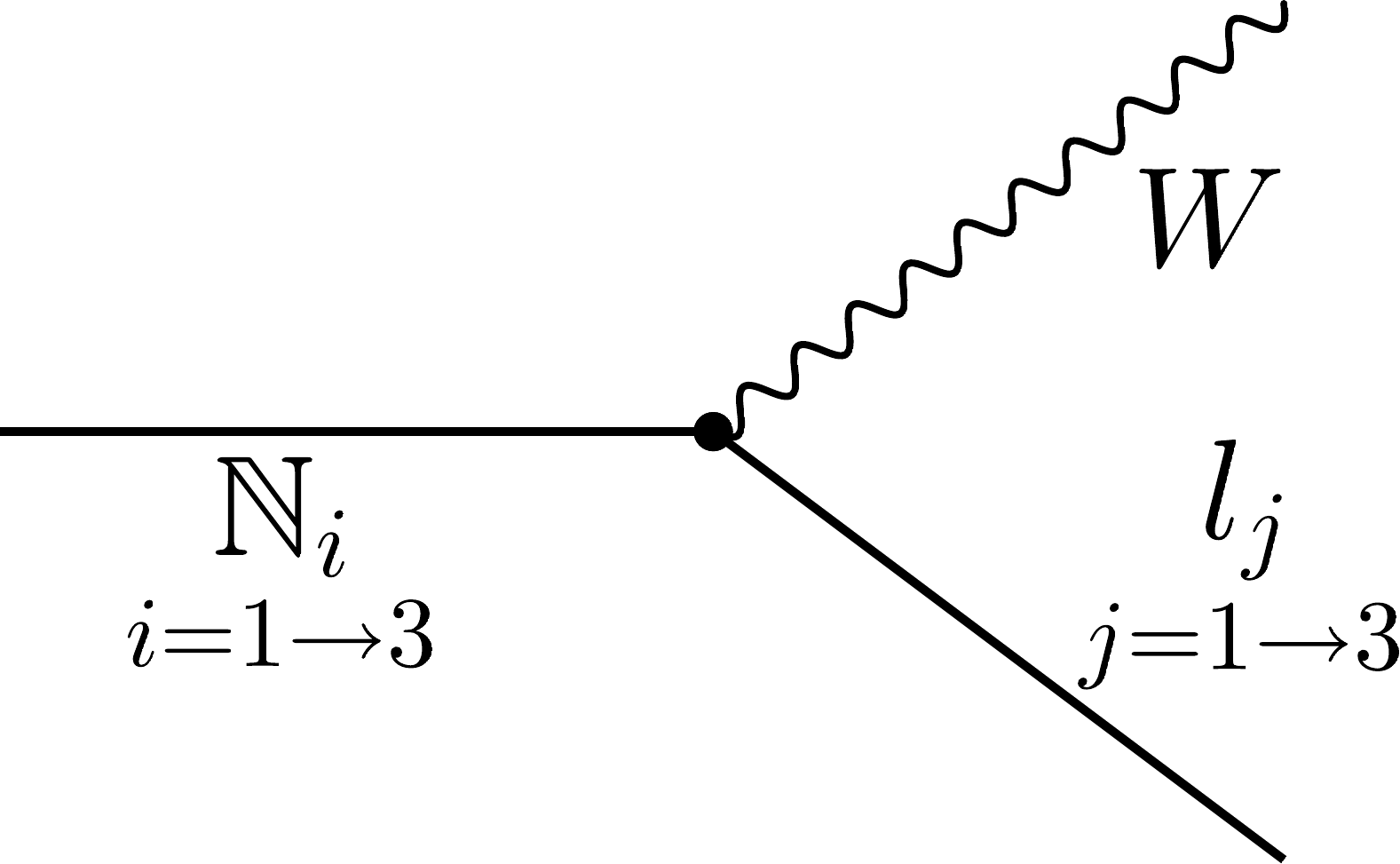}\qquad
	\includegraphics[scale=0.15]{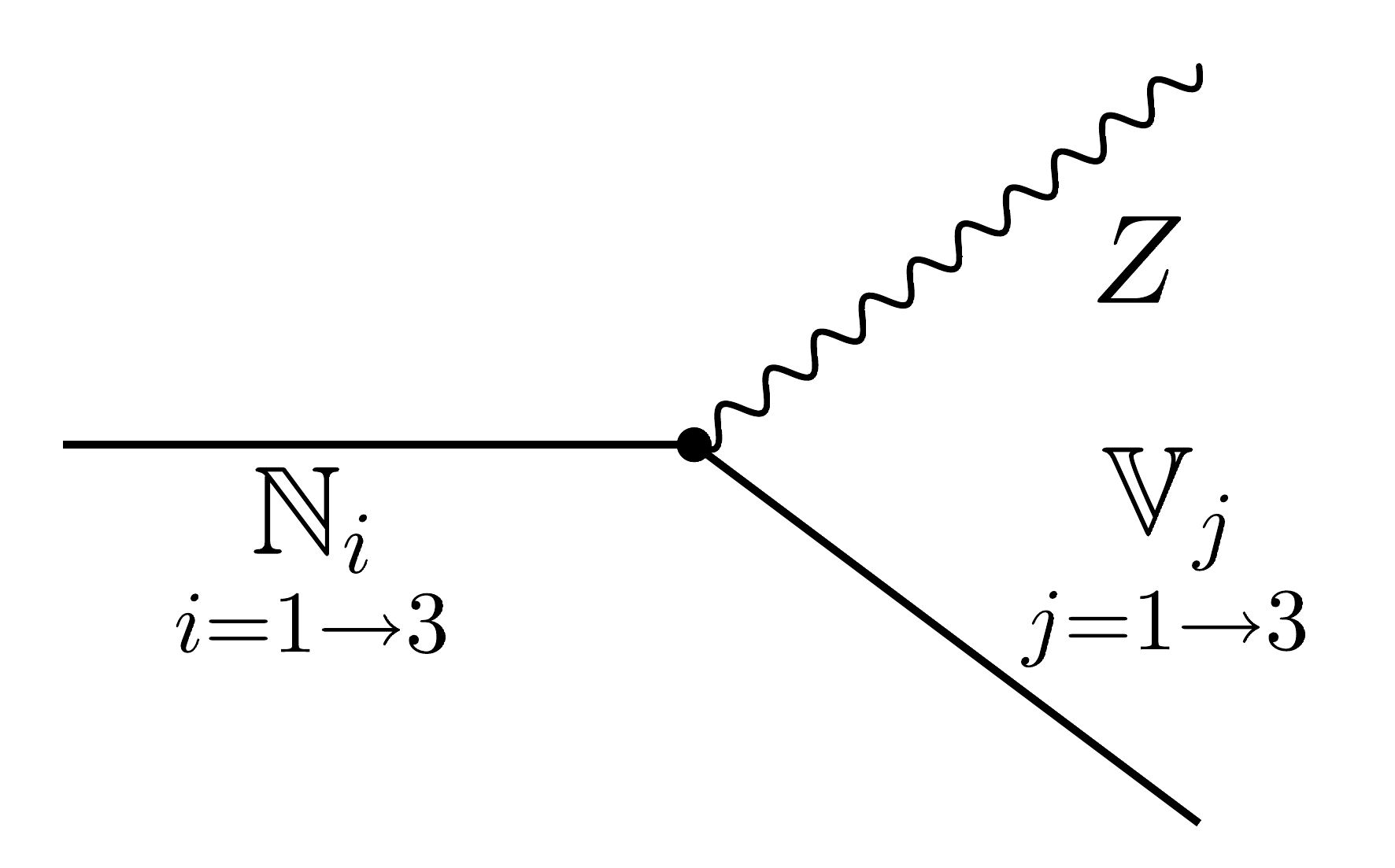}
	\caption{Feynman diagrams of the heavy neutrino $\mathbb{N}_{1,2,3}$ decaying to dark sector particles $\chi,\phi$, and to SM particles.}
	\label{Fig:FDND}
\end{figure}

\begin{figure}[h!]
	\centering
	\includegraphics[scale=0.15]{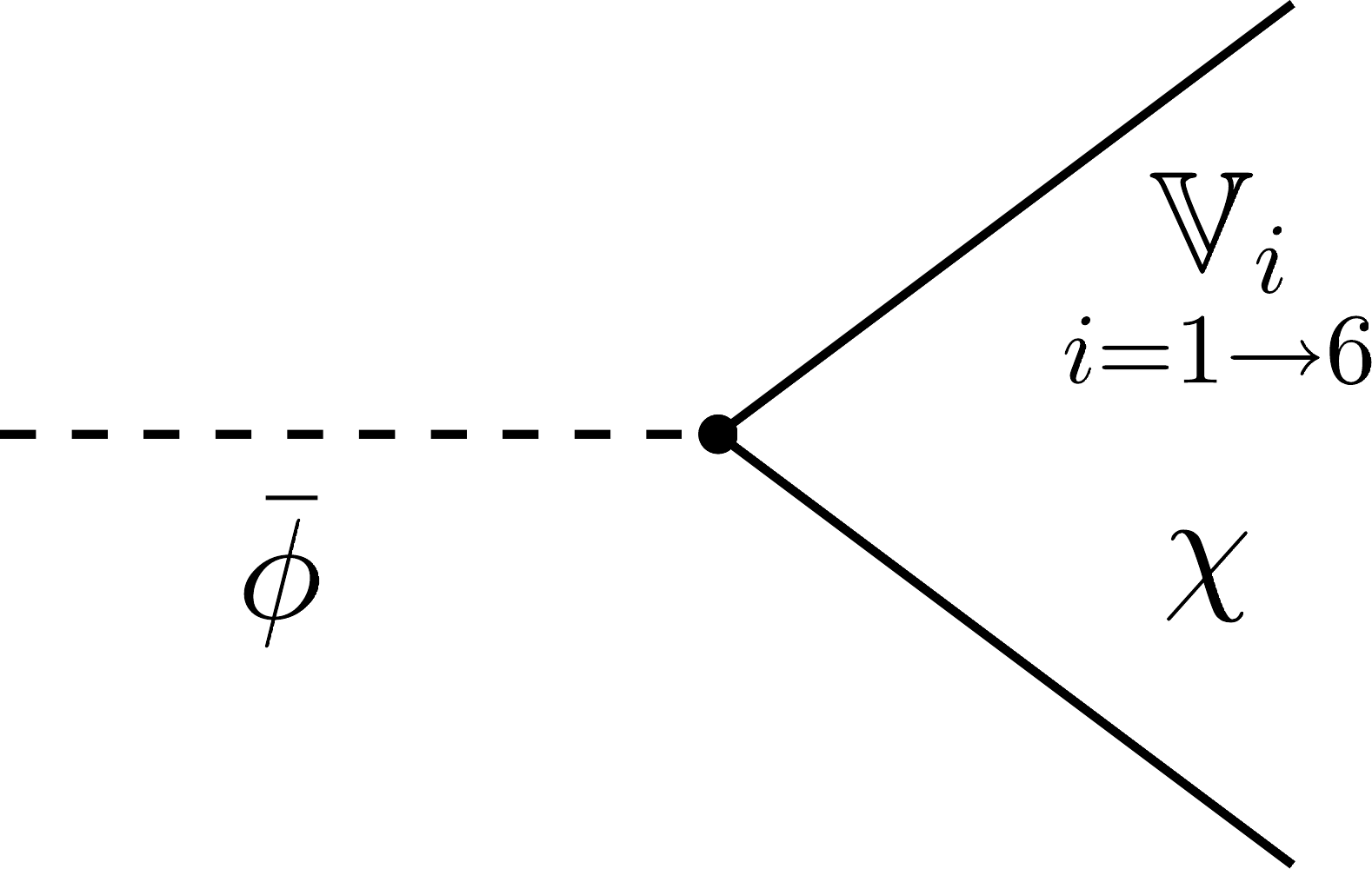}
	\caption{Feynman diagram of $\phi$ decay to $\chi$ and light (always present) or heavy (if $m_\phi > m_\chi + m_{\mathbb{N}_i}$) neutrinos. }
	\label{Fig:FDPD}
\end{figure}

Relevant Feynman diagrams involving the dark sector particles $\chi,\phi$ and the active/heavy neutrinos $\mathbb{V}_{i=1-6}$
are summarized in Figs.~\ref{Fig:FD22}, \ref{Fig:FDND}, \ref{Fig:FDPD}.

\newpage


The relevant collision terms $\mathcal{C}_{\chi},\mathcal{C}_{\phi},\mathcal{C}_{\mathbb{N}_{i}}$ are given by
\begin{align}
	\mathcal{C}_{\chi}= & \sum_{i,j=1}^{6}\bigl[Y_{\mathbb{V}_{i}}Y_{\mathbb{V}_{j}}\left\langle \sigma v\right\rangle _{\mathbb{V}_{i}\mathbb{V}_{j}\to\chi\overline{\chi}}-Y_{\chi}^{2}\left\langle \sigma v\right\rangle _{\chi\overline{\chi}\to\mathbb{V}_{i}\mathbb{V}_{j}}\nonumber \\
	& +2\bigl(Y_{h}Y_{\mathbb{V}_{i}}\left\langle \sigma v\right\rangle _{h\mathbb{V}_{i}\to\chi\phi}-Y_{\chi}Y_{\phi}\left\langle \sigma v\right\rangle _{\chi\phi\to h\mathbb{V}_{i}}\bigr)+(h\leftrightarrow Z)\nonumber \\
	& +\theta(m_{\phi}-m_{\chi}-m_{i})\bigl(\tfrac{1}{s}Y_{\phi}\left\langle \Gamma\right\rangle_{\phi\to\overline{\chi}\mathbb{V}_{i}}-Y_{\chi}Y_{\mathbb{V}_{i}}\left\langle \sigma v\right\rangle _{\overline{\chi}\mathbb{V}_{i}\to\phi}\bigr)\bigr]\nonumber \\
	& +2\sum_{i=1}^{3}\bigl(Y_{W}Y_{\ell_{i}}\left\langle \sigma v\right\rangle _{W\ell_{i}\to\chi\phi}-Y_{\chi}Y_{\phi}\left\langle \sigma v\right\rangle _{\chi\phi\to W\ell_{i}}\bigr)\nonumber \\
	& +\sum_{i=4}^{6}\theta(m_{\phi}-m_{\chi}-m_{i})\bigl(\tfrac{1}{s}Y_{\mathbb{V}_{i}}\left\langle \Gamma\right\rangle_{\mathbb{V}_{i}\to\chi\phi}-Y_{\chi}Y_{\phi}\left\langle \sigma v\right\rangle _{\chi\phi\to\mathbb{V}_{i}}\bigr)\nonumber \\
	& +Y_{\phi}^{2}\bigl(2\left\langle \sigma v\right\rangle _{\phi\phi\to\overline{\chi\chi}}+\left\langle \sigma v\right\rangle _{\phi\overline{\phi}\to\chi\overline{\chi}}\bigr)-Y_{\chi}^{2}\bigl(2\left\langle \sigma v\right\rangle _{\overline{\chi\chi}\to\phi\phi}+\left\langle \sigma v\right\rangle _{\chi\overline{\chi}\to\phi\overline{\phi}}\bigr)\,,\label{eq:Cchi}\\
	\mathcal{C}_{\phi}= & \sum_{i,j=1}^{6}\bigl[Y_{\mathbb{V}_{i}}Y_{\mathbb{V}_{j}}\left\langle \sigma v\right\rangle _{\mathbb{V}_{i}\mathbb{V}_{j}\to\phi\overline{\phi}}-Y_{\phi}^{2}\left\langle \sigma v\right\rangle _{\phi\overline{\phi}\to\mathbb{V}_{i}\mathbb{V}_{j}}\nonumber \\
	& +2\bigl(Y_{h}Y_{\mathbb{V}_{i}}\left\langle \sigma v\right\rangle _{h\mathbb{V}_{i}\to\chi\phi}-Y_{\chi}Y_{\phi}\left\langle \sigma v\right\rangle _{\chi\phi\to h\mathbb{V}_{i}}\bigr)+(h\leftrightarrow Z)\nonumber \\
	& +\theta(m_{\phi}-m_{\chi}-m_{i})\bigl(-\tfrac{1}{s}Y_{\phi}\left\langle \Gamma\right\rangle_{\phi\to\overline{\chi}\mathbb{V}_{i}}+Y_{\chi}Y_{\mathbb{V}_{i}}\left\langle \sigma v\right\rangle _{\overline{\chi}\mathbb{V}_{i}\to\phi}\bigr)\bigr]\nonumber \\
	& +2\sum_{i=1}^{3}\bigl(Y_{W}Y_{\ell_{i}}\left\langle \sigma v\right\rangle _{W\ell_{i}\to\chi\phi}-Y_{\chi}Y_{\phi}\left\langle \sigma v\right\rangle _{\chi\phi\to W\ell_{i}}\bigr)\nonumber \\
	& +\sum_{i=4}^{6}\theta(m_{i}-m_{\chi}-m_{\phi})\bigl(\tfrac{1}{s}Y_{\mathbb{V}_{i}}\left\langle \Gamma\right\rangle_{\mathbb{V}_{i}\to\chi\phi}-Y_{\chi}Y_{\phi}\left\langle \sigma v\right\rangle _{\chi\phi\to\mathbb{V}_{i}}\bigr)\nonumber \\
	& -Y_{\phi}^{2}\bigl(2\left\langle \sigma v\right\rangle _{\phi\phi\to\overline{\chi\chi}}+\left\langle \sigma v\right\rangle _{\phi\overline{\phi}\to\chi\overline{\chi}}\bigr)+Y_{\chi}^{2}\bigl(2\left\langle \sigma v\right\rangle _{\overline{\chi\chi}\to\phi\phi}+\left\langle \sigma v\right\rangle _{\chi\overline{\chi}\to\phi\overline{\phi}}\bigr)\,,\label{eq:Cphi}\\
	\mathcal{C}_{\mathbb{N}_{i}}= & \sum_{i,j=1}^{3}\bigl[c_{ij}\bigl(Y_{\chi}^{2}\left\langle \sigma v\right\rangle _{\chi\overline{\chi}\to\mathbb{N}_{i}\mathbb{N}_{j}}+Y_{\phi}^{2}\left\langle \sigma v\right\rangle _{\phi\overline{\phi}\to\mathbb{N}_{i}\mathbb{N}_{j}}\bigr)\nonumber \\
	& -c_{ij}Y_{\mathbb{N}_{i}}Y_{\mathbb{N}_{j}}\bigl(\left\langle \sigma v\right\rangle _{\mathbb{N}_{i}\mathbb{N}_{j}\to\chi\overline{\chi}}+\left\langle \sigma v\right\rangle _{\mathbb{N}_{i}\mathbb{N}_{j}\to\phi\overline{\phi}}\bigr)\nonumber \\
	& +2\bigl(Y_{\chi}Y_{\phi}\left\langle \sigma v\right\rangle _{\chi\phi\to h\mathbb{N}_{i}}-Y_{h}Y_{\mathbb{N}_{i}}\left\langle \sigma v\right\rangle _{h\mathbb{N}_{i}\to\chi\phi}\bigr)+2(h\leftrightarrow Z)\nonumber \\
	& +2\,\theta(M_{i}-m_{\chi}-m_{\phi})\bigl(-\tfrac{1}{s}Y_{\mathbb{N}_{i}}\left\langle \Gamma\right\rangle_{\mathbb{N}_{i}\to\chi\phi}+Y_{\chi}Y_{\phi}\left\langle \sigma v\right\rangle _{\chi\phi\to\mathbb{N}_{i}}\bigr)\nonumber \\
	& +2\,\theta(m_{\phi}-m_{\chi}-M_{i})\bigl(\tfrac{1}{s}Y_{\phi}\left\langle \Gamma\right\rangle_{\phi\to\overline{\chi}\mathbb{N}_{i}}-Y_{\chi}Y_{\mathbb{N}_{i}}\left\langle \sigma v\right\rangle _{\overline{\chi}\mathbb{N}_{i}\to\phi}\bigr)\nonumber \\
	& +2\,\theta(M_{i}-m_{W}-m_{\ell_{j}})\bigl(-\tfrac{1}{s}Y_{\mathbb{N}_{i}}\left\langle \Gamma\right\rangle_{\mathbb{N}_{i}\to W\ell_{j}}+Y_{W}Y_{\ell_{j}}\left\langle \sigma v\right\rangle _{W\ell_{j}\to\mathbb{N}_{i}}\bigr)\nonumber \\
	& +2\,\theta(M_{i}-m_{h}-m_{j})\bigl(-\tfrac{1}{s}Y_{\mathbb{N}_{i}}\left\langle \Gamma\right\rangle_{\mathbb{N}_{i}\to h\mathbb{V}_{j}}+Y_{h}Y_{\mathbb{V}_{j}}\left\langle \sigma v\right\rangle _{h\mathbb{V}_{j}\to\mathbb{N}_{i}}\bigr)+(h\leftrightarrow Z)\bigr]\,,\label{eq:CNi}
\end{align}
where $c_{ij}$ = 1 (for $i\neq j$), $c_{ij}$ = 2 (for $i=j$).

In the first two collision terms, we introduce the notation $\mathbb{V}_{4,5,6}\equiv\mathbb{N}_{1,2,3}$
and $m_{4,5,6}\equiv~M_{1,2,3}$ for simplicity.
The theta functions indicate whether the decay and inverse annihilation processes are kinematically allowed.

For the process $a+b\leftrightarrow c+d$, where $a,b$ are dark sector
particles and $c,d$ are SM particles kept in thermal equilibrium,
the collision term can be written as
\begin{equation}
\mathcal{C}_{ab\to cd}=-\int{\rm d}\Pi_{a}{\rm d}\Pi_{b}{\rm d}\Pi_{c}{\rm d}\Pi_{d}\,(2\pi)^{4}\left|\mathcal{M}\right|_{ab\leftrightarrow cd}^{2}\delta^{4}\big(p_{a}+p_{b}-p_{c}-p_{d}\big)\,\big(f_{a}f_{b}-f_{c}^{{\rm eq}}f_{d}^{{\rm eq}}\big)\,.
\end{equation}
Here $f_{a}$ and $f_{b}$ denote the actual dark sector distribution
functions, which in general need not coincide with the corresponding
equilibrium distributions.

For the SM particles, we use the Maxwell-Boltzmann approximation
and neglect their chemical potentials, so that
\begin{equation}
f_{i}^{{\rm eq}}={\rm e}^{-E_{i}/T}.
\end{equation}
Energy conservation enforced by the phase-space delta function implies
\begin{equation}
E_{a}+E_{b}=E_{c}+E_{d}\,,
\end{equation}
and therefore
\begin{equation}
f_{c}^{{\rm eq}}f_{d}^{{\rm eq}}={\rm e}^{-(E_{c}+E_{d})/T}={\rm e}^{-(E_{a}+E_{b})/T}=f_{a}^{{\rm eq}}f_{b}^{{\rm eq}}\,.
\end{equation}
As a result, one obtains
\begin{equation}
\langle\sigma v \rangle_{cd\to ab}Y_{c}^{\rm eq}Y_{d}^{\rm eq}
=
\langle\sigma v \rangle_{ab\to cd}Y_{a}^{\rm eq}Y_{b}^{\rm eq}\,. \label{eq:EQDB}
\end{equation}
which is the detailed balance relation in thermal equilibrium,
implying that the forward and backward reaction rates are equal at equilibrium.

Accordingly, the Boltzmann equation can be written as
\begin{equation}
\frac{{\rm d}Y_{a}}{{\rm d}T}=-\frac{s}{T\overline{H}}\mathcal{C}_{ab\to cd}=\frac{s}{T\overline{H}}\big(\langle\sigma v \rangle_{ab\to cd}Y_{a}Y_{b}- \langle\sigma v \rangle_{cd\to ab}Y_{c}^{{\rm eq}}Y_{d}^{{\rm eq}}\big)\,,
\end{equation}
where the last term may be replaced by the right-hand side of Eq.~\eqref{eq:EQDB}.
One then obtains
\begin{equation}
\frac{{\rm d}Y_{a}}{{\rm d}T}
=
\frac{s}{T\overline{H}}
\langle\sigma v \rangle_{ab\to cd}
\bigl(Y_{a}Y_{b}-Y_{a}^{\rm eq}Y_{b}^{\rm eq}\bigr)\,.
\end{equation}
For processes in which either the initial or final states consist entirely of SM particles,
it is often more convenient to evaluate the evolution through the inverse process.
For example, it is simpler to compute
\begin{equation}
\big[(Y_{\chi}^{\rm eq})^{2}-Y_{\chi}^{2}\big]
\left\langle \sigma v\right\rangle_{\chi\overline{\chi}\to\mathbb{V}_{i}\mathbb{V}_{j}}
\end{equation}
than
\begin{equation}
Y_{\mathbb{V}_{i}}Y_{\mathbb{V}_{j}}
\left\langle \sigma v\right\rangle_{\mathbb{V}_{i}\mathbb{V}_{j}\to\chi\overline{\chi}}
-
Y_{\chi}^{2}
\left\langle \sigma v\right\rangle_{\chi\overline{\chi}\to\mathbb{V}_{i}\mathbb{V}_{j}}\,.
\end{equation}

\paragraph{Factors in  Boltzmann equations}

In the collision terms, the coefficient $\pm2$ in front of certain contributions indicate that two units of particle are created or annihilated. The squared amplitude should have an overall factor of 1/2 that accounts for avoiding double counting when integrating over the phase space of two identical particles, which
is however absorbed into the definition of the cross section for the given process.
Hence, at the Boltzmann equation level,
the 1/2 factor accounting for identical particles does not appear.

The factor discussed above also appears in processes involving the Majorana neutrinos. For example, both $\mathbb{N}_{i}\leftrightarrow\chi\phi$ and its conjugate channel $\mathbb{N}_{i}\leftrightarrow\overline{\chi\phi}$ contribute to the same collision term, yielding an overall factor of $2$ in $\text{d}Y_{\mathbb{N}_{i}}/\text{d}T$. In contrast, $\chi$ and $\overline{\chi}$ (and similarly $\phi$ and $\overline{\phi}$) are treated as symmetric but distinct species with $Y_{\chi}=Y_{\overline{\chi}}$ and $Y_{\phi}=Y_{\overline{\phi}}$, and thus each channel contributes to its own species and no extra factor $2$ appears in the Boltzmann equations for $Y_{\chi}$ and $Y_{\phi}$. The same reason applies to other channels such as $\phi\leftrightarrow\overline{\chi}\mathbb{N}_{i}$
($\overline{\phi}\leftrightarrow \chi\mathbb{N}_{i}$)
and $\chi\phi\,(\overline{\chi\phi})\to Z\mathbb{N}_{i}$.
For the processes $\chi\phi\leftrightarrow h\mathbb{N}_{i}$,
this contains four related channels: $\chi\phi,\overline{\chi\phi}\leftrightarrow LH,\overline{LH}$.
The factor $2$ in front of $W\ell_{i}\leftrightarrow\chi\phi$ arises from $W^{+}\ell_{i}^{-}$ and $W^{-}\ell_{i}^{+}$.

The total relic abundance of dark matter is obtained by combining the contributions from $\chi$ and $\overline{\chi}$, i.e., $2Y_\chi$, and $\phi$ and $\overline{\phi}$ ($2Y_\phi$) if they are also dark matter candidates.

\paragraph{RIS subtraction}

In a Boltzmann system containing processes $AB \leftrightarrow G \leftrightarrow CD$, where $G$ mediates the $s$-channel exchange $AB \leftrightarrow CD$, the contribution from real intermediate states (RIS) must be subtracted from the $2 \to 2$ cross section, since the on-shell resonance at $s = M_G^2$ is already accounted for by the $1 \leftrightarrow 2$ processes $AB \leftrightarrow G$ and $G \leftrightarrow CD$.
Such subtraction was 
discussed in~\cite{Kolb:1979qa,Giudice:2003jh}.

In the narrow-width limit $\Gamma_{G}\to0$, the Breit-Wigner resonance approaches a Dirac delta function
\begin{equation}
\frac{1}{\pi}\frac{M_{G}\Gamma_{G}}{(s-M_{G}^{2})^{2} + M_{G}^{2}\Gamma_{G}^{2}}
  \xrightarrow{\Gamma_{G}\to0}\delta(s-M_{G}^{2})\,,
\end{equation}
which can be seen from
\begin{equation}
\lim_{\Gamma_{G} \to 0}\int_{-\infty}^{\infty}
\left[\frac{1}{\pi}\frac{M_{G}\Gamma_{G}}{(s-M_{G}^{2})^{2} + M_{G}^{2}\Gamma_{G}^{2}}\right]\mathrm{d}s = 1 \,,
\end{equation}
and for $s\neq M_{G}^{2}$
\begin{equation}
\lim_{\Gamma_{G} \to 0} \left[\frac{1}{\pi}\frac{M_{G}\Gamma_{G}}{(s-M_{G}^{2})^{2} + M_{G}^{2}\Gamma_{G}^{2}}\right]=0\,.
\end{equation}
The Breit-Wigner contribution is negligible away from the resonance $s \neq M_G^{2}$.
This approximation applies in the narrow-width regime $\Gamma_G \ll M_G$.

To compute the RIS-subtracted contribution $|\mathcal{M}^{{\rm RIS}}|^{2}$,
we rewrite the squared amplitude for $AB\leftrightarrow CD$ as
\begin{align}
|\mathcal{M}|^{2} & \equiv D_{F}\,M=\frac{1}{\big[(s-M_{G}^{2})^{2}+M_{G}^{2}\Gamma_{G}^{2}\big]}M \nonumber \\
 & \xrightarrow{\Gamma_{G}\ll M_{G}}\frac{\pi}{M_{G}\Gamma_{G}}\delta(s-M_{G}^{2})\,M \equiv D_{elta}\,M\\
 & \xrightarrow{\Gamma_{G}\ll M_{G}}\frac{\pi}{M_{G}\Gamma_{G}}\frac{2\epsilon^{3}/\pi}{\big[(s-M_{G}^{2})^{2}+\epsilon^{2}\big]^{2}}\,M
  \equiv D_{{\rm RIS}}\,M\,.
\end{align}
where $M$ is the numerator of the amplitude square. In the third
line we express the Dirac function using another analytic form
for $s$ away from $M_{G}^{2}$ and $\epsilon\equiv\Gamma_{G}/M_{G}$.
The RIS-subtracted amplitude square for the process $AB\leftrightarrow CD$
can thus be computed in either of the following ways:
\begin{align}
|\mathcal{M}^{{\rm RIS}-}|^{2} =|\mathcal{M}|^{2}-|\mathcal{M}^{{\rm RIS}}|^{2} &=(D_{F}-D_{elta})M\,,\\
{\rm or}\quad &=(D_{F}-D_{{\rm RIS}})M\,,
\end{align}
which is valid for $s$ away from $M_{G}^{2}$ and removes the resonant contribution.
The thermally averaged cross section is then computed using the RIS-subtracted squared amplitude $|\mathcal{M}^{{\rm RIS}-}|^{2}$.

Here are two remarks on the RIS subtraction:
\begin{itemize}
  \item RIS subtraction is required only for $s$-channel processes.
  \item For $AB \leftrightarrow CD$ with multiple intermediate states $G_i$ ($i>1$),
        the interference terms in the squared amplitude do not lead to double counting;
        hence RIS subtraction is not needed.
\end{itemize}

\section{Multi-temperature universe and evolution of hidden sector particles}\label{App:MultiT}

In this Appendix we review the derivation of the evolution of
hidden sector particles and the hidden sector temperature~\cite{Aboubrahim:2020lnr},
for the one hidden sector extension of the SM.
If the hidden sector interacts only feebly with the visible sector,
it can maintain a separate temperature $T_h$, distinct from the visible sector temperature $T$ of the observed Universe.
From Friedmann equations we have
\begin{equation}
\frac{{\rm d}\rho_{h }}{{\rm d}t}+3H\left(\rho_{h }+p_{h }\right)  =j_{h }\,,\label{EOCh}
\end{equation}
where $\rho_h (p_{h})$ is the hidden sector energy density (pressure)
determined by all of the particles inside the hidden sector.

The total energy density in the universe $\rho$ is a sum of $\rho_h$ and  $\rho_v$ (the energy density in the visible sector),
giving rise to the Hubble parameter depending on both $T$ and $T_{h}$
\begin{equation}
H^{2}=\frac{8\pi G_{N}}{3}\big[\rho_{v}(T)+\rho_{h}(T_{h})\big]
= \frac{8\pi G_{N}}{3} \left(\frac{\pi^{2}}{30}g_{{\rm eff}}^{v}T^{4} + \frac{\pi^{2}}{30}g_{{\rm eff}}^{h}T_{h}^{4} \right)\,,\label{hubble}
\end{equation}
where $g_{{\rm eff}}^{v}\,(g_{{\rm eff}}^{h})$ is the visible (hidden) effective degrees of freedom.
Further one can derive the derivative of $\rho_v,\rho_h$ w.r.t. $T_h$
\begin{align}
\rho\frac{{\rm d}\rho_{h}}{{\rm d}T_{h}}&=\left(\frac{\zeta_{h}}{\zeta}\rho_{h}-\frac{j_{h}}{4H\zeta}\right)\frac{{\rm d}\rho}{{\rm d}T_{h}}\,,\\
\frac{{\rm d}\rho_{v}}{{\rm d}T_{h}}&=\frac{\zeta\rho_{v}+\rho_{h}(\zeta-\zeta_{h})+j_{h}/(4H)}{\zeta_{h}\rho_{h}-j_{h}/(4H)}\frac{{\rm d}\rho_{h}}{{\rm d}T_{h}}\,,
\end{align}
where $\zeta=\frac{3}{4}(1+p/\rho)$. Here $\zeta=1$ is for the radiation
dominated era and $\zeta=3/4$ for the matter dominated universe.
The total entropy of the universe is conserved and thus
\begin{align}
s & =\frac{2\pi^{2}}{45}\big(h_{{\rm eff}}^{v}T^{3}+h_{{\rm eff}}^{h}T_{h}^{3}\big)\,,\label{entropy}
\end{align}
where $h_{{\rm eff}}^{v}\,(h_{{\rm eff}}^{h})$ is the visible (hidden) effective entropy degrees of freedom.

With manipulations of the above equations,
the evolution of the ratio of temperatures for the two sectors $\eta(T_h)=T/T_{h}$ can be derived as
\begin{equation}
\frac{{\rm d}\eta}{{\rm d} T_{h}}=-\frac{A_{v}}{B_{v}}+
\frac{\zeta\rho_{v}+\rho_{h}(\zeta-\zeta_{h})+j_{h}/(4H)}{B_{v} [ \zeta_{h}\rho_{h}-j_{h}/(4H) ]}
\frac{{\rm d}\rho_{h}}{{\rm d}T_{h}}\,,\label{y30}
\end{equation}
with
\begin{equation}
A_{v} =\frac{\pi^{2}}{30}\Big(\frac{{\rm d}g_{{\rm eff}}^{v}}{{\rm d}T}\eta^{5}T_{h}^{4}+4g_{{\rm eff}}^{v}\eta^{4}T_{h}^{3}\Big)\,,\qquad
B_{v} =\frac{\pi^{2}}{30}\Big(\frac{{\rm d}g_{{\rm eff}}^{v}}{{\rm d}T}\eta^{4}T_{h}^{5}+4g_{{\rm eff}}^{v}\eta^{3}T_{h}^{4}\Big)\,.
\end{equation}

The complete evolution of hidden sector particles can be obtained
by solving the combined differential equations of  the evolution equation for $\eta(T_h)$, together with
the modified (temperature dependent) Boltzmann equations for the hidden sector particles, given by
\begin{equation}
\frac{\mathrm{d}Y_{i}}{\mathrm{d}T_{h}}  =-\frac{s}{H}\frac{{\rm d}\rho_{h}/{\rm d}T_{h}}{4\rho_{h}-j_{h}/H}
\sum  \big( - Y_i^2 \langle \sigma v \rangle^{T_h}_{i \bar{i} \to \times \times} + \cdots \big)\,,
\end{equation}
where $Y_i$ is the comoving number density of the $i$th hidden sector particle
and the sum is over various scattering or decay processes involving this particle,
which may depend on $T_h$ (as shown by the example ${\bar{i} i \to \times \times}$) or depend on the visible sector $T$.
The plus (minus) sign corresponds to increasing (decreasing) the number of the $i$th particle by one unit.

The complete Boltzmann equations for hidden sector particles as well as hidden sector temperatures for the freeze-in case
are given in Section~\ref{sec:BE}.

\newpage

\section{Seesaw benchmarks}\label{App:SSBM}

\paragraph{Benchmark-RS} The regular seesaw benchmark parameters with electroweak right-handed neutrinos, c.f., Eq.~(\ref{PCaseRS}),
are given by 9 Yukawa couplings $y^\nu_{ij}$ in Eq.~(\ref{eq:RSL})
\begin{table}[!h]
		\begin{center}
			\begin{tabular}{|c|c|c|c|}
				\hline
				& $\mathbb{N}_1^0$               & $\mathbb{N}_2^0$               & $\mathbb{N}_3^0$               \\	\hline
				$\nu_e$    & $2.000\times10^{-7}$ & $1.913\times10^{-8}$ & $4.473\times10^{-8}$ \\ \hline
				$\nu_\mu$  & $5.248\times10^{-8}$ & $4.894\times10^{-7}$ & $9.101\times10^{-8}$ \\ \hline
				$\nu_\tau$ & $1.372\times10^{-7}$ & $2.560\times10^{-7}$ & $3.511\times10^{-7}$  \\ \hline
			\end{tabular}
		\end{center}
		\label{YukawaRS}
\end{table}

\noindent
and three Majorana neutrino masses 
$M_1 = 210$~GeV, $M_2 = 240$~GeV, $M_3 = 270$~GeV,
giving rise to the following seesaw mixing
\begin{table}[!h]
	\footnotesize
	\begin{center}
			\begin{tabular}{|c|c|c|c|c|c|c|}
				\hline
				& $\mathbb{V}_{1}$     & $\mathbb{V}_{2}$    & $\mathbb{V}_{3}$    & $\mathbb{N}_{1}$     & $\mathbb{N}_{2}$     & $\mathbb{N}_{3}$    \\ \hline
				$\mathbb{V}_1^0$       & 0.8276                 &-0.5416               & 0.1474                & $1.657\times10^{-7}$  & $1.386\times10^{-8}$  & $2.882\times10^{-8}$  \\ \hline
				$\mathbb{V}_2^0$       & 0.2651                 & 0.6087                & 0.7478                & $4.347\times10^{-8}$  & $3.547\times10^{-7}$  & $5.863\times10^{-8}$  \\ \hline
				$\mathbb{V}_3^0$       &-0.4947                 &-0.5798                & 0.6473                & $1.137\times10^{-7}$ & $1.855\times10^{-7}$  & $2.262\times10^{-7}$   \\  \hline
				$\mathbb{N}_1^0$        &-$9.239\times10^{-8}$  & $1.292\times10^{-7}$ & -$1.305\times10^{-7}$ & 1.000                 & $2.716\times10^{-13}$ & $1.156\times10^{-13}$  \\	\hline
				$\mathbb{N}_2^0$       &-$1.372\times10^{-8}$  &-$1.008\times10^{-7}$ & -$3.874\times10^{-7}$ &-$3.104\times10^{-13}$ & 1.000                 & $5.052\times10^{-13}$  \\ \hline
				$\mathbb{N}_3^0$       & $7.251\times10^{-8}$  & $1.111\times10^{-7}$ & -$1.945\times10^{-7}$ &-$1.487\times10^{-13}$ &-$5.684\times10^{-13}$ & 1.000                  \\ \hline
			\end{tabular}
		\end{center}
		\label{MixingRS}
\end{table}	

\paragraph{Benchmark-SC} The structure cancellation with electroweak right-handed neutrinos has in total 17 parameters, c.f., Eq.~(\ref{PCaseSC}),
with three Majorana masses $210.4014, 467.4182, 530.5053$~GeV, two Yukawa couplings $Y_1 = 1.4652\times10^{-3}$, $Y_2 = 2.3950\times10^{-3}$,
two parameters $A = 1.0162 $, $B = 2.4227$,
a fine-tuned small parameter $\epsilon = 7.5059\times 10^{-9}$,
and a $3\times 3$ matrix $X$
\begin{table}[!h]
		\begin{center}
			\begin{tabular}{|c|c|c|}
				\hline
				-0.8153    & -0.2986    & -0.7415   \\ \hline
				0.1185     & -0.4458    & 0.8574    \\ \hline
				0.1560     & -0.1994    & 0.6003 \\ \hline
			\end{tabular}
		\end{center}
		\label{YukawaSC}
	\end{table}	

\noindent
giving rise to the following fine-tuned Yukawa couplings
\begin{table}[h!]
	\setlength{\tabcolsep}{8pt}
	\begin{center}
		\begin{tabular}{|c|c|c|c|}
			\hline
			& $\mathbb{N}_1^0$               & $\mathbb{N}_2^0$               & $\mathbb{N}_3^0$                 \\ \hline
			$\nu_e$    & $1.46\times10^{-3}$  & $3.13\times10^{-3}$  & $2.39\times10^{-3}$   \\ \hline
			$\nu_\mu$  & $1.50\times10^{-3}$  & $3.18\times10^{-3}$  & $2.43\times10^{-3}$   \\ \hline
			$\nu_\tau$ & $3.55\times10^{-3}$  & $7.59\times10^{-3}$  & $5.80\times10^{-3}$   \\ \hline
		\end{tabular}
	\end{center}
	\label{YukawaSC}
\end{table}

\noindent
and the final seesaw mixings for Case-SC are given by
\begin{table}[h!]
	\footnotesize
		\begin{center}
			\begin{tabular}{|c|c|c|c|c|c|c|}
				\hline
				& $\mathbb{V}_{1}$    & $\mathbb{V}_{2}$    & $\mathbb{V}_{3}$    & $\mathbb{N}_{1}$    & $\mathbb{N}_{2}$    & $\mathbb{N}_{3}$    \\ \hline
				$\mathbb{V}_1^0$   & 0.8109                &-0.5658                & 0.1490                & 0.0012 & $7.852\times10^{-4}$ &-0.0012 \\ \hline
				$\mathbb{V}_2^0$   & 0.3390                & 0.6620                & 0.6685                & 0.0012 &$7.979\times10^{-4}$ &-0.0012 \\ \hline
				$\mathbb{V}_3^0$  &-0.4769                &-0.4916                & 0.7286                & 0.0029 &0.0019 &-0.0028 \\ \hline
				$\mathbb{N}_1^0$  &$2.866\times10^{-11}$& 0.0013 &-0.0031 & 1.000                & $4.941\times10^{-6}$ & $3.466\times10^{-6}$ \\ \hline
				$\mathbb{N}_2^0$  &-$1.377\times10^{-13}$& $8.513\times10^{-4}$ &-0.0020 &-$1.246\times10^{-5}$ & 1.000 & $3.848\times10^{-6}$		 \\ \hline
				$\mathbb{N}_3^0$  &-$1.329\times10^{-11}$&-0.0013 & 0.0030 & $7.700\times10^{-6}$ & $3.390\times10^{-6}$ & 1.000 \\ \hline
			\end{tabular}
		\end{center}
		\label{MixingSC}
	\end{table}

\paragraph{Benchmark-SS}
For Case-SS, two of the right-handed neutrinos are responsible for generating the
light neutrino masses and mixings via the seesaw mechanism, while the third
(assumed to be $\mathbb{N}_1$) couples ultraweakly to the SM.
Therefore, the couplings $y^\nu_{i1}$ are chosen by hand.
Consequently,
the $\mathbb{N}_1$ portal to the dark sector can only generate the freeze-in production of dark particles.

We take three benchmarks for the Case-SS analysis, with different magnitudes of the  $y_1^\nu$ coupling constant.
All three benchmarks have 12 input parameters (9 Yukawa couplings and 3 Majorana masses).\\

\noindent
{\bf Benchmark-SS1:} The three Majorana masses are $360, 400, 507$~GeV and 9~Yukawa couplings are given by
\begin{table}[h!]
	\begin{center}
		\begin{tabular}{|c|c|c|c|}
			\hline
			& $\mathbb{N}_{1}^{0}$ & $\mathbb{N}_{2}^{0}$ & $\mathbb{N}_{3}^{0}$\tabularnewline
			\hline
			$\nu_{e}$ & $5\times10^{-11}$ & $2.255\times10^{-7}$ & $3.418\times10^{-8}$\tabularnewline
			\hline
			$\nu_{\mu}$ & $6\times10^{-11}$ & $4.201\times10^{-9}$ & $6.552\times10^{-7}$\tabularnewline
			\hline
			$\nu_{\tau}$ & $7\times10^{-11}$ & $3.826\times10^{-7}$ & $5.456\times10^{-7}$\tabularnewline
			\hline
		\end{tabular}
	\end{center}
\end{table}

\noindent
resulting in the seesaw mixing
\begin{table}[h!]
	\footnotesize
	\begin{center}
		\begin{tabular}{|c|c|c|c|c|c|c|}
			\hline
			& $\mathbb{V}_{1}$ & $\mathbb{V}_{2}$ & $\mathbb{V}_{3}$ & $\mathbb{V}_{4}$ & $\mathbb{V}_{5}$ & $\mathbb{V}_{6}$\tabularnewline
			\hline
			$\mathbb{V}_{1}^{0}$ & $0.8034$ & $-0.5779$ & $0.1430$ & $2.416\times10^{-11}$ & $1.173\times10^{-8}$ & $9.807\times10^{-8}$\tabularnewline
			\hline
			$\mathbb{V}_{2}^{0}$ & $0.3557$ & $0.6586$ & $0.6631$ & $2.899\times10^{-11}$ & $2.248\times10^{-7}$ & $1.827\times10^{-9}$\tabularnewline
			\hline
			$\mathbb{V}_{3}^{0}$ & $-0.4774$ & $-0.4819$ & $0.7347$ & $3.823\times10^{-11}$ & $1.872\times10^{-7}$ & $1.664\times10^{-7}$\tabularnewline
			\hline
			$\mathbb{V}_{4}^{0}$ & $-1.358\times10^{-11}$ & $1.117\times10^{-11}$ & $-4.753\times10^{-11}$ & $1.000$ & $3.218\times10^{-17}$ & $7.245\times10^{-17}$\tabularnewline
			\hline
			$\mathbb{V}_{5}^{0}$ & $1.027\times10^{-15}$ & $-5.105\times10^{-8}$ & $-2.883\times10^{-7}$ & $-4.531\times10^{-17}$ & $1.000$ & $-1.551\times10^{-13}$\tabularnewline
			\hline
			$\mathbb{V}_{6}^{0}$ & $1.496\times10^{-15}$ & $1.357\times10^{-7}$ & $-1.375\times10^{-7}$ & $-8.050\times10^{-17}$ & $1.224\times10^{-13}$ & $1.000$\tabularnewline
			\hline
		\end{tabular}
	\end{center}
\end{table}

\noindent
{\bf Benchmark-SS2:} The three Majorana masses are again taken $360, 400, 507$~GeV. Yukawa couplings and the resulting seesaw mixing are given by
\begin{table}[h!]
\begin{center}
\begin{tabular}{|c|c|c|c|}
\hline
 & $\mathbb{N}_{1}^{0}$ & $\mathbb{N}_{2}^{0}$ & $\mathbb{N}_{3}^{0}$\tabularnewline
\hline
$\nu_{e}$ & $4\times10^{-13}$ & $2.255\times10^{-7}$ & $3.418\times10^{-8}$\tabularnewline
\hline
$\nu_{\mu}$ & $6\times10^{-13}$ & $4.201\times10^{-9}$ & $6.522\times10^{-7}$\tabularnewline
\hline
$\nu_{\tau}$ & $8\times10^{-13}$ & $3.826\times10^{-7}$ & $5.456\times10^{-7}$\tabularnewline
\hline
\end{tabular}
\end{center}
\end{table}

\noindent
resulting in the seesaw mixing
\begin{table}[h!]
\footnotesize
\begin{center}
\begin{tabular}{|c|c|c|c|c|c|c|}
\hline
 & $\mathbb{V}_{1}$ & $\mathbb{V}_{2}$ & $\mathbb{V}_{3}$ & $\mathbb{V}_{4}$ & $\mathbb{V}_{5}$ & $\mathbb{V}_{6}$\tabularnewline
\hline
$\mathbb{V}_{1}^{0}$ & $0.8034$ & $-0.5779$ & $0.1430$ & $1.933\times10^{-13}$ & $9.807\times10^{-8}$ & $1.1728\times10^{-8}$\tabularnewline
\hline
$\mathbb{V}_{2}^{0}$ & $0.3557$ & $0.6586$ & $0.6631$ & $2.899\times10^{-13}$ & $1.827\times10^{-9}$ & $2.248\times10^{-7}$\tabularnewline
\hline
$\mathbb{V}_{3}^{0}$ & $-0.4774$ & $-0.4819$ & $0.7347$ & $3.866\times10^{-13}$ & $1.664\times10^{-7}$ & $1.872\times10^{-7}$\tabularnewline
\hline
$\mathbb{V}_{4}^{0}$ & $-7.386\times10^{-14}$ & $1.071\times10^{-13}$ & $-5.039\times10^{-13}$ & $1.000$ & $7.543\times10^{-19}$ & $3.425\times10^{-19}$\tabularnewline
\hline
$\mathbb{V}_{5}^{0}$ & $8.154\times10^{-20}$ & $1.357\times10^{-7}$ & $-1.375\times10^{-7}$ & $-8.381\times10^{-19}$ & $1.000$ & $1.224\times10^{-13}$\tabularnewline
\hline
$\mathbb{V}_{6}^{0}$ & $6.099\times10^{-20}$ & $-5.105\times10^{-8}$ & $-2.883\times10^{-7}$ & $-4.823\times10^{-19}$ & $-1.551\times10^{-13}$ & $1.000$\tabularnewline
\hline
\end{tabular}
\end{center}
\end{table}

\newpage
\noindent
{\bf Benchmark-SS3:} The three Majorana masses are taken $225, 400, 507$~GeV and 9~Yukawa couplings are given by
\begin{table}[h!]
\begin{center}
\begin{tabular}{|c|c|c|c|}
\hline
 & $\mathbb{N}_{1}^{0}$ & $\mathbb{N}_{2}^{0}$ & $\mathbb{N}_{3}^{0}$\tabularnewline
\hline
$\nu_{e}$ & $1\times10^{-10}$ & $2.255\times10^{-7}$ & $3.418\times10^{-8}$\tabularnewline
\hline
$\nu_{\mu}$ & $2\times10^{-10}$ & $4.201\times10^{-9}$ & $6.552\times10^{-7}$\tabularnewline
\hline
$\nu_{\tau}$ & $3\times10^{-10}$ & $3.826\times10^{-7}$ & $5.456\times10^{-7}$\tabularnewline
\hline
\end{tabular}
\end{center}
\end{table}

\noindent
This gives rise to the following seesaw mixing
\begin{table}[h!]
\footnotesize
\begin{center}
\begin{tabular}{|c|c|c|c|c|c|c|}
\hline
 & $\mathbb{V}_{1}$ & $\mathbb{V}_{2}$ & $\mathbb{V}_{3}$ & $\mathbb{V}_{4}$ & $\mathbb{V}_{5}$ & $\mathbb{V}_{6}$\tabularnewline
\hline
$\mathbb{V}_{1}^{0}$ & $0.8035$ & $-0.5779$ & $0.1430$ & $7.731\times10^{-11}$ & $1.173\times10^{-8}$ & $9.807\times10^{-8}$\tabularnewline
\hline
$\mathbb{V}_{2}^{0}$ & $0.3557$ & $0.6586$ & $0.6631$ & $1.546\times10^{-10}$ & $2.248\times10^{-7}$ & $1.827\times10^{-9}$\tabularnewline
\hline
$\mathbb{V}_{3}^{0}$ & $-0.4774$ & $-0.4819$ & $0.7347$ & $2.319\times10^{-10}$ & $1.872\times10^{-7}$ & $1.664\times10^{-7}$\tabularnewline
\hline
$\mathbb{V}_{4}^{0}$ & $-6.384\times10^{-12}$ & $5.462\times10^{-11}$ & $-2.840\times10^{-10}$ & $1.000$ & $6.312\times10^{-17}$ & $5.973\times10^{-17}$\tabularnewline
\hline
$\mathbb{V}_{5}^{0}$ & $1.905\times10^{-15}$ & $-5.105\times10^{-8}$ & $-2.883\times10^{-7}$ & $-1.422\times10^{-16}$ & $1.000$ & $-1.551\times10^{-13}$\tabularnewline
\hline
$\mathbb{V}_{6}^{0}$ & $2.355\times10^{-15}$ & $1.357\times10^{-7}$ & $-1.375\times10^{-7}$ & $-1.062\times10^{-16}$ & $1.224\times10^{-13}$ & $1.000$\tabularnewline
\hline
\end{tabular}
\end{center}
\end{table}

\end{document}